\documentclass[manyauthors]{fundam}

\usepackage{tikz-cd}

\usepackage{amssymb,stmaryrd,bbm}
\newtheorem{notation}{Notation}

\usepackage{tabularray}

\usepackage{float}
\usepackage{graphicx}
\usepackage{subcaption}
\usepackage{tikz}
\usepackage{tikz-cd}
\usepackage{circuitikz}
\usetikzlibrary{petri,positioning, calc,decorations.pathmorphing,shapes,cd,arrows}

\usepackage{caption}
\newcommand{\arrowflow}{
\tikz \draw[-{to}] (-1pt,0) -- (1pt,0);
}

\newcommand{\flow}[4]{ 
	\draw[#3] (#1) to [#4] node [pos=0.5] {\arrowflow} (#2);
}
\newcommand{\flowdiag}[5]{ 
	\draw[#3] (#1) to [#4] node [#5] {\arrowflow} (#2);
}

\tikzset{fill fraction/.style n args={1}{path picture={
 \fill[black] (path picture bounding box.south west) rectangle
 ($(path picture bounding box.north west)!0.5!(path picture bounding box.north
 east)$);}}
}
\newcommand{\qmatch}[4]{ 
	\node[draw,fill=white,fill fraction={black}{0.5},inner sep=1.2pt,minimum size=3pt,label=above:{\scriptsize #4}] (#1) at (#2,#3) {};
}
\newcommand{\dmatch}[5]{ 
	\node[circle,draw,fill=white,fill fraction={black}{0.5},inner sep=1.2pt,minimum size=3pt,label=#5:{\scriptsize #4}] (#1) at (#2,#3) {};
}
\newcommand{\qinplain}[4]{ 
\node[draw=black,fill=white,inner sep=0pt,minimum size=3pt,label=left:\scriptsize #4] (#1) at (#2,#3) {};
}
\newcommand{\qin}[4]{ 
\flow{{#2,#3}}{$(#2,#3)+(0.8,0)$}{dashed}{};
\qinplain{#1}{#2}{#3}{#4};
}
\newcommand{\dinplain}[4]{ 
\node[circle,draw=black,fill=white,inner sep=0pt,minimum size=3pt,label=left:\scriptsize #4] (#1) at (#2,#3) {};
}
\newcommand{\din}[4]{ 
\flow{{#2,#3}}{$(#2,#3)+(0.8,0)$}{}{};
\dinplain{#1}{#2}{#3}{#4};
}
\newcommand{\qoutplain}[4]{ 
\node[fill=black,inner sep=0pt,minimum size=3pt,label=right:\scriptsize #4] (#1) at (#2+0.8,#3) {};
}
\newcommand{\qout}[4]{ 
\flow{{#2,#3}}{$(#2,#3)+(0.8,0)$}{dashed}{};
\qoutplain{#1}{#2}{#3}{#4};
}
\newcommand{\doutplain}[4]{ 
\node[circle,fill=black,inner sep=0pt,minimum size=3pt,label=right:\scriptsize #4] (#1) at (#2+0.8,#3) {};
}
\newcommand{\dout}[4]{ 
\flow{{#2,#3}}{$(#2,#3)+(0.8,0)$}{}{};
\doutplain{#1}{#2}{#3}{#4};
}

\newcommand{\computonPrimitive}[5]{ 
\draw[draw=black,fill=white] (#1,#2) rectangle ++(#3,#4) node[pos=0.5]{#5};
}
\newcommand{\computonComposite}[4]{ 
\draw[draw=black,fill=black!2] (#1,#2) rectangle ++(#3,#4);
}

\begin{document}
\title{Computons for Control-Driven Boolean Circuits}
\author{Damian Arellanes\\Lancaster University}
\maketitle
\address{D.Arellanes, Infolab21, Lancaster University, Bailrigg, Lancaster LA1 4WA}

\runninghead{D. Arellanes}{Computons for Control-Driven Boolean Circuits}

\begin{abstract}
Boolean circuits abstract away from hardware aspects to capture the logical structure and computational behaviour of digital components. Despite their long history of study, their compositional foundations have received comparatively little attention. Building on the theoretical foundations of computons introduced in previous work, we develop a concrete and operationally defined model of high-level computation for control-driven Boolean circuits. We describe colimit-based operators for composing such circuits into sequential, parallel, branching and iterative structures, supported by concrete operational semantics grounded in abstract computing devices. The framework refines the semantics of composition, establishes the algebraic properties of its operators and introduces atomic circuits to support more expressive computation units. We demonstrate the applicability of the proposed framework through examples from propositional logic, biomedicine and electronics, and establish an operational equivalence with respect to classical Boolean circuits through a systematic transformation that accommodates gates of arbitrary arity. By this equivalence, we show that control-driven Boolean circuits are at least as powerful as their classical counterpart. That is, they are capable of non-uniformly computing any Boolean function over inputs of arbitrary length. Implementations in Idris and Java validate the practical realisability of the framework. 
\end{abstract}

\section{Introduction}

Boolean circuits are a fundamental model of computation in which computational problems, including the recognition of formal languages, are represented by collections of circuits indexed by input length. Such collections constitute a non-uniform model of computation, as each circuit may be designed independently for inputs of a given size \cite{clote_boolean_2002}. The construction of arbitrary Boolean functions relies on a functionally complete set of logical connectives, with common examples including $\{\land,\lor,\lnot\}$, $\{\implies,\lnot\}$ and the singleton basis $\{\uparrow\}$ \cite{wegener_complexity_1987}.

Despite their computational power, classical Boolean circuit families offer limited support for a systematic algebraic treatment of compositional construction. In particular, although non-uniform circuit families can encode arbitrary language membership decisions through input-length-specific circuit descriptions, this expressive capability does not provide a principled framework for constructing circuits from smaller ones using well-defined composition operators with appropriate closure properties. Here, composition refers to the construction of complex computing structures from simpler ones while abstracting away their internal details \cite{ghica_complete_2025}. A compositionality theory would therefore provide a foundation for modular construction and reasoning, enabling circuits to be treated as reusable, self-similar computational components across distinct application domains \cite{giannakopoulou_compositional_2018}.

One important application of compositionality is that of compositional verification, which establishes properties of a composite by reasoning about constituents to demonstrate that relevant properties can be preserved under composition. Particularly, a property can be established for a composite circuit $C$ when it holds for the constituents and remains invariant under the composition operations used to construct $C$ \cite{dardik_recomposition_2024}. A formal theory of circuit composition can consequently support not only modular circuit construction but also the development of systematic techniques for formal analysis.

Motivated by the observation that every computing device exhibits implicit or explicit control flow, irrespective of how they exchange data \cite{arellanes_models_2025,arellanes_compositional_2026}, this work introduces a formal framework for compositional Boolean circuits in which control and data flow are explicitly represented. The proposed framework employs colimit-based composition operators to combine circuits by control flow. Four classes of elementary operators are introduced to capture sequential composition, parallel composition, branching and iteration, operators that respectively support the ordered execution of computing devices, their simultaneous activation, non-deterministic computational paths and repeated execution. For sequencing, parallelising and branching, we study algebraic properties concerning identity, commutativity and associativity. We further demonstrate that the proposed model retains the computational power of classical Boolean circuits, i.e., every Boolean function can be computed within this framework. We refer to the resulting model as (families of) control-driven Boolean circuits.

The remainder of this paper is organised as follows. Sections \ref{sec:model}, \ref{sec:operators} and \ref{sec:operation} introduce the notion of the class of circuits we propose, describe colimit-based operators for their compositional construction and establish operational semantics, respectively. Section \ref{sec:equivalence} examines the computational power of control-driven circuits in relation to their classical counterpart. Section \ref{sec:applications} describes a number of examples of circuits in the domains of propositional logic, biomedicine and electronics. Section \ref{sec:related-work} discusses the most relevant related approaches and situates the proposed framework within the existing literature. Finally, Section \ref{sec:conclusion} summarises the main contributions and identifies future research directions.

\section{Control-Driven Boolean Circuits}\label{sec:model}

A computon is a fundamental unit of composition, formed by finite sets $U$, $P$, $I$, $O$, $\Sigma$ and $D$ of computation units, ports, inflows, outflows, colours and abstract computing devices, respectively. A port $p \in P$ is esentially a construct assigned to a specific colour $c(p) \in\Sigma$, serving as buffer for coloured values. Inflows/ outflows serve to link ports to/from each computation unit $u \in U$, indicating how information travels into/from $u$. A unit $u$ encapsulates a collection of (potentially interrelated) computing devices which consume input values from ports attached to $u$, perform some (halting) computation and produce exactly one output on a port connected from $u$. Connectivity is determined by the spans of functions described in Definition \ref{def:computon}, namely $U\xleftarrow{\tau} I \xrightarrow{s} P$ and $U\xleftarrow{\sigma} O \xrightarrow{t} P$. The relationship between a device and its inputs/output is given by the functions $f\colon O\to D$ and $r\colon I\to O$. 

\begin{definition}\label{def:computon}
A computon $\lambda$ is a diagram in the category of finite sets and total functions, which has the following shape:
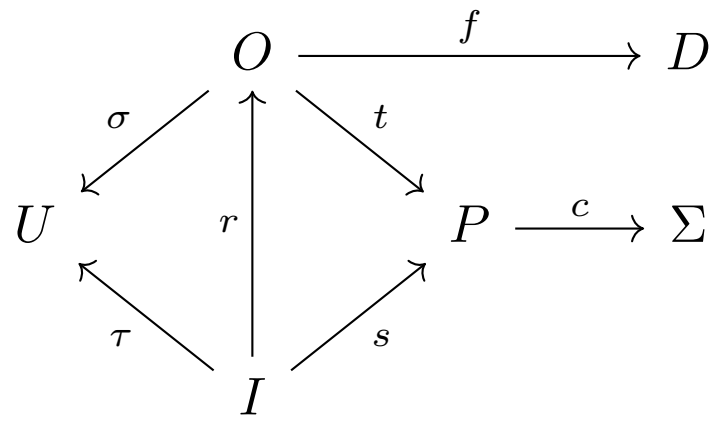
\[
\begin{tikzcd}[ampersand replacement=\&]
 \& O \arrow[dl,"\sigma"']\arrow[dr, "t"]\arrow[rr, "f"] \& \& D  \\
U \& \& P \arrow[r,"c"] \& \Sigma \\
 \& I \arrow[ul,"\tau"]\arrow[ur, "s"']\arrow[uu, "r"] \& \& \\
\end{tikzcd}
\] such that:
\begin{itemize}
\item $P\neq\emptyset$,
\item $\Sigma\subset\mathbb{N}$ with $0\in\Sigma$,
\item $\sigma$, $\tau$, $r$, ${\tau\restriction_{(c\circ s)^{-1}(0)}}$ and ${\sigma\restriction_{(c\circ t)^{-1}(0)}}$ are surjective,
\item ${\sigma\circ r = \tau}$, and
\item there are ports ${p \in P\setminus t(O)}$ and ${q \in P\setminus s(I)}$ with ${c(p)=0=c(q)}$.
\end{itemize}
\end{definition}

\begin{notation}\label{not:pre-post-sets}
Given a computation unit ${u \in U}$ and a port ${p\in P}$, we use ${\bullet u}$, ${u\bullet}$, ${\bullet p}$ and ${p\bullet}$ to denote ${s(\tau^{-1}(u))}$, ${t(\sigma^{-1}(u))}$, ${\sigma(t^{-1}(p))}$ and ${\tau(s^{-1}(p))}$, correspondingly.
\end{notation}

Particularly, the first condition in Definition \ref{def:computon} indicates that computons always have ports. The second one says that the $0$-type is always present, which represents control signals. The last condition states that there always are $0$-typed ports $p$ and $q$ with no incoming and outgoing flows, respectively, serving to indicate the control inputs and control outputs of a computon. Outflows and inflows are attached to computation units via the functions $\sigma$ and $\tau$, whereas $s$ and $t$ define the source and target ports of inflows and outflows. Computation units always have $0$-typed ports attached to them due to the surjectivity conditions on ${\tau\restriction_{(c\circ s)^{-1}(0)}}$ and ${\sigma\restriction_{(c\circ t)^{-1}(0)}}$. The function $r$ simply establishes a relationship between inflows and outflows, and the equation ${\sigma\circ r = \tau}$ constitutes an encapsulation principle for input-output relations. That is, although a device $d\in D$ can be encapsulated by different units $u_1,\ldots,u_j\in U$, the execution of $d$ within $u_k$ will take inputs from the ports in $\bullet u_k$ and place a single output in a port from $u_k\bullet$ for $k\in\{1,\ldots,j\}$.

Definition \ref{def:computon} was initially proposed in \cite{arellanes_compositional_2026} and later extended in \cite{arellanescolimit2026} and \cite{arellanesboolean2026}. In this paper, we build upon it to propose a particular class of computons referred to as \emph{control-driven Boolean circuits} (or circuits for short) whose definition is given below. Graphical examples are presented in Section \ref{sec:applications}.

\begin{definition}\label{def:circuit}
Let $\llbracket\mathbbm{1}\rrbracket$ be the set-theoretic interpretation $\{*, \bot_{\mathbbm{1}}\}$ of the unit type $\mathbbm{1}$ extended with the empty value $\bot_{\mathbbm{1}}$ and $\llbracket\mathbb{B}\rrbracket$ be the set-theoretic interpretation $\{T,F, \bot_{\mathbb{B}}\}$ of the Boolean type $\mathbb{B}$ extended with the empty Boolean value $\bot_{\mathbb{B}}$. A control-driven Boolean circuit is a computon where $\Sigma:=\{0,1\}$ and $D:=\{\kappa_m\}_{m\geq 1}\cup\{\uparrow_n\}_{n\geq 1}\cup\{\epsilon_o\}_{o\geq 1}$ such that:
\begin{itemize}
\item each $\kappa_m\colon\llbracket\mathbbm{1}\rrbracket^m\to\llbracket\mathbbm{1}\rrbracket$ is an $m$-ary function for echoing control signals, given by: \begin{align*}
  \kappa_m(x_1,\ldots,x_m) 
    &=\begin{cases}
       \bot_{\mathbbm{1}} & x_j=\bot_{\mathbbm{1}}\text{ for some }j\in\{1,\ldots,m\} \\
        * & \text{otherwise}
    \end{cases}
\end{align*}
\item each $\uparrow_n\colon\llbracket\mathbb{B}\rrbracket^n\to\llbracket\mathbb{B}\rrbracket$ is an (extended) $n$-ary NAND function, given by: \begin{align*}
  \uparrow_n(x_1,\ldots,x_n) 
    &=\begin{cases}
       \bot_{\mathbb{B}} & x_j=\bot_{\mathbb{B}}\text{ for some }j\in\{1,\ldots,n\}\\
       \neg\left(\bigwedge_{j=1}^{n} x_j\right) & \text{otherwise}
    \end{cases}
\end{align*}
\item and each $\epsilon_o\colon\llbracket\mathbbm{B}\rrbracket^o\to\llbracket\mathbbm{1}\rrbracket$ is an $o$-ary function for discarding Boolean values, defined as $\epsilon_o(x_1,\ldots,x_o)=*$.
\end{itemize}
\end{definition}

\begin{remark}
It is easy to see that functions $\kappa_m$, $\uparrow_n$ and $\kappa_o$ are idempotent in the sense that repeating an input does not change the result. This observation becomes important when describing operational semantics in Section \ref{sec:operation}.    
\end{remark}

In Section \ref{sec:operation}, we will see that the colours $0$ and $1$ are mapped to the types $\mathbbm{1}$ and $\mathbb{B}$ which represent extended control signals and extended Boolean values, respectively. We say they are extended because they support value absence to handle errors. Thus, a circuit is able to receive (empty) signals and (empty) truth values on their input interface in order to attempt triggering a number of $\kappa_m$-devices (to disregard $m$ control signals and produce new ones), $\uparrow_n$-devices (to compute $n$-ary NAND functions) and $\kappa_o$-devices (to disregard $o$ Boolean values). The result of computation is stored in a so-called output interface.

\begin{definition}\label{def:interface}
The interface of a circuit $\lambda$ is a tuple ${(P^+,P^-)}$ where ${P^+:=P\setminus t(O)}$ and ${P^-:=P\setminus s(I)}$.
\end{definition}

When there is a sequence of flows taking every inport $p\in P^+$ (and every port with an outgoing inflow) to some outport $q\in P^-$, we say that the circuit satisfies the connectivity property. This notion is jointly formalised by Definitions \ref{def:bipartite} and \ref{def:connected}.

\begin{definition}\label{def:bipartite}
The bipartite graph of a circuit $\lambda$ has $U \sqcup P$ and $I \sqcup O$ as the sets of nodes and edges, respectively. The source and target functions ${s',t'\colon I\sqcup O\to U\sqcup P}$ are given as follows:

$\begin{aligned}
  s'(e) &=
    \begin{cases}
      s(e) & e\in I \\
      \sigma(e) & e\in O
    \end{cases}
  \qquad\qquad
  t'(e) &=
    \begin{cases}
      \tau(e) & e\in I \\
      t(e) & e\in O
    \end{cases}
\end{aligned}$
\end{definition}

\begin{definition}\label{def:connected}
If $G$ is the bipartite graph of a circuit $\lambda$, we say that $\lambda$ is connected if, for each ${p\in s(I)\cup P^+}$, there is a finite path ${\langle e_1,e_2,\ldots,e_k \rangle}$ in $G$ of length ${k\geq 2}$, satisfying ${s'(e_1)=p}$ and ${t'(e_k)\in P^-}$.
\end{definition}

A circuit that is not connected, has no flows, no computation units and all ports in $P^+\cap P^-$, is called trivial; whereas a connected circuit with exactly one computation unit is called atomic. These structures are described in Definitions \ref{def:trivial}, \ref{def:unit} and \ref{def:atomic}.

\begin{definition}\label{def:trivial}
A circuit $\lambda$ is trivial if its set-valued diagram has the following shape:
\[
\begin{tikzcd}[ampersand replacement=\&]
 \& \emptyset \arrow[dl,"\sigma"']\arrow[dr, "t"]\arrow[rr, "f"] \& \& D  \\
\emptyset \& \& P \arrow[r,"c"] \& \Sigma \\
 \& \emptyset \arrow[ul,"\tau"]\arrow[ur, "s"']\arrow[uu, "r"] \& \& \\
\end{tikzcd}
\]
\end{definition}

\begin{definition}\label{def:unit}
A unit circuit is a trivial circuit with $|P|=1$.
\end{definition}

\begin{definition}\label{def:atomic}
An atomic circuit $\lambda$ is a circuit with $|U|=1$. 
\end{definition}

\section{Colimit-Based Composition Operators}\label{sec:operators}

In this section, we present operators for composing control-driven circuits, whose behaviour is characterised as finite colimit constructions. Before describing colimits, we present the notion of a circuit morphism which intuitively is an embedding of a circuit into a (potentially) more complex one, with the restriction that new computation units can only be attached at the interface, as captured by the constraint presented at the bottom of Definition \ref{def:morphism}, which results in Lemma \ref{lem:interface-preservation}. When a morphism identifies all the inports of a circuit, we call it \emph{in-marker}; and when it identifies all the outports we refer to it as \emph{out-marker}. These notions are formalised in Definition \ref{def:markers}.

\begin{definition}\label{def:morphism}
A circuit morphism ${\alpha\colon\lambda_1\rightarrow\lambda_2}$ consists of six functions, namely ${\alpha_U\colon U_1\rightarrow U_2}$, ${\alpha_P\colon P_1\rightarrow P_2}$, ${\alpha_I\colon I_1\rightarrow I_2}$, $\alpha_O\colon O_1\rightarrow O_2$, ${\alpha_\Sigma\colon\Sigma_1\hookrightarrow\Sigma_2}$ and ${\alpha_D\colon D_1\hookrightarrow D_2}$, which make the following diagrams commute:\[
\begin{tikzcd}
I_1 \arrow[r,"\tau_1"]\arrow[d, "\alpha_I"']
& U_1 \arrow[d, "\alpha_U"] 
& O_1 \arrow[l, "\sigma_1"']\arrow[d, "\alpha_O"] \\
I_2 \arrow[r, "\tau_2"']
& U_2 
& O_2 \arrow[l, "\sigma_2"]
\end{tikzcd}
\begin{tikzcd}
I_1 \arrow[r, "s_1"]\arrow[d, "\alpha_I"']
& P_1 \arrow[d, "\alpha_P"] 
& O_1 \arrow[l, "t_1"']\arrow[d, "\alpha_O"] \\
I_2 \arrow[r, "s_2"']
& P_2 
& O_2 \arrow[l, "t_2"]
\end{tikzcd}
\]
\[
\begin{tikzcd}
D_1 \arrow[d, hook, "\alpha_D"'] & O_1 \arrow[d, "\alpha_O"]\arrow[l,"f_1"'] & I_1 \arrow[d, "\alpha_I"]\arrow[l, "r_1"'] \\
D_2 & O_2 \arrow[l,"f_2"] & I_2 \arrow[l, "r_2"]
\end{tikzcd}
\quad
\begin{tikzcd}
P_1 \arrow[r, "c_1"]\arrow[d, "\alpha_P"']
& \Sigma_1 \arrow[d, hook, "\alpha_\Sigma"] \\
P_2 \arrow[r, "c_2"']
& \Sigma_2
\end{tikzcd}
\]
Given ${\vec{i}(\alpha):=\{p\in P_1\mid\bullet\alpha_P(p)\setminus\alpha_P(\bullet p)\neq\emptyset\}}$ and ${\vec{o}(\alpha):=\{p\in P_1\mid\alpha_P(p)\bullet\setminus\alpha_P(p\bullet)\neq\emptyset\}}$, we also have that $\alpha$ satisfies ${\vec{i}(\alpha)\cup\vec{o}(\alpha)\subseteq P_1^+\cup P_1^-}$.  
\end{definition}

\begin{lemma}[\cite{arellanes_compositional_2026}]\label{lem:interface-preservation}
If ${\alpha\colon\lambda_1\rightarrow\lambda_2}$ is a circuit morphism, then ${\alpha^{-1}(P_2^-)\subseteq P_1^-}$ and ${\alpha^{-1}(P_2^+)\subseteq P_1^+}$.
\end{lemma}

\begin{definition}\label{def:markers}
Given $\Box\in\{+,-\}$, the marker ${\lambda^\Box}$ of a circuit $\lambda$ is a circuit monomorphism ${\lambda_0\rightarrow\lambda}$ where $\lambda_0$ is a trivial circuit and ${\lambda^\Box(P_0)=P^\Box}$. When $\Box=+$, we refer to ${\lambda^\Box}$ as an in-marker; otherwise, we call it out-marker.
\end{definition}

To combine simpler circuits into more complex ones, we use colimit operations in the category of circuits built upon pushouts and coproduct constructions.

\begin{definition}[Coproduct]\label{def:coproduct}
The coproduct of $\lambda_1$ and $\lambda_2$, written ${\lambda_1+\lambda_2}$, is given as follows:

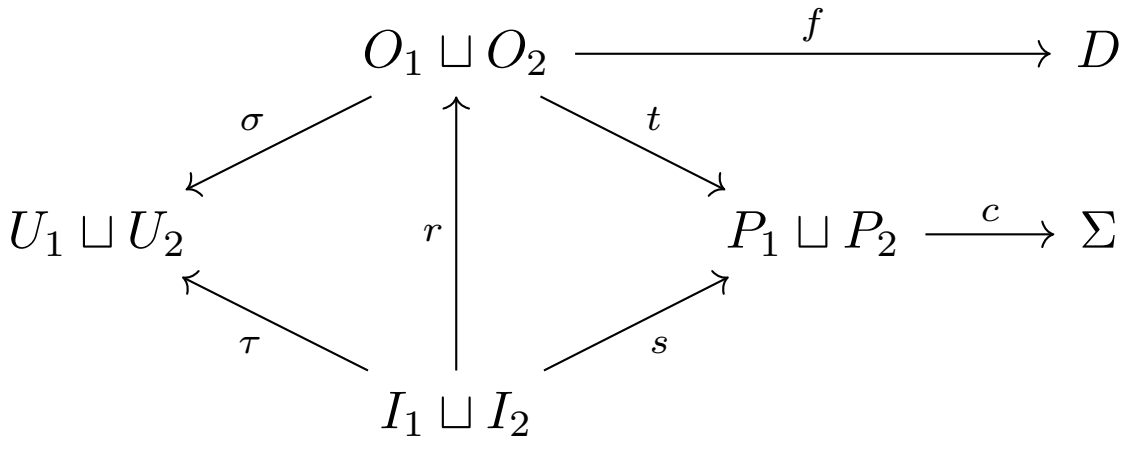
\[
\begin{tikzcd}[ampersand replacement=\&]
 \& O_1\sqcup O_2 \arrow[dl,"\sigma"']\arrow[dr, "t"]\arrow[rr, "f"] \& \& D  \\
U_1\sqcup U_2 \& \& P_1\sqcup P_2 \arrow[r,"c"] \& \Sigma \\
 \& I_1\sqcup I_2 \arrow[ul,"\tau"]\arrow[ur, "s"']\arrow[uu, "r"] \& \& \\
\end{tikzcd}
\]
with the functions of ${\lambda_1+\lambda_2}$ computed in the canonical manner via the canonical coproduct monomorphisms ${\beta_1\colon\lambda_1\to\lambda_1+\lambda_2}$ and ${\beta_2\colon\lambda_2\to\lambda_1+\lambda_2}$.
\end{definition}

\begin{definition}[Pushout]\label{def:pushout}
The pushout of a span ${\lambda_1\xleftarrow{\alpha_1}\lambda_0\xrightarrow{\alpha_2}\lambda_2}$ of circuit morphisms, written ${\lambda_1+_{\lambda_0}\lambda_2}$, is given as follows:

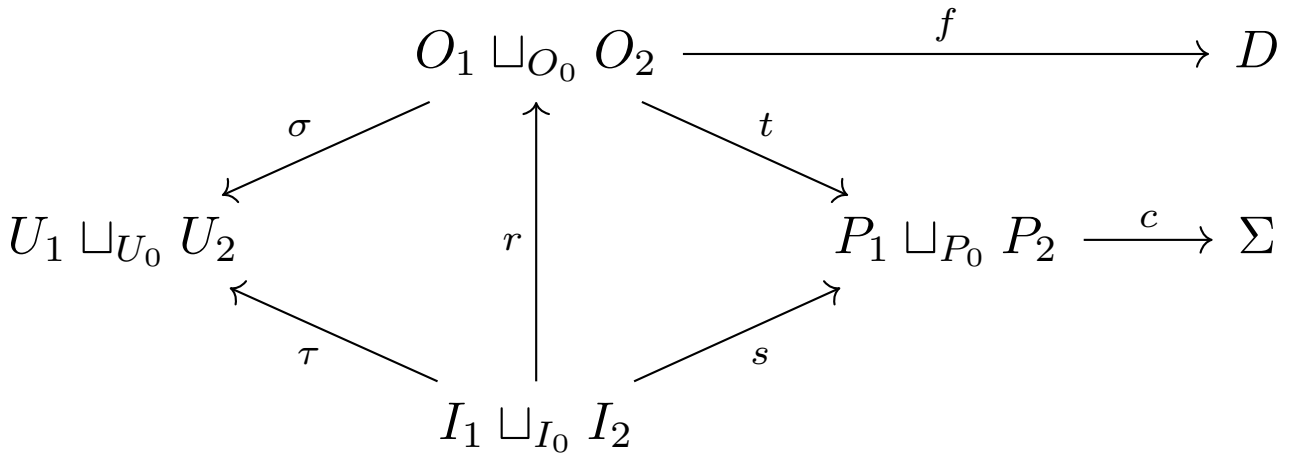
\[
\begin{tikzcd}[ampersand replacement=\&]
 \& O_1\sqcup_{O_0} O_2 \arrow[dl,"\sigma"']\arrow[dr, "t"]\arrow[rr, "f"] \& \& D  \\
U_1\sqcup_{U_0} U_2 \& \& P_1\sqcup_{P_0} P_2 \arrow[r,"c"] \& \Sigma \\
 \& I_1\sqcup_{I_0} I_2 \arrow[ul,"\tau"]\arrow[ur, "s"']\arrow[uu, "r"] \& \& \\
\end{tikzcd}
\]
with the functions of ${\lambda_1+_{\lambda_0}\lambda_2}$ computed in the canonical manner via the pushout-induced circuit morphisms ${\beta_1\colon\lambda_1\to\lambda_1+_{\lambda_0}\lambda_2}$ and ${\beta_2\colon\lambda_2\to\lambda_1+_{\lambda_0}\lambda_2}$.
\end{definition}

Although the category of circuits has all coproducts \cite{arellanes_compositional_2026}, Theorem \ref{th:pushout-existence} says that such a category is not co-complete since pushouts can only be computed for spans that satisfy Definition \ref{def:pushable-span}.

\begin{definition} [Pushable Span] \label{def:pushable-span}
A span ${\lambda_1 \xleftarrow{\alpha_1} \lambda_0 \xrightarrow{\alpha_2} \lambda_2}$ of circuit morphisms is pushable if ${\alpha_1(\vec{i}(\alpha_2)) \cup \alpha_1(\vec{o}(\alpha_2)) \subseteq P_1^+ \cup P_1^-}$ and ${\alpha_2(\vec{i}(\alpha_1)) \cup \alpha_2(\vec{o}(\alpha_1)) \subseteq P_2^+ \cup P_2^-}$.
\end{definition}

\begin{theorem}[\cite{arellanes_compositional_2026}]\label{th:pushout-existence}
The pushout of a span $\lambda_1\xleftarrow{\alpha_1}\lambda_0\xrightarrow{\alpha_2}$ of circuit morphisms exists if and only if the span is pushable. 
\end{theorem}

Using coproducts and computable pushouts, we define colimit-based operators for composing circuits into (partial or total) sequential, parallel, (open or closed) branching or iterative composites.

\subsection{Sequential Composition}

Control-driven circuits are composed sequentially through so-called sequentiable spans whose morphisms identify some or all the outports of one circuit with some or all the inports of another. Partial identification is given by partial sequentiable spans, whereas full identification by total sequentiable spans. This is formalised in Definition \ref{def:sequentiable-span}.

\begin{definition}[Sequentiable Span]\label{def:sequentiable-span}
We say that a span ${\lambda_1\xleftarrow{\alpha_1}\lambda_0\xrightarrow{\alpha_2}\lambda_2}$ is sequentiable when $\lambda_0$ is a trivial circuit and the legs $\alpha_1$ and $\alpha_2$ are monomorphisms that satisfy ${\alpha_1(P_0)\subseteq P_1^-}$ and ${\alpha_2(P_0)\subseteq P_2^+}$. If ${\alpha_1(P_0)=P_1^-}$ and ${\alpha_2(P_0)=P_2^+}$, the span is total sequential; otherwise, it is partial sequential.
\end{definition}

Naturally, Definition \ref{def:sequential-computon} and Lemma \ref{prop:sequencing-mono} collectively state that the pushout of a span that satisfies Definition \ref{def:sequentiable-span} yields a sequential circuit with pushout-induced monomorphisms. Such pushouts can always be computed in the category of circuits as per Theorem \ref{th:sequentiable-is-pushable} and, interestingly, any two circuits can always be composed sequentially by Theorem \ref{th:sequential-always}. Examples for the construction of partial and total sequential circuits are described in Section \ref{sec:applications}.

\begin{definition}[Sequential Circuit]\label{def:sequential-computon}
The pushout of a sequentiable span ${\rho:=\lambda_1\xleftarrow{\alpha_1}\lambda_0\xrightarrow{\alpha_2}\lambda_2}$ is a total sequential circuit ${\lambda_1\unrhd_{\rho}\lambda_2}$ if $\rho$ is total. Otherwise, it is a partial sequential circuit ${\lambda_1\rhd_{\rho}\lambda_2}$.
\end{definition}

\begin{lemma}\label{prop:sequencing-mono}
If ${\lambda_1\xrightarrow{\beta_1}\lambda_3\xleftarrow{\beta_2}\lambda_2}$ is the cospan induced by the pushout of a sequentiable span ${\lambda_1\xleftarrow{\alpha_1}\lambda_0\xrightarrow{\alpha_2}\lambda_2}$, then $\beta_1$ and $\beta_2$ are circuit monomorphisms.
\end{lemma}
\begin{proof}
We prove only for $\beta_1$, since the proof of the other is analogous. For this, first note that the $\Sigma$- and $D$-components of ${\beta_1}$ are always injective because Definition \ref{def:morphism} says they are inclusion functions. For the $I$-component, $I_1$-elements are never identified with $I_2$-elements by the fact that ${\lambda_0}$ is a trivial circuit with no flows at all, i.e., ${I_0=\emptyset}$. Thus, by Definition \ref{def:pushout}, ${I_3=I_1\sqcup_{I_0}I_2}$ has an equivalence class for each $I_1$-element, i.e., the $I$-component of $\beta_1$ is injective. The same reasoning holds for the $O$- and $U$-components of ${\beta_1}$ considering ${O_0=\emptyset=U_0}$.

Now, assume for contradiction that the $P$-component of $\beta_1$ is not injective so there are ports ${p_1,q_1\in P_1}$ with ${\beta_1(p_1)=\beta_1(q_1)}$. As ${P_3=P_1 \sqcup_{P_0} P_2}$, we have two possible scenarios: 
\begin{enumerate}
\item There is no ${p_2\in P_2}$ with ${\beta_2(p_2)=\beta_1(p_1)=\beta_1(q_1)}$. In this case, $p_1$ and $q_1$ form individual equivalence classes in $P_3$ for satisfying the reflexivity of the equivalence relation for ${P_1 \sqcup_{P_0} P_2}$. Thus, contradicting our assumption ${\beta_1(p_1)=\beta_1(q_1)}$. \label{prop:sequencing-mono-1}
\item There is some ${p_2\in P_2}$ where ${\beta_2(p_2)=\beta_1(p_1)=\beta_1(q_1)}$. In this case, the commutativity property of pushouts entails that there are ${p_0,q_0\in P_0}$ such that ${\alpha_1(p_0)=p_1}$, ${\alpha_1(q_0)=q_1}$ and ${\alpha_2(p_0)=p_2=\alpha_2(q_0)}$. Clearly contradicting that ${\alpha_2}$ is injective (see Definition \ref{def:sequentiable-span}). \label{prop:sequencing-mono-2}
\end{enumerate}
As all its components are injective functions, we conclude that $\beta_1$ is a circuit monomorphism, as required.
\end{proof}

\begin{theorem}\label{th:sequentiable-is-pushable}
A sequentiable span of circuit morphisms is pushable.
\end{theorem}
\begin{proof}
The proof follows directly from the facts $\alpha_1(P_0)\subseteq P_1^-$ and $\alpha_2(P_0)\subseteq P_2^+$ from Definition \ref{def:sequentiable-span}.
\end{proof}

\begin{theorem}\label{th:sequential-always}
For any pair of circuits $\lambda_1$ and $\lambda_2$, there always is a sequentiable span $\rho$.
\end{theorem}
\begin{proof}
Definition \ref{def:computon} states that every circuit has at least one inport and at least one outport. So, for any two circuits $\lambda_1$ and $\lambda_2$, we can always form a sequentiable span ${\rho:=\lambda_1\xleftarrow{\alpha_1}\lambda_0\xrightarrow{\alpha_2}\lambda_2}$ where $\lambda_0$ is a trivial circuit with ${\alpha_1(P_0)\subseteq P_1^-}$ and ${\alpha_2(P_0)\subseteq P_2^+}$. As the pushability conditions from Definition \ref{def:pushable-span} follow trivially, the pushout of $\rho$ exists which, by Definition \ref{def:sequential-computon}, must be either ${\lambda_1\rhd_\rho\lambda_2}$ or ${\lambda_1\unrhd_\rho\lambda_2}$.
\end{proof}

The interface of a sequential circuit is obtained from the interfaces of the composed objects, as captured by Lemmas \ref{prop:interface-sequential} and \ref{prop:interface-sequential-total}. The operations for constructing total and partial sequential circuits are called \emph{total sequencing} and \emph{partial sequencing}, respectively, which by Theorems \ref{th:sequencing-identity} and \ref{th:sequencing-commutative} are not commutative despite of satisfying identity. Theorem \ref{th:sequencing-associative} states that only total sequencing exhibits associativity.

\begin{lemma}\label{prop:interface-sequential}
Assuming ${\lambda_1\xrightarrow{\beta_1}\lambda_3\xleftarrow{\beta_2}\lambda_2}$ is the cospan induced by the pushout of a sequentiable span ${\lambda_1\xleftarrow{\alpha_1}\lambda_0\xrightarrow{\alpha_2}\lambda_2}$, we have ${\beta_1(P_1^+)\subseteq P_3^+}$ and ${\beta_2(P_2^-)\subseteq P_3^-}$.
\end{lemma}
\begin{proof}
We just show ${\beta_1(P_1^+)\subseteq P_3^+}$ since the proof of ${\beta_2(P_2^-)\subseteq P_3^-}$ is analogous. For this, assume for contrapositive ${p_3\notin P_3^+}$ so there is some ${o_3\in O_3}$ with ${t_3(o_3)=p_3}$. As ${O_3=O_1\sqcup_{O_0}O_2}$ according to Definition \ref{def:pushout}, we have three possible cases:
\begin{enumerate}
\item There exclusively is some ${o_1 \in O_1}$ with ${\beta_1(o_1)=o_3}$ so, by the commutativity squares of $\beta_1$, ${\beta_1(t_1(o_1))=t_3(\beta_1(o_1))=t_3(o_3)=p_3}$. As ${t_1(o_1)\notin P_1^+}$ as per Definition \ref{def:interface}, ${p_3\notin\beta_1(P_1^+)}$.\label{prop:interface-sequential-1}
\item There exclusively is some ${o_2 \in O_2}$ with ${\beta_2(o_2)=o_3}$. In this case, the proof is analogous to \ref{prop:interface-sequential-1}.\label{prop:interface-sequential-2}
\item There are ${o_1\in O_1}$ and ${o_2\in O_2}$ for which ${\beta_1(o_1)=o_3=\beta_2(o_2)}$. This case never holds since the apex of $\rho$ is a trivial circuit with no flows at all.\label{prop:interface-sequential-3}
\end{enumerate}
Proving that \ref{prop:interface-sequential-3} does not hold and that ${p_3\notin P_3^+\implies p_3\notin\beta_1(P_1^+)}$ for \ref{prop:interface-sequential-1} and \ref{prop:interface-sequential-2} entail ${\beta_1(P_1^+)\subseteq P_3^+}$, as required.
\end{proof}

\begin{lemma}\label{prop:interface-sequential-total}
Assuming ${\lambda_1\xrightarrow{\beta_1}\lambda_3\xleftarrow{\beta_2}\lambda_2}$ is the cospan induced by the pushout of a total sequentiable span ${\lambda_1\xleftarrow{\alpha_1}\lambda_0\xrightarrow{\alpha_2}\lambda_2}$, we have ${\beta_1(P_1^+)=P_3^+}$ and ${\beta_2(P_2^-)=P_3^-}$.
\end{lemma}
\begin{proof}
We only prove ${\beta_1(P_1^+)=P_3^+}$ since the proof of the other is completely analogous. For this, let ${p_3\in P_3^+}$. Using Lemma \ref{lem:interface-preservation} and the fact that $P_3$ is given by ${P_1\sqcup_{P_0}P_2}$ (see Definition \ref{def:pushout}), we have three cases:
\begin{enumerate}
\item $p_3$ is exclusively identified with $\lambda_1$-ports so ${p_3\in\beta_1(P_1^+)}$.\label{prop:interface-sequential-total-1}
\item $p_3$ is exclusively identified with $\lambda_2$-ports. Here, let ${p_2\in P_2^+}$ with ${\beta_2(p_2)=p_3}$. As $\rho$ satisfies ${\alpha_2(P_0)=P_2^+}$ (by Definition \ref{def:sequential-computon}), ${p_2\in P_2^+\iff p_2\in\alpha_2(P_0)}$. Clearly, a contradiction to the fact that $p_3$ is not identified with any $\lambda_1$-port.\label{prop:interface-sequential-total-2}
\item there exist ${p_1\in P_1^+}$ and ${p_2\in P_2^+}$ where ${\beta_1(p_1)=p_3=\beta_2(p_2)}$. As ${p_1\in P_1^+}$, ${p_3\in\beta_1(P_1^+)}$ directly.\label{prop:interface-sequential-total-3}
\end{enumerate}
Disproving \ref{prop:interface-sequential-total-2} and proving ${p_3\in P_3^+\implies p_3\in\beta_1(P_1^+)}$ for \ref{prop:interface-sequential-total-1} and \ref{prop:interface-sequential-total-3} imply ${P_3^+\subseteq\beta_1(P_1^+)}$. Simply using Lemma \ref{prop:interface-sequential} we conclude ${\beta_1(P_1^+)=P_3^+}$.
\end{proof}

\begin{theorem}[\cite{arellanescolimit2026}]\label{th:sequencing-identity}
The unit circuit is the left- and right-identity for total and partial sequencing.
\end{theorem}

\begin{theorem}\label{th:sequencing-associative}
Total sequencing is associative up to isomorphism, but partial sequencing is not. 
\end{theorem}
\begin{proof}
Disproving the associativity of partial sequencing can be done via the counterexample presented in \cite{arellanes_compositional_2026}. Here, we just prove that total sequencing is associative up to isomorphism. For this, assume that ${\lambda_1 \unrhd_{\rho_1}\lambda_2}$ and ${\lambda_2 \unrhd_{\rho_2}\lambda_3}$ are total sequential circuits with ${\rho_1:=\lambda_1 \xleftarrow{\alpha_1} \lambda_0 \xrightarrow{\alpha_2} \lambda_2}$ and ${\rho_2:=\lambda_2 \xleftarrow{\alpha_3} \lambda_4 \xrightarrow{\alpha_4} \lambda_3}$. So, we have the following commutative diagram:
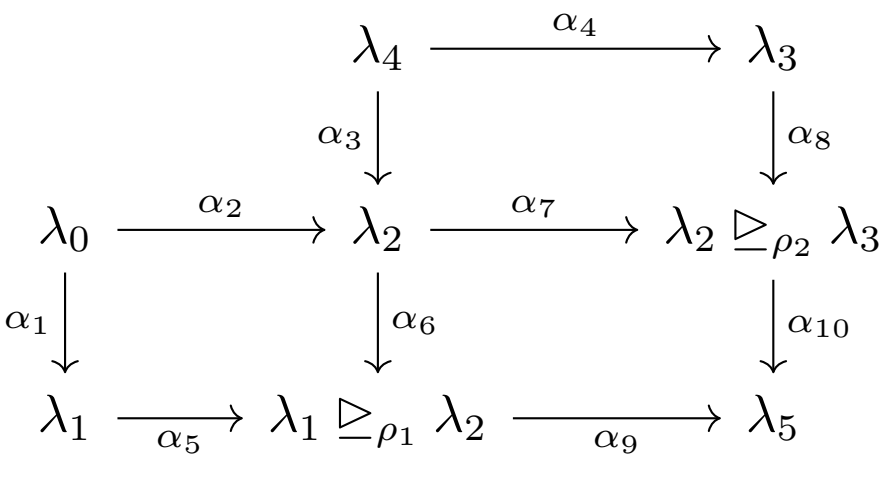
\[
\begin{tikzcd}
 & \lambda_4 \arrow[d, "\alpha_3"']\arrow[r, "\alpha_4"] & \lambda_3 \arrow[d, "\alpha_{8}"] \\
\lambda_0 \arrow[d, "\alpha_1"']\arrow[r, "\alpha_2"] & \lambda_2 \arrow[d, "\alpha_6"]\arrow[r, "\alpha_7"] & \lambda_2 \unrhd_{\rho_2} \lambda_3 \arrow[d, "\alpha_{10}"] \\
\lambda_1 \arrow[r, "\alpha_5"'] & \lambda_1 \unrhd_{\rho_1} \lambda_2 \arrow[r, "\alpha_{9}"'] & \lambda_5 
\end{tikzcd}
\]
where the cospans $(\alpha_5,\alpha_6)$ and $(\alpha_7,\alpha_8)$ are induced by the pushouts of $\rho_1$ and $\rho_2$, respectively. Composing circuit morphisms horizontally and vertically, we obtain the following commutative diagrams:
\[
\begin{tikzcd}
\lambda_0 \arrow[d, "\alpha_1"']\arrow[r, "\alpha_7\circ\alpha_2"] & \lambda_2 \unrhd_{\rho_2} \lambda_3 \arrow[d, "\alpha_{10}"] & & \lambda_4 \arrow[d, "\alpha_6\circ\alpha_3"']\arrow[r, "\alpha_4"] & \lambda_3 \arrow[d, "\alpha_{10}\circ\alpha_8"] \\
\lambda_1 \arrow[r, "\alpha_{9}\circ\alpha_5"'] & \lambda_5 & & \lambda_1 \unrhd_{\rho_1} \lambda_2 \arrow[r, "\alpha_{9}"'] & \lambda_5
\end{tikzcd}
\]

To show that every $\lambda_0$-port is mapped to an inport of $\lambda_2 \unrhd_{\rho_2} \lambda_3$, consider the following chain of equivalences: $p \in\alpha_7(\alpha_2(P_0))\iff p \in \alpha_7(P_2^+)$ (because $\alpha_2(P_0)=P_2^+$ by the fact that the pushout of $\rho_1$ is a total sequential circuit) $\iff p$ is an inport of $\lambda_2 \unrhd_{\rho_2} \lambda_3$ (as per Lemma \ref{prop:interface-sequential-total}).

With Lemma \ref{prop:sequencing-mono} and the fact that the pushout of $\rho_1$ is total sequential, we have that $\alpha_2$ and $\alpha_7$ are monomorphisms so ${\alpha_7\circ\alpha_2}$ is also a monomorphism. Having ${\alpha_1(P_0)=P_1^-}$, because the pushout of $\rho_1$ is total sequential, Definition \ref{def:sequentiable-span} says that ${\rho_3:=\lambda_1 \xleftarrow{\alpha_1} \lambda_0 \xrightarrow{\alpha_7\circ\alpha_2} \lambda_2 \unrhd_{\rho_2} \lambda_3}$ must be sequentiable. Here, $\lambda_0$ is the apex circuit whilst $\lambda_1$ and $\lambda_2\unrhd_{\rho_2}\lambda_3$ are the left and right operands, respectively. Using Definition \ref{def:sequential-computon}, we establish that the pushout of $\rho_3$ is ${\lambda_1 \unrhd_{\rho_3}(\lambda_2\unrhd_{\rho_2}\lambda_3)}$. A a similar reasoning allows us to deduce that ${(\lambda_1 \unrhd_{\rho_1}\lambda_2)\unrhd_{\rho_4}\lambda_3}$ is the pushout of the span ${\rho_4:=\lambda_1 \unrhd_{\rho_1}\lambda_2\xleftarrow{\alpha_6\circ\alpha_3} \lambda_4 \xrightarrow{\alpha_4} \lambda_3}$. Hence: 
\[
{\lambda_1 \unrhd_{\rho_3}(\lambda_2\unrhd_{\rho_2}\lambda_3)\cong\lambda_5\cong (\lambda_1 \unrhd_{\rho_1}\lambda_2)\unrhd_{\rho_4}\lambda_3}
\]
\end{proof}

\begin{theorem}\label{th:sequencing-commutative}
Total and partial sequencing are not commutative. 
\end{theorem}
\begin{proof}
Disproving the commutativity of both total and partial sequencing can be done through the counterexamples presented in \cite{arellanes_compositional_2026}.
\end{proof}

\subsection{Parallel Composition}

Control-driven circuits can be composed into coproduct-based structures to enable their parallel execution. Such structures can always be formed in the category of circuits as per Theorem \ref{th:p-async-exists}. Examples are shown in Section \ref{sec:applications}.

\begin{definition}\label{def:p-async-computon}
A parallel circuit ${\lambda_1+\lambda_2}$ is given by the coproduct of $\lambda_1$ and $\lambda_2$.
\end{definition}

\begin{theorem}\label{th:p-async-exists}
A parallel circuit always exists in a category of circuits for any pair of circuits.
\end{theorem}
\begin{proof}
It is trivial to verify that coproduct can always be computed in the category of control-driven circuits because such a colimit construction is computed componentwise in the category of finite sets (which has all coproducts). A full proof is analogous to that from \cite{arellanes_compositional_2026}.
\end{proof}

Theorem \ref{th:p-async-identity} says that the operation for forming a parallel circuit, called parallelising, does not satisfy the identity law because this would require the existence of a circuit with no ports; thus, contradicting Definition \ref{def:computon}. Such an operation is associative and commutative by Theorem \ref{th:p-async-associativity-commutativity}.

\begin{theorem}[\cite{arellanescolimit2026}]\label{th:p-async-identity}
Parallelising has no identity.
\end{theorem}

\begin{theorem}\label{th:p-async-associativity-commutativity}
Parallelising is associative and commutative up to isomorphism.
\end{theorem}
\begin{proof}
The proof follows directly from the well-known fact that coproduct in the category of finite sets is both associative and commutative up to isomorphism.
\end{proof}

Apart from asynchronous structures, composites can also be formed to synchronise the execution of multiple circuits. This is achieved through the use of so-called \emph{glue circuits} for joining and forking control signals. Examples are shown in Section \ref{sec:applications}.

\begin{definition}\label{def:glue}
A glue is an atomic circuit $\lambda$ which satisfies $c(p)=0$ for all $p\in P$ and $f(o)=\kappa_m$ for all $o\in O$ with $m\geq 1$.    
\end{definition}

A glance at the examples presented in Section \ref{sec:applications} reveals that a synchronous parallel circuit is constructed by performing a sequencing operation of a fork glue with a parallel composite, before sequencing the result with a join glue. The order of these two operations is irrelevant, yielding different ways of constructing the same synchronous structure.

\subsection{Branching Composition}

Branching composition is the process of constructing a control-driven circuit that is able to decide whether to execute one circuit or another in a non-deterministic manner. There are two different operators to achieve this. The first allows the construction of a branching circuit (called open) that matches inports only, whereas the second (called closed) matches both inports and outports. More formally, \emph{open branching} is characterised as the pushout of a span of in-markers (as per Definition \ref{def:branching-open}), whilst \emph{closed braching} is the finite colimit of a diagram formed by a span of in-markers and a span of out-markers (see Definition \ref{def:branching-closed}). By Lemma \ref{lem:adjoints-for-closed-branching}, the in-marker of a closed branching circuit is the in-marker of the composed circuits and the same holds with respect to out-markers. An example of a closed braching circuit is described in Section \ref{sec:applications}. 

\begin{definition}\label{def:branching-open}
The pushout of a span ${\rho:=\lambda_1\xleftarrow{\lambda_1^+}\lambda_0\xrightarrow{\lambda_2^+}\lambda_2}$ is called an open branching circuit ${\lambda_1?_{\rho}\lambda_2}$.
\end{definition}

\begin{definition}\label{def:branching-closed}
If $\rho$ is a diagram
\begin{tikzcd}
\lambda_1 & \lambda_0 \arrow[l,"\lambda_1^+"']\arrow[r,"\lambda_2^+"] & \lambda_2 \\
 & \lambda_3\arrow[ul,bend left=30,"\lambda_1^-"]\arrow[ur,bend right=30,"\lambda_2^-"'] & 
\end{tikzcd}
where $\lambda_1$ and $\lambda_2$ are connected circuits, then the finite colimit of $\rho$ is the closed branching circuit $\lambda_1??_{\rho}\lambda_2$, computed as ${\lambda_1+_{\lambda_0+\lambda_3}\lambda_2}$.
\end{definition}

\begin{lemma}\label{lem:adjoints-for-closed-branching}
If $\rho$ is the diagram 
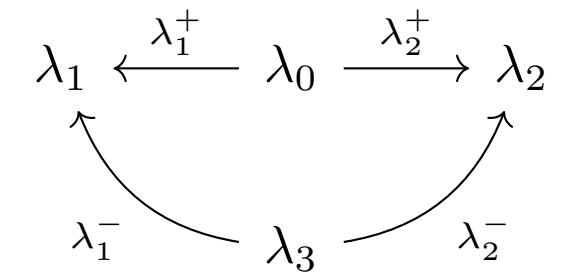
\begin{tikzcd}
\lambda_1 & \lambda_0 \arrow[l,"\lambda_1^+"']\arrow[r,"\lambda_2^+"] & \lambda_2 \\
 & \lambda_3\arrow[ul,bend left=30,"\lambda_1^-"]\arrow[ur,bend right=30,"\lambda_2^-"'] & 
\end{tikzcd} then $\lambda_0$ and $\lambda_3$ are the domains of the in- and out-markers of ${\lambda_1??_{\rho}\lambda_2}$, respectively.
\end{lemma}
\begin{proof}
Assuming that ${\alpha\colon\lambda_0\rightarrow\lambda_0+\lambda_3}$ and ${\beta\colon\lambda_3\rightarrow\lambda_0+\lambda_3}$ are the canonical injections induced by the coproduct ${\lambda_0+\lambda_3}$, we obtain the following diagram:
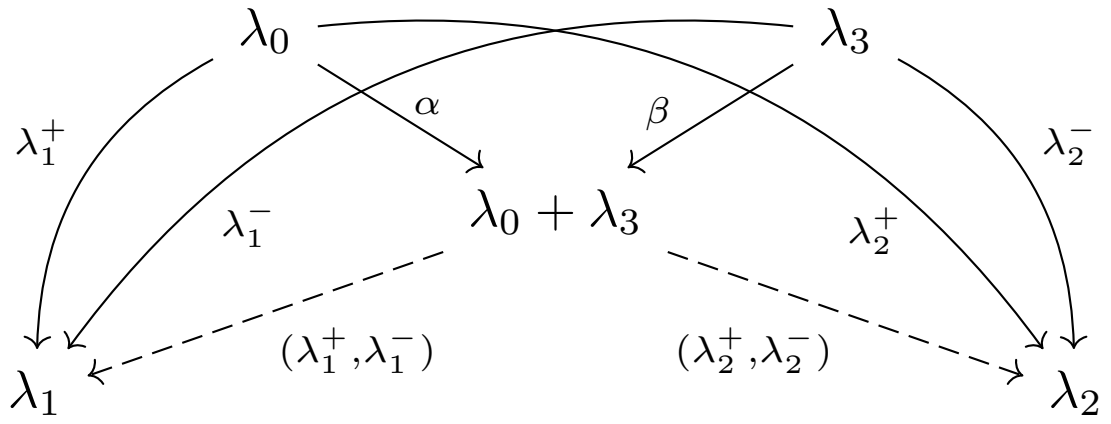
\[
\begin{tikzcd}
 & \lambda_0\arrow[ddl,bend right=30,"\lambda_1^+"']\arrow[ddrrr,bend left=30,"\lambda_2^+"{xshift=2em,yshift=-3.5em}]\arrow[dr,"\alpha"{yshift=-0.2em}] & & \lambda_3\arrow[ddr,bend left=30,"\lambda_2^-"]\arrow[ddlll,bend right=30,"\lambda_1^-"{xshift=-3.3em,yshift=-2em}]\arrow[dl,"\beta"'{xshift=-0.5em,yshift=-0.5em}] & \\
 & & \lambda_0+\lambda_3\arrow[dll,dash pattern=on 4pt off 2pt,"{(\lambda_1^+,\lambda_1^-)}"]\arrow[drr,dash pattern=on 4pt off 2pt,"{(\lambda_2^+,\lambda_2^-)}"'] & & \\
 \lambda_1 & & & & \lambda_2 
\end{tikzcd}
\]
where ${(\lambda_1^+,\lambda_1^-)}$ and ${(\lambda_2^+,\lambda_2^-)}$ are the unique morphisms derived from the universal property of coproducts. This property follows from the fact that coproduct is computed in the category of finite sets and total functions. 

Now, by Definition \ref{def:branching-closed}, we know that the colimit ${\lambda_4:=\lambda_1??_{\rho}\lambda_2}$ of the above diagram corresponds to the pushout of ${(\lambda_1^+,\lambda_1^-)}$ and ${(\lambda_2^+,\lambda_2^-)}$. If ${\gamma\colon\lambda_1\rightarrow\lambda_4}$ and ${\theta\colon\lambda_2\rightarrow\lambda_4}$ are the morphisms induced by such a pushout, the following equalities must hold by commutativity: 

\[
\gamma\circ\lambda_1^+=\gamma\circ(\lambda_1^+,\lambda_1^-)\circ\alpha=\theta\circ(\lambda_2^+,\lambda_2^-)\circ\alpha=\theta\circ\lambda_2^+
\]
\[
\theta\circ\lambda_2^-=\gamma\circ(\lambda_1^+,\lambda_1^-)\circ\beta=\theta\circ(\lambda_2^+,\lambda_2^-)\circ\beta=\gamma\circ\lambda_1^-
\]

With the above in mind, we now prove as follows: ${p_4\in P_4^+ \iff }$ there is some ${p_1\in P_1^+}$ and some ${p_2\in P_2^+}$ where ${\gamma_P(p_1)=p_4=\theta_P(p_2)}$ (see Lemma \ref{lem:interface-preservation}) $\iff$ there is some ${p\in P_0+P_3}$ where ${(\lambda_1^+,\lambda_1^-)(p)=p_1}$ and ${(\lambda_2^+,\lambda_2^-)(p)=p_2}$ (as per the commutativity property of pushouts). As ${p_1\in P_1^+}$ and ${p_2\in P_2^+}$, Definition \ref{def:markers} entails that commutativity only holds if ${p_1=\lambda_1^+(p_0)}$ and ${p_2=\lambda_2^+(p_0)}$ for some ${p_0\in P_0}$. Using the above equations, we obtain:
\begin{small}
\[
(\gamma_P\circ\lambda_1^+)(p_0)=(\gamma_P\circ(\lambda_1^+,\lambda_1^-)\circ\alpha_P)(p_0)=(\theta_P\circ(\lambda_2^+,\lambda_2^-)\circ\alpha_P)(p_0)=(\theta_P\circ\lambda_2^+)(p_0)=p_4
\]
\end{small}

Thus, proving that ${\gamma\circ\lambda_1^+\colon\lambda_0\rightarrow\lambda_4}$, ${\gamma\circ(\lambda_1^+,\lambda_1^-)\circ\alpha\colon\lambda_0\rightarrow\lambda_4}$, ${\theta\circ(\lambda_2^+,\lambda_2^-)\circ\alpha\colon\lambda_0\rightarrow\lambda_4}$ and ${\theta\circ\lambda_2^+\colon\lambda_0\rightarrow\lambda_4}$ are the in-markers of ${\lambda_4}$. Following a similar reasoning, we can show that the morphisms ${\theta\circ\lambda_2^-\colon\lambda_3\rightarrow\lambda_4}$, ${\gamma\circ(\lambda_1^+,\lambda_1^-)\circ\beta\colon\lambda_3\rightarrow\lambda_4}$, ${\theta\circ(\lambda_2^+,\lambda_2^-)\circ\beta\colon\lambda_3\rightarrow\lambda_4}$ and ${\gamma\circ\lambda_1^-\colon\lambda_3\rightarrow\lambda_4}$ are the out-markers of $\lambda_4$.
\end{proof}

In the case of open branching, given that the apex of an in-marker injectively covers all inports, the identity corresponds to a trivial circuit that is isomorphic to the in-marker apex (see Theorem \ref{th:open-branching-identity}). Closed branching does not have any identity since that would require a trivial circuit matching the inports and outports of both $\lambda_1$ and $\lambda_2$, potentially deviating from the condition imposed by Definition \ref{def:computon}, which states that every circuit must have ports at the interface. Theorem \ref{th:closed-branching-identity} captures this truth. By Theorems \ref{th:closed-branching-associativity}, \ref{th:closed-branching-commutativity} and \ref{th:open-branching-associative-commutative}, closed and open branching are both associative and commutative.

\begin{theorem}[\cite{arellanescolimit2026}]\label{th:open-branching-identity}
Given a circuit $\lambda_1$, there is a circuit $\lambda_2$ and a span $\rho$ of in-markers where $\lambda_1?_{\rho}\lambda_2\cong\lambda_1$.
\end{theorem}

\begin{theorem}[\cite{arellanescolimit2026}]\label{th:closed-branching-identity}
Closed branching does not satisfy the identity law.
\end{theorem}

\begin{theorem}\label{th:closed-branching-associativity}
Closed branching is associative up to isomorphism.
\end{theorem}
\begin{proof}
Assume that ${\lambda_2??_{\rho_1}\lambda_3}$ and ${\lambda_3??_{\rho_2}\lambda_4}$ are closed branching circuits where:
\begin{itemize}
\item $\rho_1$ is the diagram formed by the spans ${\lambda_2\xleftarrow{\lambda_2^+}\lambda_0\xrightarrow{\lambda_3^+}\lambda_3}$ and ${\lambda_2\xleftarrow{\lambda_2^-}\lambda_1\xrightarrow{\lambda_3^-}\lambda_3}$, and
\item $\rho_2$ is the diagram formed by the spans ${\lambda_3\xleftarrow{\lambda_3^+}\lambda_0\xrightarrow{\lambda_4^+}\lambda_4}$ and ${\lambda_3\xleftarrow{\lambda_3^-}\lambda_1\xrightarrow{\lambda_4^-}\lambda_4}$.
\end{itemize}
By Definition \ref{def:branching-closed}, ${\lambda_2??_{\rho_1}\lambda_3}$ and ${\lambda_3??_{\rho_2}\lambda_4}$ are constructed via the operations ${\lambda_2+_{\lambda_0+\lambda_1}\lambda_3}$ and ${\lambda_3+_{\lambda_0+\lambda_1}\lambda_4}$, respectively. Accordingly, we have the following commutative diagram:

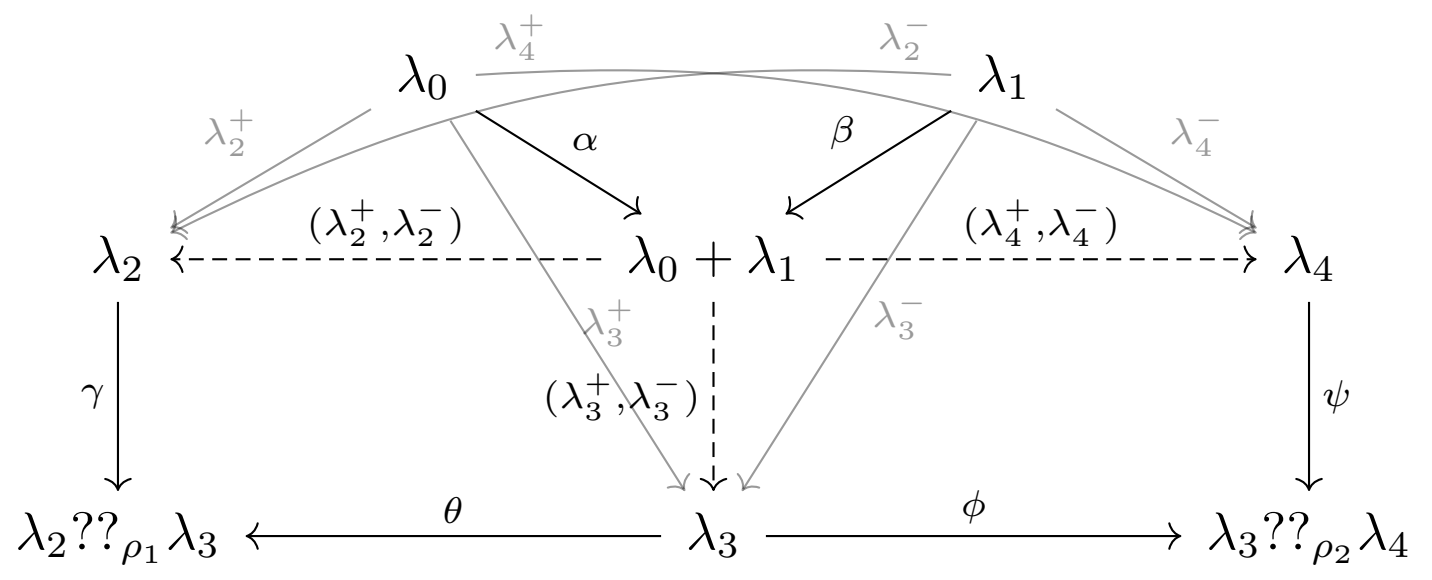
\begin{center}
\begin{tikzcd}
 & \lambda_0 \arrow[dr, "\alpha"]\arrow[dl, opacity=0.4, "\lambda_2^+"']\arrow[drrr, bend left=15, opacity=0.4,"\lambda_4^+"{yshift=4pt,xshift=-75pt}]\arrow[dddr, opacity=0.4, "\lambda_3^+"{yshift=-10pt,xshift=0pt}] & & \lambda_1 \arrow[dl, "\beta"']\arrow[dr, opacity=0.4, "\lambda_4^-"]\arrow[dlll, bend right=15, opacity=0.4,"\lambda_2^-"'{yshift=4pt,xshift=75pt}]\arrow[dddl, opacity=0.4, "\lambda_3^-"{yshift=5pt}] & \\
 \lambda_2 \arrow[dd, "\gamma"'] & & \lambda_0+\lambda_1 \arrow[ll, dashed, "{(\lambda_2^+,\lambda_2^-)}"']\arrow[rr, dashed, "{(\lambda_4^+,\lambda_4^-)}"]\arrow[dd, dashed, "{(\lambda_3^+,\lambda_3^-)}"'] & & \lambda_4 \arrow[dd, "\psi"] \\
 & & & & \\
\lambda_2??_{\rho_1}\lambda_3 & & \lambda_3 \arrow[ll, "\theta"']\arrow[rr, "\phi"] & & \lambda_3??_{\rho_2}\lambda_4
\end{tikzcd}
\end{center}

Here, $\alpha$ and $\beta$ are the canonical injections into ${\lambda_0+\lambda_1}$, and the cospan ${(\gamma,\theta)}$ is induced by the pushout of ${(\lambda_2^+,\lambda_2^-)}$ and ${(\lambda_3^+,\lambda_3^-)}$ which, in turn, are unique morphisms induced by the universal property of coproducts. Similarly, the cospan ${(\phi,\psi)}$ derives from the pushout of ${(\lambda_3^+,\lambda_3^-)}$ and ${(\lambda_4^+,\lambda_4^-)}$ which are also induced from the universal property of coproducts. 

Now, by pushout commutativity, we have ${f\colon\lambda_0+\lambda_1\rightarrow\lambda_2??_{\rho_1}\lambda_3}$ and ${g\colon\lambda_0+\lambda_1\rightarrow\lambda_3??_{\rho_2}\lambda_4}$ given by ${f=\gamma\circ(\lambda_2^+,\lambda_2^-)=\theta\circ(\lambda_3^+,\lambda_3^-)}$ and ${g=\phi\circ(\lambda_3^+,\lambda_3^-)=\psi\circ(\lambda_4^+,\lambda_4^-)}$, respectively. Thus, yielding the following diagrams:\\

\noindent
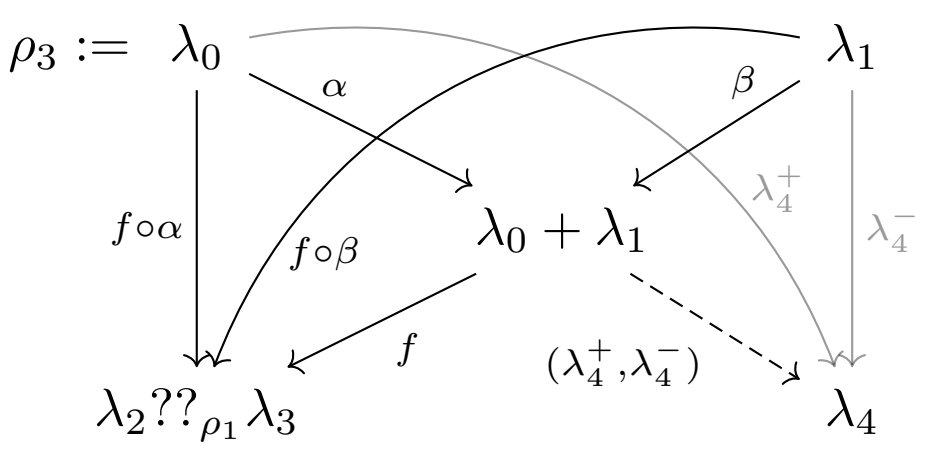
\begin{minipage}{0.35\textwidth}
\centering
\begin{tikzcd}
 \rho_3:= & [-1.4cm] \lambda_0 \arrow[dr, "\alpha"{yshift=4pt,xshift=-10pt}]\arrow[dd, "f\circ\alpha"']\arrow[ddrr, opacity=0.4, bend left=40, "\lambda_4^+"{yshift=-25pt,xshift=25pt}] & & \lambda_1 \arrow[dl, "\beta"'{yshift=4pt,xshift=10pt}]\arrow[dd, opacity=0.4, "\lambda_4^-"]\arrow[ddll, bend right=40, "f\circ\beta"{yshift=-25pt,xshift=-32pt}] \\
 & & \lambda_0+\lambda_1 \arrow[dl,"f"]\arrow[dr, dashed,"{(\lambda_4^+,\lambda_4^-)}"'] & \\
 & \lambda_2??_{\rho_1}\lambda_3 & & \lambda_4
\end{tikzcd}
\end{minipage}
\hfill
\begin{minipage}{0.6\textwidth}
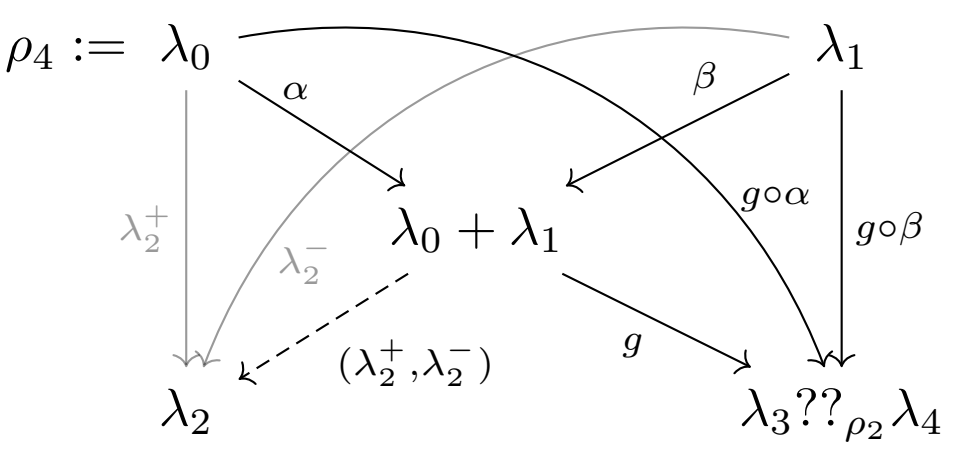
\centering
\begin{tikzcd}
 \rho_4:= & [-0.95cm] \lambda_0 \arrow[dr, "\alpha"{yshift=4pt,xshift=-10pt}]\arrow[dd, opacity=0.4, "\lambda_2^+"']\arrow[ddrr, bend left=40, "g\circ\alpha"{yshift=-25pt,xshift=25pt}] & & \lambda_1 \arrow[dl, "\beta"'{yshift=4pt,xshift=10pt}]\arrow[dd, "g\circ\beta"]\arrow[ddll, bend right=40, opacity=0.4, "\lambda_2^-"{yshift=-25pt,xshift=-32pt}] \\
 & & \lambda_0+\lambda_1 \arrow[dl, dashed, "{(\lambda_2^+,\lambda_2^-)}"]\arrow[dr, "g"'] & \\
 & \lambda_2 & & \lambda_3??_{\rho_2}\lambda_4
\end{tikzcd}
\end{minipage}

For the diagram $\rho_3$, we use Lemma \ref{lem:adjoints-for-closed-branching} to deduce that the morphisms ${f\circ\alpha\colon\lambda_0\rightarrow\lambda_2??_{\rho_1}\lambda_3}$ and ${f\circ\beta\colon\lambda_1\rightarrow\lambda_2??_{\rho_1}\lambda_3}$ are the in- and out-markers of ${\lambda_2??_{\rho_1}\lambda_3}$, respectively. Simply applying Definition \ref{def:branching-closed}, we determine that ${(\lambda_2??_{\rho_1}\lambda_3)??_{\rho_3}\lambda_4}$ is the pushout of $f$ and ${(\lambda_4^+,\lambda_4^-)}$, computed as ${(\lambda_2+_{\lambda_0+\lambda_1}\lambda_3)+_{\lambda_0+\lambda_1}\lambda_4}$. A similar reasoning allows us to deduce that ${\lambda_2??_{\rho_4}(\lambda_3??_{\rho_2}\lambda_4)}$ is the pushout of ${(\lambda_2^+,\lambda_2^-)}$ and $g$, computed as ${\lambda_2+_{\lambda_0+\lambda_1}(\lambda_3+_{\lambda_0+\lambda_1}\lambda_4)}$.

Given that ${(\lambda_j^+,\lambda_j^-)\colon\lambda_0+\lambda_1\rightarrow\lambda_j}$ is a monomorphism for all ${j\in\{2,3,4\}}$ (by Definition \ref{def:markers}) and that pushouts are computed in the category of finite sets, associativity of pushouts holds up to unique isomorphism. That is,  ${(\lambda_2+_{\lambda_0+\lambda_1}\lambda_3)+_{\lambda_0+\lambda_1}\lambda_4}\cong{\lambda_2+_{\lambda_0+\lambda_1}(\lambda_3+_{\lambda_0+\lambda_1}\lambda_4)}$. Or, equivalently, ${(\lambda_2??_{\rho_1}\lambda_3)??_{\rho_3}\lambda_4}\cong{\lambda_2??_{\rho_4}(\lambda_3??_{\rho_2}\lambda_4)}$. Thus, concluding that closed branching is associative up to unique isomorphism.
\end{proof}

\begin{theorem}\label{th:closed-branching-commutativity}
Closed branching is commutative up to isomorphism.
\end{theorem}
\begin{proof}
If ${\lambda_1??_{\rho_1}\lambda_2}$ and ${\lambda_2??_{\rho_2}\lambda_1}$ are closed branching circuits, Definition \ref{def:branching-closed} says that each of them must be the pushout of circuit morphisms. As these morphisms are built upon total functions between finite sets (see Definition \ref{def:morphism}), it is clear that pushout is computed in the category of finite sets and total functions. In this category, it is well-known that pushout is commutative up to unique isomorphism. Hence, ${\lambda_1??_{\rho_1}\lambda_2\cong\lambda_2??_{\rho_2}\lambda_1}$ must hold trivially.
\end{proof}

\begin{theorem}\label{th:open-branching-associative-commutative}
Open branching is associative and commutative up to isomorphism.
\end{theorem}
\begin{proof}
Associativity and commutativity follow from the well-establish fact that pushout in the category of finite sets and total functions satisfies those properties up to isomorphism. The proof that the pushout of in-markers can be computed in a category of circuits is analogous to the proof presented in \cite{arellanes_compositional_2026}.
\end{proof}

\subsection{Iterative Composition}

Although classical Boolean circuits do not support iteration, control-driven circuits offer operators for constructing iterative composites that decide cycle termination non-deterministically. The first operator performs the decision before executing the circuit being iterated over (as in \emph{while} constructs), whereas the other makes the decision after executing the circuit (as in \emph{do-while} constructs). Accordingly, these operators define head- and tail-iterative composites.

Particularly, the operation to form a \emph{head-iterative} circuit is called \emph{head iteration} and is characterised as the colimit of a so-called \emph{h-diagram} $\rho$ which, apart from the circuit $\lambda$ being iterated over, requires three more objects to specify the entry and exit to the iterative structure $*_\rho\lambda$ and what to do after executing $\lambda$. These additional objects are needed due to the interface restrictions imposed by Definition \ref{def:morphism}. The corresponding formalisation is given below.

\begin{definition}\label{def:head}
An h-diagram $\rho$ has the following shape in the category of circuits:
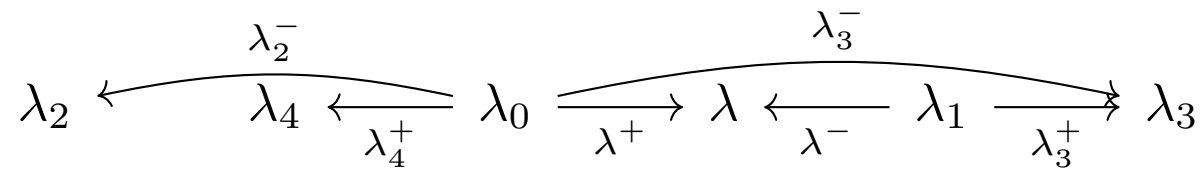
\[
\begin{tikzcd}
\lambda_2 & \lambda_4 & \lambda_0 \arrow[l, "\lambda_4^+"]\arrow[r, "\lambda^+"']\arrow[ll, bend right=12, "\lambda_2^-"']\arrow[rrr, bend left=12, "\lambda_3^-"] & \lambda & \lambda_1 \arrow[l, "\lambda^-"]\arrow[r, "\lambda_3^+"'] & \lambda_3
\end{tikzcd}
\]
with $\lambda$, $\lambda_2$, $\lambda_3$ and $\lambda_4$ being connected. A head-iterative circuit $*_\rho\lambda$ is the colimit of $\rho$, computed as ${(\lambda_4+_{\lambda_0}\lambda)+_{(\lambda_0+\lambda_1)}(\lambda_2+_{\lambda_0}\lambda_3)}$.
\end{definition}

Like, head-iteration, tail-iteration requires additional objects for composition. The difference fundamentally lies in the position where cycle termination is decided, so a different colimit operation is required on a so-called \emph{t-diagram}. The corresponding formalisation is given in Definition \ref{def:tail} and an example is presented in Section \ref{sec:applications}.

\begin{definition}\label{def:tail}
A t-diagram $\rho$ has the following shape in the category of circuits:
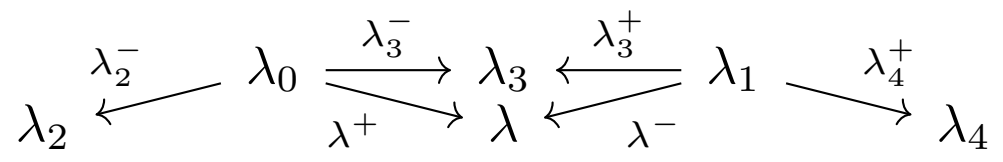
\[
\begin{tikzcd}[row sep=0.9em]
 & \lambda_0 \arrow[dl, "\lambda_2^-"']\arrow[r, "\lambda_3^-"]\arrow[dr, "\lambda^+"'] & \lambda_3 & \lambda_1 \arrow[l, "\lambda_3^+"']\arrow[dr, "\lambda_4^+"]\arrow[dl, "\lambda^-"] \\
 [-0.5cm]
\lambda_2 & & \lambda & & \lambda_4
\end{tikzcd}
\]
with $\lambda$, $\lambda_2$, $\lambda_3$ and $\lambda_4$ being connected. A tail-iterative circuit $\lambda*_\rho$ is the colimit of $\rho$, computed as ${((\lambda_2+_{\lambda_0}\lambda_3)+_{\lambda_3}(\lambda_3+_{\lambda_1}\lambda_4))+_{(\lambda_0+\lambda_1)}\lambda}$.
\end{definition}

Evidently, head- and tail-iteration are unary operations built upon coproduct and pushout constructions. Accordingly, we do not study identity, associativity or commutativity properties. These operators are just described for the completeness of our theory.

\section{Operational Semantics}\label{sec:operation}

In Section \ref{sec:operators}, we presented colimit-based composition semantics for circuit construction. We now describe a particular transition system for the execution of arbitrary circuits. 

The general idea is to visualise ports as constructs that hold information which is consumed and produced by either the external world or by computing devices within units. The execution of a circuit $\lambda$ starts by placing values on all the ports in $P^+$. As control ports are always required in $P^+$, as per Definition \ref{def:computon}, control signals are always present within circuits to drive execution. Although data values are optional, they are required to simulate Boolean functions by which Boolean data is transformed (see Section \ref{sec:model}). In any case, control and data independently flow. In fact, there are functors to orthogonally extract the control and data flow graphs of any arbitrary circuit \cite{arellanes_compositional_2026}. 

Since ports store data or control signals, the state of a circuit is a function that maps each port to a specific value from a fixed type system. Definition \ref{def:circuit} mentions that circuits rely on a system that contains the types $\mathbbm{1}$ and $\mathbb{B}$ to represent control signals and Boolean values, respectively, with options to express value absence. Accordingly, Definition \ref{def:typing-function} states that there is a typing function for circuits for mapping colours to types as follows: $0\mapsto\mathbbm{1}$ and $1\mapsto\mathbb{B}$. On this basis, Definition \ref{def:computon-state} formalises the concept of a (discrete-time) circuit state, having in mind that $\llbracket\mathbbm{1}\rrbracket$ and $\llbracket\mathbb{B}\rrbracket$ are the set-theoretic interpretations of $\mathbbm{1}$ and $\mathbb{B}$ (as mentioned in Definition \ref{def:circuit}).

\begin{definition}\label{def:typing-function}
The typing function $T\colon\Sigma\to\{\mathbbm{1},\mathbb{B}\}$ for a circuit $\lambda$ is a function defined as follows: \begin{align*}
  T(x) 
    &=\begin{cases}
       \mathbbm{1} & x=0\\
       \mathbb{B} & x=1
    \end{cases} &\!\!\!\!\!\!\!\!\! \text{ for every }{x\in\Sigma}
\end{align*}
\end{definition}

\begin{definition}\label{def:computon-state}
The state $\delta^j$ of a circuit $\lambda$ at time ${j \geq 0}$ is a function $P\to\llbracket\mathbbm{1}\rrbracket\cup\llbracket\mathbb{B}\rrbracket$ where, for all $p\in P$, $\delta^j(p)\in\llbracket T(c(p))\rrbracket$. Particularly, $\delta^j$ is called initial when: 
\begin{itemize}
\item $j=0$,
\item for all $p\in P^+$, $\delta^j(p)\neq\bot_{\mathbbm{1}}$ if $c(p)=0$ and $\delta^j(p)\neq\bot_{\mathbbm{B}}$ if $c(p)=1$, and
\item for all $q\notin P^+$, $\delta^j(q)=\bot_{\mathbbm{1}}$ if $c(q)=0$ and $\delta^j(q)=\bot_{\mathbbm{B}}$ if $c(q)=1$.
\end{itemize}
\end{definition}

In a given state $\delta^j$, devices are executed only if the unit they belong to is \emph{enabled}. As per Definition \ref{def:unit-status}, a unit is enabled under $\delta^j$ if every port attached to it has values of the corresponding type; otherwise, the unit is \emph{idle} under $\delta^j$. When all its units are idle, a circuit reaches termination, as described in Definition \ref{def:termination}. 

\begin{definition}\label{def:unit-status}
The computation unit $u\in U$ of a circuit $\lambda$ is enabled under a state $\delta^j$ if $\delta^j(p)\neq\bot_{\mathbbm{1}}$ for every $p\in\bullet u$ with $c(p)=0$ and $\delta^j(q)\neq\bot_{\mathbbm{B}}$ for every $q\in\bullet u$ with $c(p)=1$. Otherwise, $u$ is idle under $\delta^j$.
\end{definition}

\begin{definition}\label{def:termination}
The state $\delta^j$ of a circuit $\lambda$ is final if every $u\in U$ is idle under $\delta^j$.   
\end{definition}

For the execution of branching circuits, we handle the case when multiple units are enabled. We do this by establishing a partition of enabled units that share ports: ${u_1\sim u_2\iff\bullet u_1=\bullet u_2}$. With this, we simply invoke the axiom of choice to non-deterministically choose a representative unit from each partition. The resulting collection is called the set of \emph{ready units}, and it is formalised in Definition \ref{def:ready-units}.

\begin{definition}\label{def:ready-units}
If $E^j$ is the partition induced by the equivalence relation ${u_1\sim u_2\iff\bullet u_1=\bullet u_2}$ on the set of enabled units under $\delta^j$ and $\Omega$ is the random function on $E^j$, the set $R^j$ of computation units that are ready under $\delta^j$ is $\{\Omega(E)\mid E\in E^j\}$.
\end{definition}

After building a set $R^j$, each computing device $f(o)$ of each $R^j$-unit is evaluated according to Definition \ref{def:evaluation} by taking input values from the ports connected to the inflows in $r^{-1}(o)$ and producing a result in the port $t(o)$. Values of ports not attached to ready units simply remain untouched and they are passed onto the next state. To indicate value consumption from the ports used by triggered devices, an empty value of the corresponding type is placed. Collectively, this defines a state transition, which is formalised in Definition \ref{def:transition}.

\begin{definition}\label{def:evaluation}
Let ${o\in O}$ be an outflow of a circuit $\lambda$. The result of evaluating a computing device ${f(o)}$ under ${\delta^j}$ is denoted $f(o)^j$ and is given by $f(o)(\delta^j(s(i_1)),\ldots,\delta^j(s(i_k)))$ where $\{i_1,\ldots,i_k\}=r^{-1}(o)$ with $i_m\neq i_n$ for $m\neq n$ and $f(o)\in\{\kappa_{|r^{-1}(o)|},\uparrow_{|r^{-1}(o)|},\epsilon_{|r^{-1}(o)|}\}$. The value $f(o)^j$ is well-typed if and only if $f(o)^j\in\llbracket T(c(t(o)))\rrbracket$.
\end{definition}

\begin{definition}\label{def:transition}
If ${\delta^j}$ is the state of a circuit $\lambda$ at time ${j\geq 0}$, the next state ${\delta^{j+1}}$ is given as follows for all ${p\in P}$:
\[\delta^{j+1}(p) = 
    \begin{cases}
       f(o)^j & (\exists u \in R^j)(\exists o \in O)[\sigma(o)=u\text{ and }t(o)=p]\\       
       \delta^j(p) & (\nexists u \in R^j)[p\in\bullet u\cup u\bullet]\\ 
       \bot_{T(c(p))} & \text{otherwise}
    \end{cases}
\]
\end{definition} 

\section{On the Relationship Between Control-Driven Boolean Circuits and Classical NAND Circuits}\label{sec:equivalence}

In this section, we study the relationship between our proposal and classical NAND circuits. Although there are multiple functionally complete sets, we decide to use $\{\uparrow\}$ since NAND is invariant under input permutation by its commutative and associative properties. Using such a set also simplifies the treatment of computing devices so that only one logical operator is required. Having a functionally complete set means that we can construct any fundamental logic gate to perform a Boolean function on an input of fixed length, just as supported by classical NAND circuits.

\begin{definition}\label{def:classical-circuit}
A classical NAND circuit $C$ is an acyclic digraph ${(V,E)}$ where $V$ and $E$ are non-empty finite sets of vertices and edges, respectively. Each vertex ${v\in V}$ exclusively corresponds to: 
\begin{enumerate}
\renewcommand\labelenumi{\theenumi}
\renewcommand{\theenumi}{(\roman{enumi})}
\item an input variable if ${v^+=0}$ and ${v^-\geq 1}$,
\item an output variable if ${v^+\geq 1}$ and ${v^-=0}$, or
\item a NAND gate if ${v^+ \geq 1}$ and ${v^-=1}$.
\end{enumerate}
Each edge ${e\in E}$ expresses information that flows from an input variable to a gate, between gates or from a gate to an output variable.
\end{definition}

\begin{notation}
If $v\in V$, we use $\overrightarrow{v}$ and $\overleftarrow{v}$ to denote $\{(x,v)\in E\}$ and $\{(v,w)\in E\}$, noticing that $\overleftarrow{v}$ is a singleton by the equality $v^-=1$.
\end{notation}

\begin{remark}\label{rem:classical-circuits}
Notice in Definition \ref{def:classical-circuit} that ${v^+= 0=v^-}$ never holds so there are no vertices with no incoming and no outgoing edges. Also, observe that there are no NAND circuits with variables only, i.e., gates are always present.
\end{remark}

\begin{lemma}\label{lem:path-to-output}
If $C$ is a classical NAND circuit, for every $v_1\in V$ satisfying $v_1^-\neq 0$, there is a finite path $\langle (v_1,w_1),(v_2,w_2),\ldots,(v_n,w_n) \rangle$ of length $n\geq 1$ where $w_n^-=0$ and $(v_i,w_i)\in E$ for $i=1,\ldots,n$.   
\end{lemma}
\begin{proof}
Let $v_1\in V$ be an arbitrary vertex with $v_1^-\neq 0$ so there is some edge $(v_1,w_1)\in E$. If $w_1^-=0$, the lemma follows directly; otherwise, we apply the same argument repeatedly to produce a sequence $\langle (v_1,w_1),(v_2,w_2),\ldots,(v_n,w_n) \rangle$ of edges. This process must terminate after finitely many steps with some $w_n^-=0$, given that $C$ is finite and acyclic. 
\end{proof}

To transform a classical NAND circuit into its corresponding control-driven version, we use an enriched digraph structure which provides useful information for the transformation process. This structure, referred to as \emph{annotated NAND circuit}, is formalised below. 

\begin{definition}\label{def:annotated-circuit}
An annotated NAND circuit ${{\mathcal{C}}}$ is a 7-tuple $(C,G,E_0,E_1,g_0,g_1,\Xi)$ where:
\begin{itemize}
\item $C$ is a classical NAND circuit $(V,E)$,
\item ${G:=\{v\in V\mid v^+\geq 1\land v^-=1\}}$,
\item ${E_0:=E\setminus\{(v,w)\in E\mid v^+=0\land w^+\geq 1\land w^-=1\}}$,
\item ${E_1:=E\setminus\{(x,y)\in E\mid x^+\geq 1\land x^-=1\land y^-=0\}}$,
\item ${g_0\colon E_0\sqcup E_0\rightarrow G}$ is a function ${g_0((v,w),j)=v}$ for ${j\in\{0,1\}}$, 
\item ${g_1\colon E_1\sqcup E_1\rightarrow G}$ is a function ${g_1((x,y),k)=y}$ for ${k\in\{0,1\}}$, and
\item $\Xi\colon E_1\sqcup E_1\rightarrow E_0\sqcup E_0$ is a function $\Xi((x,y),l)=((y,w),l)$ where $(y,w)\in\overleftarrow{y}$ and ${l\in\{0,1\}}$.
\end{itemize}
\end{definition}

A glance at Definition \ref{def:annotated-circuit} reveals that $G$ contains all the gates of a given NAND circuit, and that each ${(v,w)\in E_0}$ is an edge that runs from a gate $v$ to a gate/output $w$ while each ${(x,y)\in E_1}$ goes from a gate/input $x$ to a gate $y$. The functions ${g_0}$ and ${g_1}$ just extract the gate from the edge they operate on. These functions are well-defined because edges cannot have multiple source or target nodes. Lemma \ref{lem:g-surjective} indicates that $g_0$ is surjective in addition. The function $\Xi$ simply chooses the unique outgoing edge of the target gate of an $E_1$-edge, while respecting tagging, which by Lemma \ref{lem:xi-surjective} is well-defined and surjective. By convention, tagging is realised via the natural numbers $0$ and $1$. 

\begin{lemma}\label{lem:g-surjective}
The functions $g_0\colon E_0\sqcup E_0\rightarrow G$ and $g_1\colon E_1\sqcup E_1\rightarrow G$ of an annotated NAND circuit are surjective.    
\end{lemma}
\begin{proof}
Given a gate $g\in G$, we use the fact $g^-=1$ from Definition \ref{def:annotated-circuit} to deduce that there is an edge $(g,w)\in E$. As this means $(g,w)\in E_0$, there must also be some $((g,w),j)\in E_0\sqcup E_0$ with $j\in\{0,1\}$. Simply applying $g_0$, we obtain $g_0((g,w),j)=g$, as required. The proof of surjectivity for $g_1$ is completely analogous.
\end{proof}

\begin{lemma}\label{lem:xi-surjective}
The function $\Xi\colon E_1\sqcup E_1\rightarrow E_0\sqcup E_0$ of an annotated NAND circuit is surjective.    
\end{lemma}
\begin{proof}
To show that $\Xi$ is well-defined, assume for contradiction that there are $((v_0,w_0),m),((v_1,w_1),n)\in E_0\sqcup E_0$ where $\Xi(e,k)=((v_0,w_0),m)=((v_1,w_1),n)$ for some $e\in E_1$ so that $v_0=v_1$ by the definition of $\Xi$. If $w_0\neq w_1$, $v_0^-,v_1^-\geq 2$ which contradicts the fact $v_0^-=v_1^-=1$ from Definition \ref{def:classical-circuit}. Since $m\neq n$ cannot hold under $\Xi$, we have that $\Xi$ is a well-defined function. 

To prove surjectivity, let $((v,w),j)\in E_0\sqcup E_0$ and use Definition \ref{def:annotated-circuit} to deduce that there is some edge $(x,v)\in E$, given that $v$ is a gate. As $x$ must be an input or a gate, $(x,v)\in E_1$ and hence $((x,v),j)\in E_1\sqcup E_1$ for $j\in\{0,1\}$. Using the definition of $\Xi$, we obtain $\Xi((x,v),j)=((v,y),j)$ for some $(v,y)\in E_0$. Since $v$ is a gate with $v^-=1$, $y=w$ must hold, i.e., $\Xi((x,v),j)=((v,y),j)=((v,w),j)$.
\end{proof}

The state of an annotated NAND circuit $\mathcal{C}$ is a function that maps each edge to some $\llbracket\mathbbm{B}\rrbracket$-value. Unlike control-driven circuits, the execution of $\mathcal{C}$ is persistent in the sense that truth values are retained once produced by gates \cite{AroraBarak2009}. This notion is formalised in Definition \ref{def:state-classical}.

\begin{definition}\label{def:state-classical}
The state of an annotated NAND circuit $\mathcal{C}$ is a function $\delta^j_\mathcal{C}\colon E\rightarrow\llbracket\mathbbm{B}\rrbracket$ for $j\geq 0$. When $j=0$, we say that $\delta^j_{\mathcal{C}}$ is the initial state of $\mathcal{C}$ which satisfies $\delta^j_{\mathcal{C}}(v,w)\in\{T,F\}$ for all $(v,w)\notin E_0$ and $\delta^j_{\mathcal{C}}(x,y)\in\{\bot_{\mathbbm{B}}\}$ for all $(x,y)\in E_0$. A state transition is defined as follows for ${(v,w)\in E}$ and $j\geq 0$:
\[\delta^{j+1}_C(v,w) = 
    \begin{cases}
       \uparrow_{v^+}(\delta^{j}_C(x_1,v),\ldots,\delta^{j}_C(x_{v^+},v)) & v\in G \text{ and } \delta^j_{\mathcal{C}}(v,w)=\bot_{\mathbb{B}} \text{ and } \\ 
       & \delta^{j}_C(x,v)\in\{T,F\}\text{ for all }(x,v)\in \vec{v} \\
       \delta^j_{C}(v,w)   & \text{otherwise}
    \end{cases}
\]
\end{definition}

\begin{definition}\label{def:gate-firing}
We say that a gate $v\in G$ fires at $j\geq1$ if $\delta^{j-1}_C(v,w)=\bot_{\mathbb{B}}$ for the unique $(v,w)\in\overleftarrow{v}$ and $\delta^{j-1}_C(x,v)\in\{T,F\}$ for all $(x,v)\in \vec{v}$.
\end{definition}

The procedure to transform an annotated NAND circuit into its corresponding control-driven counterpart is given in Definition \ref{def:transformation}.

\begin{definition}\label{def:transformation}
Given an annotated NAND circuit ${\mathcal{C}}$, we construct its corresponding control-driven circuit $\lambda^{{\mathcal{C}}}$ as follows:
\begin{itemize}
\item ${U\cong G}$,
\item ${P\cong E\sqcup E}$,
\item ${I\cong E_1\sqcup E_1}$,
\item ${O\cong E_0\sqcup E_0}$,
\item ${\Sigma:=\{0,1\}}$, and
\item $D:=\{\kappa_m\}_{m\geq 1}\cup\{\uparrow_m\}_{m\geq 1}\cup\{\epsilon_m\}_{m\geq 1}$.
\end{itemize}
If ${\psi\colon U\rightarrow G}$, ${\phi\colon P\rightarrow E\sqcup E}$, ${\gamma\colon I\rightarrow E_1\sqcup E_1}$ and ${\theta\colon O\rightarrow E_0\sqcup E_0}$ are the corresponding isomorphisms, and ${\iota_n\colon E_n\sqcup E_n\rightarrow E\sqcup E}$ is the obvious inclusion function for ${n\in\{0,1\}}$, we define the functions ${\sigma\colon O\rightarrow U}$, ${t\colon O\rightarrow P}$, ${\tau\colon I\rightarrow U}$, ${s\colon I\rightarrow P}$, ${c\colon P\rightarrow\Sigma}$, ${r\colon I\rightarrow O}$ and ${f\colon O\rightarrow D}$ as follows:
\begin{itemize}
\item ${\sigma(o)=(\psi^{-1}\circ g_0\circ\theta)(o)}$,
\item ${t(o)=(\phi^{-1}\circ\iota_0\circ\theta)(o)}$,
\item ${\tau(i)=(\psi^{-1}\circ g_1\circ\gamma)(i)}$,
\item ${s(i)=(\phi^{-1}\circ\iota_1\circ\gamma)(i)}$,
\item $c(p) = 
    \begin{cases}
       0 & \phi(p)=(e,0)\text{ for } e\in E\\
       1 & \phi(p)=(e,1)\text{ for } e\in E
    \end{cases}$
\item ${r(i)=(\theta^{-1}\circ\Xi\circ\gamma)(i)}$,
\item $f(o) = 
    \begin{cases}
       \kappa_{|(\Xi^{-1}\circ\theta)(o)|} & \theta(o)=((v,w),0)\text{ for } (v,w)\in E_0\\
       \uparrow_{|(\Xi^{-1}\circ\theta)(o)|} & \theta(o)=((v,w),1)\text{ for } (v,w)\in E_0
    \end{cases}$
\end{itemize}
The initial state of $\lambda^{{\mathcal{C}}}$ is given as follows for every $(v,w)\in E$:
\[
\begin{aligned}
\delta^0(\phi^{-1}((v,w),0)) &= \begin{cases}
       * & \delta^0_C(v,w)\in\{T,F\} \\
       \bot_{\mathbbm{1}} & \text{otherwise}
    \end{cases} \\
\delta^0(\phi^{-1}((v,w),1)) &= \delta^0_C(v,w)    
\end{aligned}
\]
Subsequent states are given according to the transition function from Definition \ref{def:transition}.
\end{definition}

\begin{remark}
Although the inverses of the isomorphisms presented in Definition \ref{def:transformation} yield a singleton set because they are injective, in this paper we treat them as producing individual elements.  
\end{remark}

\begin{remark}\label{rem:ins-by-r}
Notice in Definition \ref{def:transformation} that: 
\[
\begin{aligned}
r^{-1}(\theta^{-1}((v,w), j)) &= (\gamma^{-1}\circ \Xi^{-1}\circ \theta \circ \theta^{-1})((v,w), j)) \\
 &= (\gamma^{-1}\circ \Xi^{-1})((v,w), j)) \\
 &= \gamma^{-1}(\{((x,v),j)\}) \\
 &= \{\gamma^{-1}((x_1,v),j),\ldots,\gamma^{-1}((x_{v^+},v),j)\}
\end{aligned}
\] for $(x,v)\in\vec{v}$ and $j\in\{0,1\}$.
\end{remark}

The transformation procedure from Definition \ref{def:transformation} defines a one-to-one mapping between gates and computation units, and produces two ports for every classical edge: one for representing control and another for expressing Boolean data. Accordingly, if $g\in G$, we have exactly $2g^+$ ports in $\bullet\psi^{-1}(g)$ and exactly $2$ ports in $\psi^{-1}(g)\bullet$ (see Theorem \ref{th:two-unit-outputs}). More precisely, the ports in $\bullet\psi^{-1}(g)$ and $\psi^{-1}(g)\bullet$ correspond to the duplication of edges in $\overrightarrow{g}$ and $\overleftarrow{g}$, respectively, as formalised in Theorem \ref{th:transformation-structure-gates}. Accordingly, if a unit $\psi^{-1}(g)$ is enabled at time $j$, then it is also ready at that time (see Lemma \ref{lem:enabled-ready}). Evaluating $\psi^{-1}(g)$ produces the following at $j+1$: (i) a control signal and (ii) the result of applying a $g^+$-ary NAND function on the $\overrightarrow{g}$-edge values. The values from (i) and (ii) are assigned to the unique control port and the unique Boolean port in $\psi^{-1}(g)\bullet$, respectively. This assignment is formalised in Corollaries \ref{cor:ready-then-signals} and \ref{cor:ready-then-values} which, by Theorem \ref{th:types-preserved}, preserves the correct type.\footnote{Notice that, even though $\epsilon$-devices belong to the set $D$, they are never used in the corresponding control-driven circuit. The purpose of these devices is merely to enable more expressive computations, i.e., they are not needed for simulating classical circuits.}

\begin{theorem}\label{th:two-unit-outputs}
If $g\in G$, the computation unit $\psi^{-1}(g)$ has exactly one control port in $\psi^{-1}(g)\bullet$ and exactly one Boolean port in $\psi^{-1}(g)\bullet$.
\end{theorem}
\begin{proof}
Since $g^-=1$ by Definition \ref{def:classical-circuit}, there is only one edge $(g,w)\in E$ for some $w\in V$. Consequently, $((g,w),0),((g,w),1)\in E\sqcup E$ so there are unique ports $\phi^{-1}((g,w),0),\phi^{-1}((g,w),1)\in P$ by the injectivity of $\phi$. Applying the functions $c$ and $T$ from Definitions \ref{def:transformation} and \ref{def:typing-function} allows us to obtain $T(c(\phi^{-1}((g,w),0)))=\mathbbm{1}$ and $T(c(\phi^{-1}((g,w),1)))=\mathbb{B}$ directly. 
\end{proof}

\begin{theorem}\label{th:transformation-structure-gates}
If $(g,w)\in E$ with $g\in G$ and $j\in\{0,1\}$, $\phi^{-1}((g,w),j)\in\psi^{-1}(g)\bullet$ for the unique $(g,w)\in\overleftarrow{g}$ and $\phi^{-1}((x,g),j)\in\bullet\psi^{-1}(g)$ for each $(x,g)\in\vec{g}$.
\end{theorem}
\begin{proof}
By Definition \ref{def:transformation}, we know there is a port $\phi^{-1}((x,g),j)\in P$ for each $(x,g)\in\overrightarrow{g}$ and $j\in\{0,1\}$. As $g\in G$, Definition \ref{def:annotated-circuit} says $((x,g),j)\in E_1\sqcup E_1$ so there is a unique inflow $\gamma^{-1}((x,g),j)\in I$ (recall that $\gamma$ is injective because it is an isomorphism). Considering that $\iota_1$ is an inclusion, we apply $s$ and $\tau$ from Definition \ref{def:transformation} to obtain:
\[
\begin{aligned}
s(\gamma^{-1}((x,g),j)) &= (\phi^{-1}\circ\iota_1\circ\gamma\circ\gamma^{-1})((x,g),j) =(\phi^{-1}\circ\iota_1)((x,g),j) = \phi^{-1}((x,g),j) & & \\
\tau(\gamma^{-1}((x,g),j)) &= (\psi^{-1}\circ g_1\circ\gamma\circ\gamma^{-1})((x,g),j) = (\psi^{-1}\circ g_1)((x,g),j) = \psi^{-1}(g) & & 
\end{aligned}
\] So, $\phi^{-1}((x,g),j)\in\bullet\psi^{-1}(g)$ by Notation \ref{not:pre-post-sets}. A similar reasoning can be used to show the claim $\phi^{-1}((g,w),j)\in\psi^{-1}(g)\bullet$ for the unique $(g,w)\in\overleftarrow{g}$.
\end{proof}

\begin{corollary}\label{cor:ready-then-signals}
If $\phi^{-1}((g,w),0)\in\psi^{-1}(g)\bullet$ and $\psi^{-1}(g)\in R^j$, $\delta^{j+1}(\phi^{-1}((g,w),0))=*$.
\end{corollary}
\begin{proof}
By Theorem \ref{th:transformation-structure-gates}, we know there is a unique outflow $\theta^{-1}((g,w),0)\in O$ where $\phi^{-1}((g,w),0)\in\psi^{-1}(g)\bullet$. Consequently, $f(\theta^{-1}((g,w),0))=\kappa_{|(\Xi^{-1}\circ\theta\circ\theta^{-1})((g,w),0)|}=\kappa_{|\Xi^{-1}((g,w),0)|}$ as per Definition \ref{def:transformation}. Since $\Xi^{-1}((g,w),0)=\{(((x_1,g),0),\ldots,((x_{g^+},g),0))\}$ is evident from Definition \ref{def:annotated-circuit}, we have $f(\theta^{-1}((g,w),0))=\kappa_{g^+}$.

Using Theorem \ref{th:transformation-structure-gates}, we deduce the existence of inflows $\gamma^{-1}((x_1,g),0),\ldots,\gamma^{-1}((x_{g^+},g),0)\in I$ with $\phi^{-1}((x_1,g),0),\ldots,\phi^{-1}((x_{g^+},g),0)\in\bullet\psi^{-1}(g)$. Applying $s$ from Definition \ref{def:transformation}, we obtain $s(\gamma^{-1}((x_k,g),0))=\phi^{-1}((x_k,g),0)$ for $k=1,\ldots,g^+$. 

Assuming $\psi^{-1}\in R^j$, Definition \ref{def:ready-units} says $\delta^j(\phi^{-1}((x_k,g),0))=*$. Having $r^{-1}(\theta^{-1}((g,w),k))=\{\gamma^{-1}((x_1,g),k),\ldots,\gamma^{-1}((x_{g^+},g),k)\}$ from Remark \ref{rem:ins-by-r}, we simply use Definitions \ref{def:circuit} and \ref{def:evaluation} to obtain:
\begin{align*}
f(\theta^{-1}((g,w),0))^j &= \kappa_{g^+}(\delta^j(s(\gamma^{-1}((x_1,g),0)),\ldots,\delta^j(s(\gamma^{-1}((x_{g^+},g),0))))) \\
      &= \kappa_{v^+}(\delta^j(\phi^{-1}((x_1,g),0)),\ldots,\delta^j(\phi^{-1}((x_{g^+},g),0))) \\
      &= \kappa_{g^+}(*,\ldots,*)=*
\end{align*}
Considering that $\sigma(\theta^{-1}((g,w),0))=\psi^{-1}(g)$ and $t(\theta^{-1}((g,w),0))=\phi^{-1}((g,w),0)$ follow from Theorem \ref{th:transformation-structure-gates}, we apply Definition \ref{def:transition} to obtain: \[\delta^{j+1}(\phi^{-1}((g,w),0))=f(\theta^{-1}((g,w),0))^j=*\]
\end{proof}

\begin{corollary}\label{cor:ready-then-values}
If $\phi^{-1}((g,w),1)\in\psi^{-1}(g)\bullet$ and $\psi^{-1}(g)\in R^j$:
\[
\delta^{j+1}(\phi^{-1}((g,w),1))=\uparrow_{g^+}(\delta^j(\phi^{-1}((x_1,g),1)),\ldots,\delta^j(\phi^{-1}((x_{g^+},g),1)))\in\{T,F\}
\]
for $(x_n,g)\in\overrightarrow{v}$ and $n=1,\ldots,g^+$.
\end{corollary}
\begin{proof}
The proof is analogous to that of Corollary \ref{cor:ready-then-signals}.
\end{proof}

\begin{lemma}\label{lem:enabled-ready}
If $\psi^{-1}(g)$ is a computation unit of $\lambda^{\mathcal{C}}$ that is enabled under $\delta^j$, then $\psi^{-1}(g)\in R^j$.
\end{lemma}
\begin{proof}
Assume that there are distinct units $u_1,u_2\in U$ with $\bullet u_1=\bullet u_2$ so there are distinct inflows $i_1,i_2\in I$ with $s(i_1)=s(i_2)$. Using $s$ from Definition \ref{def:transformation}, while considering that $\iota_1$ is an inclusion, we obtain:
\[
\begin{aligned}
\phi^{-1}(\iota_1(\gamma(i_1))) &= \phi^{-1}(\iota_1(\gamma(i_2))) \\
\phi^{-1}(\gamma(i_1)) &= \phi^{-1}(\gamma(i_2))
\end{aligned}
\]

But $\phi^{-1}$ is injective because it is an isomorphism so $\gamma(i_1)=\gamma(i_2)$, which implies that $\bullet u_1=\bullet u_2$ only holds when $u_1=u_2$. That is, if $g_1\in G$, $\bullet \psi^{-1}(g_1)\neq\bullet \psi^{-1}(g_2)$ for all $\psi^{-1}(g_2)\in U$ with $g_1\neq g_2$. Using Definition \ref{def:ready-units} and assuming $\psi^{-1}(g)$ is enabled under $\delta^j$, we use the reflexivity property of equivalence relations to deduce that the singleton set $\{\psi^{-1}(g_1)\}$ is in the partition induced by $u_1\sim u_2\iff\bullet u_1=\bullet u_2$. Hence, by the same definition, $\psi^{-1}(g_1)\in R^j$.
\end{proof}

\begin{theorem}\label{th:types-preserved}
If $(v,w)\in E$, then $\delta^j(\phi^{-1}((v,w),k))\in\llbracket T(c(\phi^{-1}((v,w),k)))\rrbracket$ for every $j\geq 0$ and $k\in\{0,1\}$.
\end{theorem}
\begin{proof}
We prove by induction on discrete time with the observation that, for an edge $(v,w)\in E$, we have $c(\phi^{-1}((v,w),0))=0$ and $c(\phi^{-1}((v,w),1))=1$ by Definition \ref{def:transformation}. As Definition \ref{def:typing-function} states $T(0)=\mathbbm{1}$ and $T(1)=\mathbb{B}$, it suffices to show $\delta^j(\phi^{-1}((v,w),0))\in\llbracket\mathbbm{1}\rrbracket$ and $\delta^j(\phi^{-1}((v,w),1))\in\llbracket\mathbb{B}\rrbracket$ for $j\geq 0$.

\textbf{Base case}. When $j=0$, we immediately have $\delta^0(\phi^{-1}((v,w),0))\in\{*,\bot_{\mathbbm{1}}\}=\llbracket\mathbbm{1}\rrbracket$ and $\delta^0(\phi^{-1}((v,w),1))=\delta^0_\mathcal{C}(v,w)\in\{T,F,\bot_{\mathbb{B}}\}=\llbracket\mathbb{B}\rrbracket$ as per Definitions \ref{def:state-classical} and \ref{def:transformation}.

\textbf{Inductive step}. Assuming for inductive hypothesis that the claim holds for every $k\geq 0$, we prove it also holds for $k+1$ by considering the three cases from Definition \ref{def:transition} with $m\in\{0,1\}$:
\begin{enumerate}
\item If $\delta^{k+1}(\phi^{-1}((v,w),m))=f(\theta^{-1}((v,w),m))^k$ for some outflow $\theta^{-1}((v,w),m)\in O$ that satisfies $t(\theta^{-1}((v,w),m))=\phi^{-1}((v,w),m)$, we have: 
\[
\delta^{k+1}(\phi^{-1}((v,w),0))=\kappa_{|\Xi^{-1}((v,w),0)|}=*\in\llbracket\mathbbm{1}\rrbracket \text{ by Corollary \ref{cor:ready-then-signals} and}
\]
\[
\delta^{k+1}(\phi^{-1}((v,w),1))\in\{T,F\}\subset\llbracket\mathbb{B}\rrbracket \text{ by Corollary \ref{cor:ready-then-values}.}
\]
\item If $\delta^{k+1}(\phi^{-1}((v,w),m))=\delta^{k}(\phi^{-1}((v,w),m))$, we simply use the inductive hypothesis to derive $\delta^{k+1}(\phi^{-1}((v,w),m))\in\llbracket T(c(\phi^{-1}((v,w),m)))\rrbracket$.
\item If $\delta^{k+1}(\phi^{-1}((v,w),m))=\bot_{T(c(\phi^{-1}((v,w),m)))}$, we use again the inductive hypothesis to obtain $\delta^{k+1}(\phi^{-1}((v,w),m))\in\llbracket T(c(\phi^{-1}((v,w),m)))\rrbracket$.
\end{enumerate}
\end{proof}

The transformation procedure described in Definition \ref{def:transformation} produces a valid control-driven circuit, which satisfies the constraints imposed by Definition \ref{def:circuit}, namely surjectivity on $\sigma$, $\tau$, ${\sigma\restriction_{(c\circ t)^{-1}(0)}}$, ${\tau\restriction_{(c\circ s)^{-1}(0)}}$ and $r$, as well as the satisfaction of the equation $\sigma\circ r=\tau$ and the existence of control ports at the interface. These restrictions follow from Lemmas \ref{lem:sigma-surjective}, \ref{lem:tau-restrictions-surjective}, \ref{r-surjective}, \ref{lem:encapsulation-satisfied} and \ref{lem:interface-control-satisfied}. Thus, for every classical NAND circuit, Theorems \ref{th:classical-to-control} and \ref{th:sound-transformation} state that there is a corresponding (connected) control-driven circuit. Theorem \ref{th:gate-firing} and Corollary \ref{cor:computability} entail that such a corresponding circuit is able to simulate the same Boolean function as its classical counterpart.

\begin{lemma}\label{lem:sigma-surjective}
The functions $\sigma\colon O\rightarrow U$ and $\tau\colon I\rightarrow U$ of a control-driven circuit $\lambda^{\mathcal{C}}$ are surjective.
\end{lemma}
\begin{proof}
To show ${\tau=\psi^{-1}\circ g_1\circ\gamma}$ is onto, let us pick some ${u \in U}$ by leveraging the fact $U\neq\emptyset$ derived from $G\neq\emptyset$ (see Remark \ref{rem:classical-circuits} and Definition \ref{def:transformation}). Since ${\psi^{-1}}$ is surjective because it is an isomorphism, there must be some gate ${g \in G}$ where ${\psi^{-1}(g)=u}$. With the surjectivity of ${g_1}$, stated in Lemma \ref{lem:g-surjective}, we deduce the existence of ${((x, y), k)\in E_1\sqcup E_1}$ where ${g_1((x, y), k)=g}$ and ${k \in \{0,1\}}$. Given that $\gamma$ is also an isomorphism, there must be some ${i \in I}$ with ${\gamma(i)=((x, y), k)}$. That is, $\psi^{-1}(g_1(\gamma(i)))=\psi^{-1}(g_1((x,y),k))=\psi^{-1}(g)=u$. Proving the surjectivity of $\sigma$ follows analogously.
\end{proof}

\begin{lemma}\label{lem:tau-restrictions-surjective}
The functions ${\sigma\restriction_{(c\circ t)^{-1}(0)}}$ and ${\tau\restriction_{(c\circ s)^{-1}(0)}}$ of a control-driven circuit $\lambda^{\mathcal{C}}$ are surjective.
\end{lemma}
\begin{proof}
We just prove for ${\sigma\restriction_{(c\circ t)^{-1}(0)}}$ since the onto property of ${\tau\restriction_{(c\circ s)^{-1}(0)}}$ can be showed in a similar manner. For this, pick some $u\in U$ by leveraging the fact $U\neq\emptyset$ derived from $G\neq\emptyset$ (see Remark \ref{rem:classical-circuits} and Definition \ref{def:transformation}). As $\sigma$ is surjective by Lemma \ref{lem:sigma-surjective}, there must be some $o\in O$ where $\sigma(o)=u$. By the totality of $t$, we have that the port $t(o)$ exists. Applying  $t=\phi^{-1}\circ\iota_0\circ\theta$ from Definition \ref{def:transformation} allows us to deduce $t(o)=\phi^{-1}(\iota_0(\theta(o)))$ in addition. 

Now, considering $\theta(o)\in E_0\sqcup E_0$, we have two possibilities: $\theta(o)=((v,w),0)$ or $\theta(o)=((v,w),1)$. Picking the former while considering $\iota_0((v,w),0)=((v,w),0)\in E\sqcup E$ allows us to derive: 
\[
t(o)=\phi^{-1}(\iota_0(\theta(o)))=\phi^{-1}(\iota_0((v,w),0))=\phi^{-1}((v,w),0)
\]
Finally, applying $c$ from Definition \ref{def:transformation} with the observation $\phi(\phi^{-1}((v,w),0))=((v,w),0)$, we obtain:
\[
c(t(o))=c(\phi^{-1}(\iota_0(\theta(o))))=c(\phi^{-1}(\iota_0((v,w),0)))=c(\phi^{-1}((v,w),0))=0
\]
By the above, $o\in(c\circ t)^{-1}(0)$ holds clearly and, consequently, $\sigma\restriction_{(c\circ t)^{-1}(0)}(o)=\sigma(o)=u$. That is, ${\sigma\restriction_{(c\circ t)^{-1}(0)}}$ is surjective, as required.
\end{proof}

\begin{lemma}\label{r-surjective}
The function $r$ of a control-driven circuit $\lambda^{\mathcal{C}}$ is surjective.
\end{lemma}
\begin{proof}
Definition \ref{def:transformation} states that ${r=\theta^{-1}\circ\Xi\circ\gamma}$. So, the proposition holds by Lemma \ref{lem:xi-surjective} and by the surjectivity of the isomorphisms $\theta$ and $\gamma$.
\end{proof}

\begin{lemma}\label{lem:encapsulation-satisfied}
Given a control-driven circuit $\lambda^{\mathcal{C}}$, the equation $\sigma\circ r=\tau$ holds.
\end{lemma}
\begin{proof}
We first prove $g_0\circ\Xi=g_1$ as follows: $g_0(\Xi((x,y),k))=g_0((y,w),k)=y=g_1((x,y),k)$. Then:
\begin{align}
\sigma\circ r &= \psi^{-1}\circ g_0\circ\theta\circ\theta^{-1}\circ\Xi\circ\gamma & \text{by Definition \ref{def:transformation}} \\
      &= \psi^{-1}\circ g_0\circ\Xi\circ\gamma & \text{by the identity law} \\
      &= \psi^{-1}\circ g_1\circ\gamma & \text{by $g_0\circ\Xi=g_1$} \\
      &= \tau & \text{by Definition \ref{def:transformation}}
\end{align}    
\end{proof}

\begin{lemma}\label{lem:interface-control-satisfied}
If $\lambda^{\mathcal{C}}$ is a control-driven circuit, there are ports ${p\in P\setminus t(O)}$ and ${q\in P\setminus s(I)}$ where ${c(p)=0=c(q)}$.
\end{lemma}
\begin{proof}
We just show that there is some port ${p\in P\setminus t(O)}$ with ${c(p)=0}$ since the proof of the other is symmetric. To do this, we observe that gates always exist as per Remark \ref{rem:classical-circuits} and that they always have fan-in $\geq 1$ as per Definition \ref{def:classical-circuit}. So, there always is some ${(v,w)\in E}$ running from an input $v$ to a gate $w$. Consequently, ${((v,w),0),((v,w),1)\in E\sqcup E}$ so that ${\phi^{-1}((v,w),0)\in P}$. Supposing for contradiction the existence of some ${o\in O}$ with ${t(o)=\phi^{-1}(\iota_0(\theta(o)))=\phi^{-1}((v,w),0)}$, we have ${((v, w), 0)\in E_0\sqcup E_0}$ because ${\iota_0}$ is an inclusion, i.e., ${(v,w)\in E_0}$. By the definition of ${E_0}$, we have ${v^+\neq 0}$, clearly a contradiction to the fact ${v^+=0}$ imposed by Definition \ref{def:classical-circuit}. Hence, ${\phi^{-1}((v,w),0)\notin t(O)}$. By Definition \ref{def:transformation}, we conclude ${c(\phi^{-1}((v,w),0))=0}$, as required.   
\end{proof}

\begin{theorem}\label{th:classical-to-control}
For every annotated NAND circuit, there is an equivalent control-driven circuit.
\end{theorem}
\begin{proof}
Definition \ref{def:transformation} says that, given an annotated NAND circuit ${\mathcal{C}}$, there is a corresponding control-driven circuit $\lambda^{\mathcal{C}}$. Considering that the set of ${\mathcal{C}}$-nodes is never empty (as per Definition \ref{def:classical-circuit}), we have that the set of ports of $\lambda^{\mathcal{C}}$ is not empty either. By Definition \ref{def:transformation}, $0\in\Sigma$ in addition. The rest of the conditions imposed by Definition \ref{def:computon} follow from Lemmas \ref{lem:sigma-surjective}, \ref{lem:tau-restrictions-surjective}, \ref{r-surjective}, \ref{lem:encapsulation-satisfied} and \ref{lem:interface-control-satisfied}. That is, $\lambda^{\mathcal{C}}$ is a well-formed circuit.
\end{proof}

\begin{theorem}\label{th:sound-transformation}
If ${\mathcal{C}}$ is an annotated NAND circuit, then $\lambda^{\mathcal{C}}$ is connected.
\end{theorem}
\begin{proof}
Let ${p\in s(I)\cup P^+}$ so that ${\phi(p)=((v_1,w_1),j)\in E\sqcup E}$ for ${j\in \{0,1\}}$. Since $v_1^-\neq 0$, we use Lemma \ref{lem:path-to-output} to deduce the existence of a non-empty finite path ${\langle (v_1,w_1),\ldots,(v_n,w_n) \rangle}$ where ${w_n^-=0}$. Supposing for contradiction ${\phi^{-1}((v_n,w_n),j)\notin P^-}$ for ${j\in \{0,1\}}$, there must be some ${i\in I}$ with ${s(i)=\phi^{-1}(\iota_1(\gamma(i)))=\phi^{-1}((v_n,w_n),j)}$. As ${\phi^{-1}}$ is injective because it is an isomorphism, we have ${\iota_1(\gamma(i))=((v_n,w_n),j)}$ so that ${(v_n,w_n)\in E_1}$ (considering that $\iota_1$ is an inclusion). By the definition of $E_1$, this means that $w_n$ must be a gate with $w_n^-=1$; thus, contradicting the fact ${w_n^-=0}$ from Lemma \ref{lem:path-to-output}.
\end{proof}

\begin{theorem}\label{th:gate-firing}
If $g\in G$ is a gate that fires at $j\geq 1$ during the execution of an annotated Boolean circuit $\mathcal{C}$, then $\delta^j(\phi^{-1}((g,w),0))=*$ and $\delta^j(\phi^{-1}((g,w),1))=\delta^j_{\mathcal{C}}(g,w)$ in $\lambda^{\mathcal{C}}$.
\end{theorem}
\begin{proof}
As the digraph structure of $\mathcal{C}$ is acyclic, we observe that there is a topological ordering of gates. So, we prove by induction on gate depth, i.e., the maximum number of gates on a path from an input variable to a given gate.

\textbf{Base case}. By Definition \ref{def:classical-circuit}, we know that a gate $g\in G$ at depth $0$ is necessarily connected from input variables $x_1,\ldots,x_{g^+}$, i.e., there is an edge $(x_n,g)\in\overrightarrow{v}$ for $n=1,\ldots,g^+$. Using Theorem \ref{th:transformation-structure-gates}, we know $\phi^{-1}((x_n,g),0),\phi^{-1}((x_n,g),1)\in\bullet\psi^{-1}(g)$. Since $\delta^0_C(x_n,g)\in\{T,F\}$ by Definition \ref{def:state-classical}, we have $\delta^0(\phi^{-1}((x_n,g),0))=*$ and $\delta^0(\phi^{-1}((x_n,g),1))=\delta^0_C(x_n,g)\in\{T,F\}$ as per Definition \ref{def:transformation}. That is, $\psi^{-1}(g)$ is enabled under $\delta^0$ and, consequently, $\psi^{-1}(g)\in R^0$ as per Lemma \ref{lem:enabled-ready}.

Since $\psi^{-1}(g)\in R^0$ and Theorem \ref{th:transformation-structure-gates} entails $\phi^{-1}((g,w),0)\in\psi^{-1}(v)\bullet$ for the unique edge $(g,w)\in\overleftarrow{v}$, we use Corollary \ref{cor:ready-then-signals} to deduce $\delta^1(\phi^{-1}((g,w),0))=*$ and Corollary \ref{cor:ready-then-values} to obtain $\delta^{1}(\phi^{-1}((g,w),1))=\uparrow_{v^+}(\delta^0(\phi^{-1}((x_1,g),1)),\ldots,\delta^0(\phi^{-1}((x_{g^+},g),1)))\in\{T,F\}$.

Considering that Definition \ref{def:gate-firing} states that $g$ fires at time $1$ only if $\delta^0_C(g,w)=\bot_{\mathbb{B}}$ and $\delta^0_C(x_n,g)\in\{T,F\}$, we use Definition \ref{def:state-classical} to deduce $\delta^1_C(g,w)=\uparrow_{g^+}(\delta^{0}_C(x_1,g),\ldots,\delta^{0}_C(x_n,g))\in\{T,F\}$. That is, $\delta^1_C(g,w)=\delta^{1}(\phi^{-1}((g,w),1))$, as required.

\textbf{Inductive step}. Assuming that the claim holds for gates at depth $k$, we now prove it also holds for a gate $g\in G$ at depth $k+1$, using Definition \ref{def:classical-circuit} to observe that $g$ can only be connected from gates. Considering $g^+\geq 1$ and $n=1,\ldots,g^+$, we know $(x_n,g)\in\overrightarrow{g}$ satisfies $x_n\in G$ with $\phi^{-1}((x_n,g),0),\phi^{-1}((x_n,g),1)\in\bullet\psi^{-1}(g)$. Thus, if $x_n$ fires at $j$, we use the inductive hypothesis to establish $\delta^j(\phi^{-1}((x_n,g),0))=*$ and $\delta^j_{\mathcal{C}}(x_n,g)=\delta^{j}(\phi^{-1}((x_n,g),1))\in\{T,F\}$. Since $\phi^{-1}((x_n,g),0),\phi^{-1}((x_n,g),1)\in\bullet\psi^{-1}(g)$ as per Theorem \ref{th:transformation-structure-gates}, we have $\psi^{-1}(g)\in R^j$ by Lemma \ref{lem:enabled-ready}. Using Corollaries \ref{cor:ready-then-signals} and \ref{cor:ready-then-values}, we deduce:
\[
\delta^{j+1}(\phi^{-1}((g,w),0))=*
\]
\[
\delta^{j+1}(\phi^{-1}((g,w),1))=\uparrow_{g^+}(\delta^j(\phi^{-1}((x_1,g),1)),\ldots,\delta^j(\phi^{-1}((x_{g^+},g),1)))\in\{T,F\}
\]

Now, if $g$ fires at time $j+1$, Definition \ref{def:gate-firing} states $\delta^j_{\mathcal{C}}(g,w)=\bot_{\mathbb{B}}$ and $\delta^j_{\mathcal{C}}(x_n,g)\in\{T,F\}$. Simply using Definition \ref{def:state-classical}, we deduce $\delta^{j+1}_C(g,w)=\uparrow_{g^+}(\delta^{j}_C(x_1,g),\ldots,\delta^{j}_C(x_n,g))\in\{T,F\}$. That is, $\delta^{j+1}_C(g,w)=\delta^{j+1}(\phi^{-1}((g,w),1))$, as required.
\end{proof}

\begin{corollary}\label{cor:computability}
Every Boolean function that is computable by a classical Boolean circuit $C$ is also computable by $\lambda^{\mathcal{C}}$.        
\end{corollary}
\begin{proof}
We know that the output of $\mathcal{C}$ is the value $\delta^j_{\mathcal{C}}(v,w)$ resulting from firing the gate $v$ with the highest depth within $\mathcal{C}$ \cite{AroraBarak2009}. Since $\delta^j(\phi^{-1}((v,w),1))=\delta^j_{\mathcal{C}}(v,w)$ by Theorem \ref{th:gate-firing}, we conclude that the claim is true.
\end{proof}

Corollary \ref{cor:computability} clearly implies that control-driven circuits are at least as powerful as their classical counterpart in terms of what can be computed. But they are more expressive since they provide operators for branching and iteration, which are control flow constructs that cannot be explicitly represented within classical circuits. Classical circuits only support gate sequencing and gate parallelising implicitly. 

Although individual control-driven circuits are more expressive, they are not sufficient to simulate Boolean functions of arbitrary length. To address this, we require an extension to support non-uniform computation, as described by Theorem \ref{th:computons-any-boolean}.

\begin{theorem}\label{th:computons-any-boolean}
For any Boolean function, there is a family of control-driven Boolean circuits able to compute it.
\end{theorem}
\begin{proof}
Given a Boolean function ${f\colon\{0,1\}^*\rightarrow\{0,1\}}$, we know that we can always define a family $\{C_j\}_{j\in\mathbb{N}}$ of classical NAND circuits where, for all ${j\in\mathbb{N}}$ and all ${b\in\{0,1\}^j}$, $C_j$ computes ${f(b)}$. Since Definition \ref{def:annotated-circuit} does not alter structure but just adds extra information, there also is a family $\{\mathcal{C}_j\}_{j\in\mathbb{N}}$ where each ${\mathcal{C}_j}$ is the corresponding annotated circuit of $C_j$. By Theorem \ref{th:classical-to-control} and Corollary \ref{cor:computability}, there consequently is a family ${\{\lambda^{\mathcal{C}_j}\}_{j\in\mathbb{N}}}$ where each $\lambda^{\mathcal{C}_j}$ computes ${f(b)}$.  
\end{proof}

\section{Applications}\label{sec:applications}

\begin{figure}[H]
 \captionsetup{position=above}
 \centering 
 \subcaptionbox{NOT circuit}
 {
 \begin{tikzpicture}
 \computonPrimitive{1}{0}{0.6}{1.2}{}

 \node at (1.3,1){\scriptsize $\kappa_1$};\draw[densely dotted] (1,1) -- (1.15,1);\draw[densely dotted] (1.4,1) -- (1.6,1);
 \flow{{0.2,1}}{$(0.2,1)+(0.8,0)$}{dashed}{};
 \node[draw=black,fill=white,inner sep=0pt,minimum size=3pt] (1q0) at (0.2,1) {};  
 \din{1i1}{0.2}{0.6}{};

 \node at (1.3,0.6){\scriptsize $\lnot$};\draw[densely dotted] (1,0.6) -- (1.2,0.6);\draw[densely dotted] (1.4,0.6) -- (1.6,0.6);
 \flow{{1.6,1}}{$(1.6,1)+(0.8,0)$}{dashed}{};
 \node[fill=black,inner sep=0pt,minimum size=3pt] (2q1) at (2.4,1) {};
 \dout{1o1}{1.6}{0.6}{};
 \end{tikzpicture}
 }
 \subcaptionbox{AND circuit\label{fig:composite-and}}
 {
 \begin{tikzpicture}
\computonComposite{0.4}{-0.2}{4.4}{1.6};
\qmatch{3q}{2.5}{1}{};\flow{{1.45,1}}{{2.5,1}}{dashed}{};\flow{{2.6,1}}{{3.6,1}}{dashed}{};
\dmatch{3d}{2.5}{0.4}{}{above};\flow{{1.5,0.4}}{{2.5,0.4}}{}{};\flow{{2.56,0.4}}{{3.6,0.4}}{}{};

\computonPrimitive{1}{0}{0.6}{1.2}{}
\node at (1.3,1){\scriptsize $\kappa_1$};\draw[densely dotted] (1,1) -- (1.15,1);\draw[densely dotted] (1.4,1) -- (1.6,1);
\node at (1.3,0.4){\scriptsize $\uparrow_2$};\draw[densely dotted] (1,0.6) -- (1.2,0.4);\draw[densely dotted] (1,0.2) -- (1.2,0.4);\draw[densely dotted] (1.4,0.4) -- (1.6,0.4);
\qin{1q1}{0.2}{1}{};
\din{1i1}{0.2}{0.6}{};
\din{1i2}{0.2}{0.2}{};

\computonPrimitive{3.6}{0}{0.6}{1.2}{}
\node at (3.9,1){\scriptsize $\kappa_1$};\draw[densely dotted] (3.6,1) -- (3.75,1);\draw[densely dotted] (4,1) -- (4.2,1);
\node at (3.9,0.4){\scriptsize $\lnot$};\draw[densely dotted] (3.6,0.4) -- (3.8,0.4);\draw[densely dotted] (4,0.4) -- (4.2,0.4);
\qout{2q1}{4.2}{1}{};
\dout{2o1}{4.2}{0.4}{};
 \end{tikzpicture}
 }
 \subcaptionbox{OR circuit\label{fig:composite-or}}
 {
 \begin{tikzpicture}
\computonComposite{0.4}{-0.4}{4.2}{3.4};
\computonComposite{0.8}{-0.2}{1.1}{3};

\computonPrimitive{1}{0}{0.6}{1.2}{}
\node at (1.3,0.6){\scriptsize $\lnot$};\draw[densely dotted] (1,0.6) -- (1.2,0.6);\draw[densely dotted] (1.4,0.6) -- (1.6,0.6);
\node at (1.3,1){\scriptsize $\kappa_1$};\draw[densely dotted] (1,1) -- (1.2,1);\draw[densely dotted] (1.4,1) -- (1.6,1);
\qin{1q1}{0.2}{1}{};
\din{1i1}{0.2}{0.6}{};

\computonPrimitive{1}{1.4}{0.6}{1.2}{}
\node at (1.3,1.9){\scriptsize $\lnot$};\draw[densely dotted] (1,1.9) -- (1.2,1.9);\draw[densely dotted] (1.4,1.9) -- (1.6,1.9);
\node at (1.3,2.4){\scriptsize $\kappa_1$};\draw[densely dotted] (1,2.4) -- (1.2,2.4);\draw[densely dotted] (1.4,2.4) -- (1.6,2.4);
\qin{2q1}{0.2}{2.4}{};
\din{2i1}{0.2}{1.9}{};

\qmatch{q1}{2.5}{1.5}{};\flow{{1.6,1.5}}{{2.5,1.5}}{dashed}{};\flow{{2.6,1.5}}{{3.6,1.5}}{dashed}{};
\dmatch{d1}{2.5}{1.9}{}{above};\flow{{1.6,1.9}}{{2.5,1.9}}{}{};\flowdiag{{2.56,1.9}}{{3.4,0.8}}{}{}{pos=0.55,rotate=315};
\qmatch{q2}{2.5}{1}{};\flow{{1.6,1}}{{2.5,1}}{dashed}{};\flow{{2.6,1}}{{3.6,1}}{dashed}{};
\dmatch{d2}{2.5}{0.6}{}{above};\flow{{1.6,0.6}}{{2.5,0.6}}{}{};\flow{{2.56,0.6}}{{3.4,0.6}}{}{};

\computonPrimitive{3.4}{0.4}{0.6}{1.6}{}
\node at (3.7,1.25){\scriptsize $\kappa_2$};\draw[densely dotted] (3.4,1.5) -- (3.55,1.25);\draw[densely dotted] (3.4,1) -- (3.55,1.25);\draw[densely dotted] (3.8,1.25) -- (4,1.25);
\node at (3.7,0.6){\scriptsize $\uparrow_2$};\draw[densely dotted] (3.4,0.8) -- (3.55,0.6);\draw[densely dotted] (3.4,0.6) -- (3.55,0.6);\draw[densely dotted] (3.8,0.6) -- (4,0.6);
\qout{3q1}{4}{1.25}{};
\dout{3o1}{4}{0.6}{};
 \end{tikzpicture}
 }
 {
 \begin{tikzpicture}
\begin{scope}
\draw[draw=black,fill=white] (0,0.4) rectangle ++(0.12,0.25);
\node[inner sep=0pt,minimum size=0pt,label=right:{\scriptsize Computation unit}] at (0.2,0.5) {};
\node[draw=black,fill=white,inner sep=0pt,minimum size=3pt,label=right:{\scriptsize Control inport}] at (3.5,0.5) {};
\node[circle,draw=black,fill=white,inner sep=0pt,minimum size=3pt,label={right:{\scriptsize Boolean inport}}] at (6.8,0.5) {};
\end{scope}

\begin{scope}
\draw[dashed] (0,0) to node[pos=0.5]{\arrowflow} (0.6,0);
\node[inner sep=0pt,minimum size=3pt,label=right:{\scriptsize Control flow}] at (0.6,0) {};
\node[fill=black,inner sep=0pt,minimum size=3pt,label={right:{\scriptsize Control outport}}] at (3.5,0) {};
\node[circle,fill=black,inner sep=0pt,minimum size=3pt,label={right:{\scriptsize Boolean outport}}] at (6.8,0) {};
\end{scope}

\begin{scope}[yshift=-0.5cm]
\draw (0,0) to node[pos=0.5]{\arrowflow} (0.5,0);
\node[inner sep=0pt,minimum size=3pt,label=right:{\scriptsize Boolean flow}] at (0.5,0){};
\node[draw,fill=white,fill fraction={black}{0.5},inner sep=0pt,minimum size=3pt,label=right:{\scriptsize Control inoutport}] at (3.5,0) {};
\node[circle,draw,fill=white,fill fraction={black}{0.5},inner sep=0pt,minimum size=3pt,label=right:{\scriptsize Boolean inoutport}] at (6.8,0) {};
\end{scope}
\end{tikzpicture}
 }
 \caption{Elementary logical connectives as control-driven Boolean circuits.} 
 \label{fig:elementary-circuits}
\end{figure}
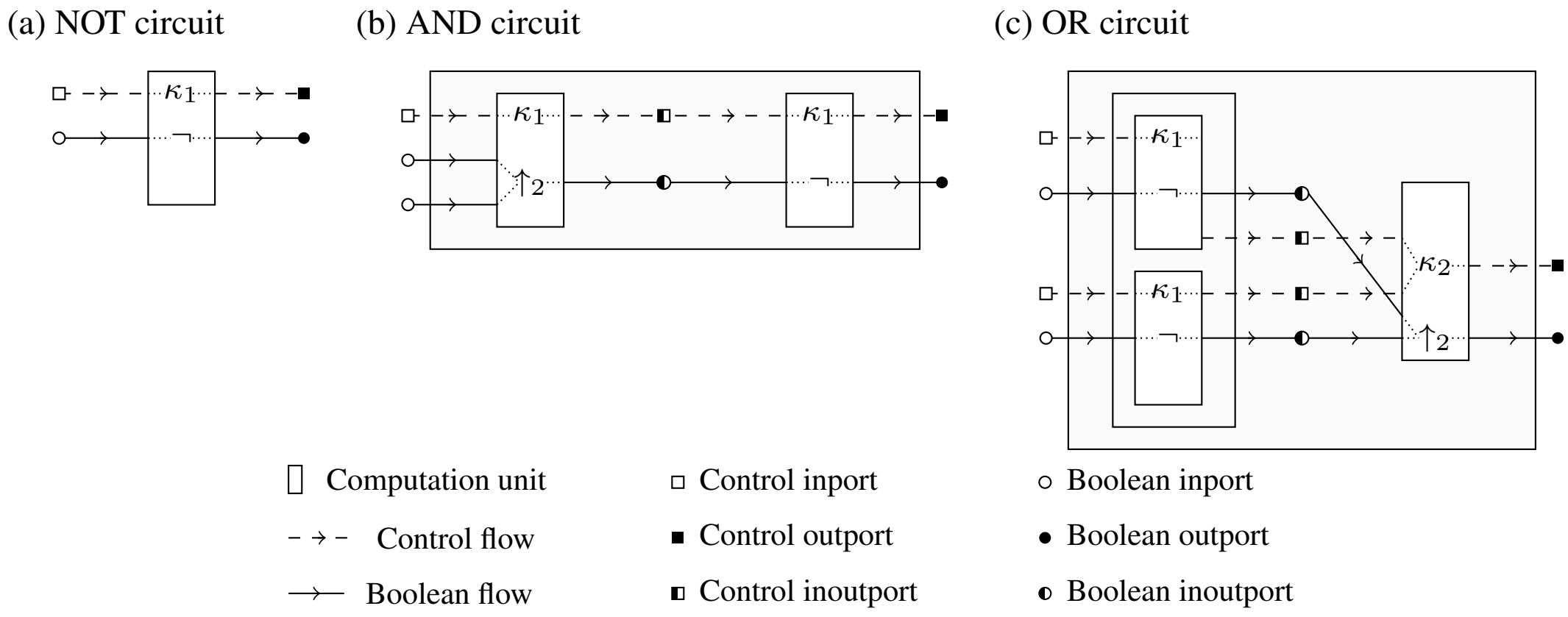

In this section, we present concrete examples of control-driven Boolean circuits, with the aim of demonstrating their applicability. As a starting point, Figure \ref{fig:elementary-circuits} shows the equivalent circuits for the NOT, AND and OR functions. 

In Definition \ref{def:evaluation}, it is stated that the $\uparrow_1$ operator is chosen from a family when there is exactly one Boolean input. As $\uparrow_1$ corresponds to NOT (see Definition \ref{def:circuit}), the computing devices corresponding to that operation are annotated with the $\lnot$ symbol. Computing devices with ${m \geq 2}$ Boolean inputs are simply labelled as $\uparrow_m$, effectively denoting an $m$-ary NAND function. As a reminder, $m$-ary devices absorbing control and Boolean data are denoted $\kappa_m$ and $\epsilon_m$, respectively. For simplicity, whenever an atomic circuit has a $\kappa$-devices and a single $\uparrow$ operator, we refer to it as the operator name.

In Figure \ref{fig:elementary-circuits}, it is clear that AND and OR take the form of composites, whereas NOT is just an atomic circuit consisting of a single computation unit with exactly one Boolean inport and one Boolean outport. More precisely, AND is formed by composing a $2$-ary NAND circuit and a NOT one into a total sequential composite. The OR composite is formed in two stages: First, two atomic NOTs are put in a parallel structure and then a total sequential composite is formed out of it using a $2$-ary NAND function.

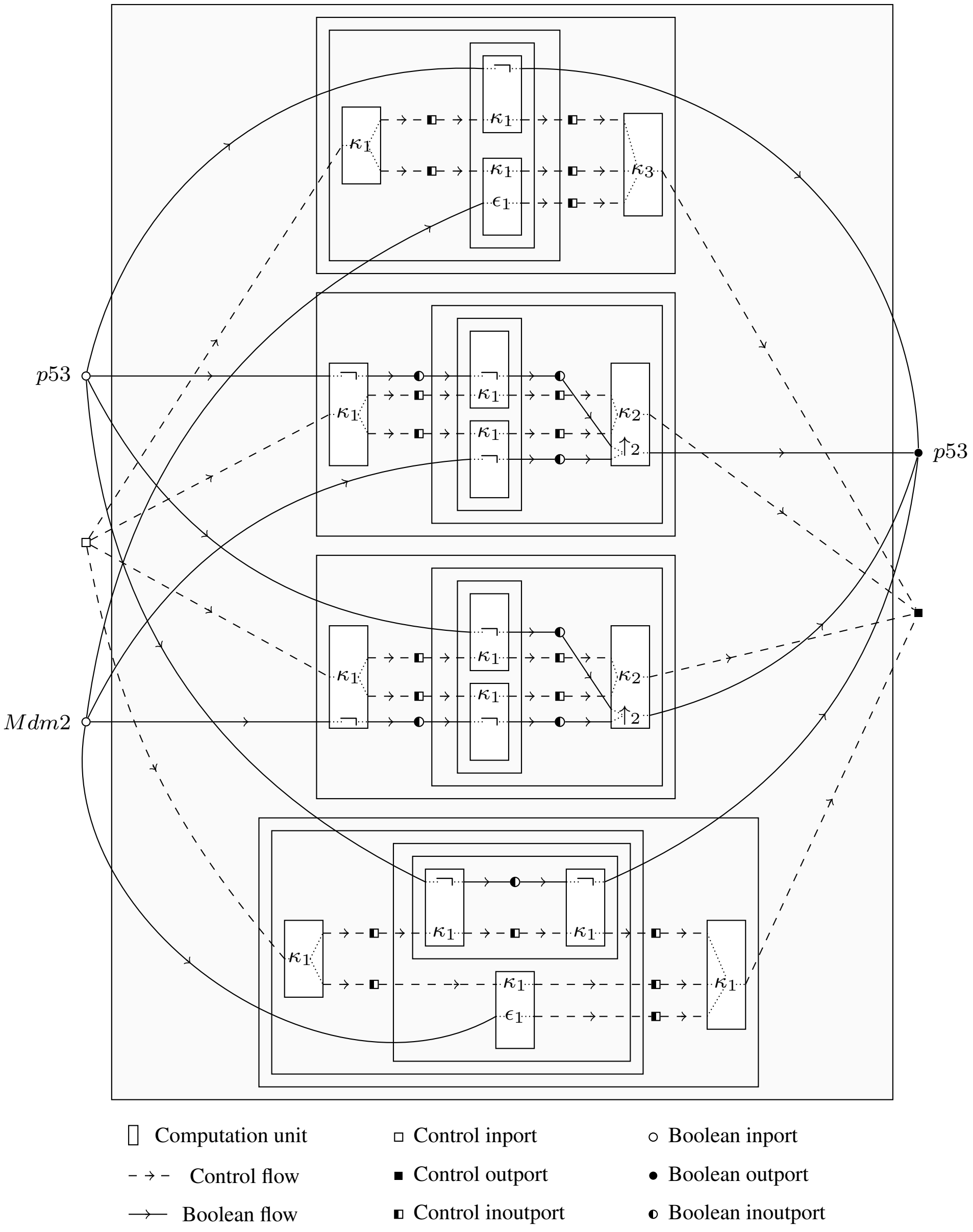
\begin{figure}[H]
 \centering 
 \begin{tikzpicture}[scale=0.83]

 \computonComposite{-2.6}{-0.6}{12.2}{17.1};

 
 \computonComposite{0.6}{12.3}{5.6}{4};
 \computonComposite{0.8}{12.5}{3.6}{3.6};
 \computonComposite{3}{12.7}{1}{3.2};
 
 \computonPrimitive{1}{13.7}{0.6}{1.2}{} 
 \node at (1.3,14.3){\scriptsize $\kappa_1$};\draw[densely dotted] (1,14.3) -- (1.15,14.3);\draw[densely dotted] (1.4,14.3) -- (1.6,14.7);\draw[densely dotted] (1.4,14.3) -- (1.6,13.9);
 \qmatch{f2o1}{2.4}{14.7}{};\flow{{1.6,14.7}}{f2o1}{dashed}{};\flow{f2o1}{{3.2,14.7}}{dashed}{};
 \qmatch{f2o2}{2.4}{13.9}{};\flow{{1.6,13.9}}{f2o2}{dashed}{};\flow{f2o2}{{3.2,13.9}}{dashed}{};
 \computonPrimitive{3.2}{14.5}{0.6}{1.2}{}
 \node at (3.5,14.7){\scriptsize $\kappa_1$};\draw[densely dotted] (3.2,14.7) -- (3.35,14.7);\draw[densely dotted] (3.6,14.7) -- (3.8,14.7);
 \node at (3.5,15.5){$\lnot$};\draw[densely dotted] (3.2,15.5) -- (3.4,15.5);\draw[densely dotted] (3.65,15.5) -- (3.8,15.5);
 \computonPrimitive{3.2}{12.9}{0.6}{1.2}{} 
 \node at (3.5,13.9){\scriptsize $\kappa_1$};\draw[densely dotted] (3.2,13.9) -- (3.35,13.9);\draw[densely dotted] (3.6,13.9) -- (3.8,13.9);
 \node at (3.5,13.4){\scriptsize $\epsilon_1$};\draw[densely dotted] (3.2,13.4) -- (3.35,13.4);\draw[densely dotted] (3.6,13.4) -- (3.8,13.4);
 \qmatch{n6o1}{4.6}{14.7}{};\flow{{3.8,14.7}}{n6o1}{dashed}{};\flow{n6o1}{{5.45,14.7}}{dashed}{};
 \qmatch{e2o1}{4.6}{13.9}{};\flow{{3.8,13.9}}{e2o1}{dashed}{};\flow{e2o1}{{5.45,13.9}}{dashed}{};
 \qmatch{e2o2}{4.6}{13.4}{};\flow{{3.8,13.4}}{e2o2}{dashed}{};\flow{e2o2}{{5.45,13.4}}{dashed}{};
 \computonPrimitive{5.4}{13.2}{0.6}{1.6}{} 
 \node at (5.7,13.9){\scriptsize $\kappa_3$};\draw[densely dotted] (5.4,13.9) -- (5.5,13.9);\draw[densely dotted] (5.4,14.7) -- (5.6,14);\draw[densely dotted] (5.4,13.4) -- (5.6,13.8);\draw[densely dotted] (5.8,13.9) -- (6,13.9);
 
 
 \computonComposite{0.6}{8.2}{5.6}{3.8};
 \computonComposite{2.4}{8.4}{3.6}{3.4};
 \computonComposite{2.8}{8.6}{1}{3};

 \computonPrimitive{0.8}{9.3}{0.6}{1.6}{}
 \node at (1.1,10.7){$\lnot$};\draw[densely dotted] (0.8,10.7) -- (1,10.7);\draw[densely dotted] (1.25,10.7) -- (1.4,10.7);
 \node at (1.1,10.1){\scriptsize $\kappa_1$};\draw[densely dotted] (0.8,10.1) -- (0.9,10.1);\draw[densely dotted] (1.25,10.1) -- (1.4,9.8);\draw[densely dotted] (1.25,10.1) -- (1.4,10.4);
 
 \qmatch{g2o1}{2.2}{10.4}{};\flow{{1.4,10.4}}{g2o1}{dashed}{};\flow{g2o1}{{3,10.4}}{dashed}{};
 \dmatch{g2o2}{2.2}{10.7}{}{above};\flow{{1.4,10.7}}{g2o2}{}{};\flow{g2o2}{{3,10.7}}{}{};
 \qmatch{g2o3}{2.2}{9.8}{};\flow{{1.4,9.8}}{g2o3}{dashed}{};\flow{g2o3}{{3,9.8}}{dashed}{}; 

 \computonPrimitive{3}{10.2}{0.6}{1.2}{}
 \node at (3.3,10.7){$\lnot$};\draw[densely dotted] (3,10.7) -- (3.2,10.7);\draw[densely dotted] (3.45,10.7) -- (3.6,10.7);
 \node at (3.3,10.4){\scriptsize $\kappa_1$};\draw[densely dotted] (3,10.4) -- (3.1,10.4);\draw[densely dotted] (3.45,10.4) -- (3.6,10.4);
\computonPrimitive{3}{8.8}{0.6}{1.2}{} 
\node at (3.3,9.8){\scriptsize $\kappa_1$};\draw[densely dotted] (3,9.8) -- (3.1,9.8);\draw[densely dotted] (3.45,9.8) -- (3.6,9.8);
\node at (3.3,9.4){$\lnot$};\draw[densely dotted] (3,9.4) -- (3.2,9.4);\draw[densely dotted] (3.45,9.4) -- (3.6,9.4);

 \qmatch{n4o1}{4.4}{10.4}{};\flow{{3.6,10.4}}{n4o1}{dashed}{};\flow{n4o1}{{5.2,10.4}}{dashed}{};
 \dmatch{n4o2}{4.4}{10.7}{}{above};\flow{{3.6,10.7}}{n4o2}{}{};\flowdiag{n4o2}{{5.2,9.6}}{}{}{pos=0.55,rotate=315};
 \qmatch{n5o1}{4.4}{9.8}{};\flow{{3.6,9.8}}{n5o1}{dashed}{};\flow{n5o1}{{5.2,9.8}}{dashed}{};
 \dmatch{n5o2}{4.4}{9.4}{}{above};\flow{{3.6,9.4}}{n5o2}{}{};\flow{n5o2}{{5.2,9.4}}{}{};

 \computonPrimitive{5.2}{9.3}{0.6}{1.6}{}
 \node at (5.5,10.1){\scriptsize $\kappa_2$};\draw[densely dotted] (5.2,10.4) -- (5.3,10.1);\draw[densely dotted] (5.2,9.8) -- (5.3,10.1);\draw[densely dotted] (5.7,10.1) -- (5.8,10.1);
 \node at (5.5,9.6){\scriptsize $\uparrow_2$};\draw[densely dotted] (5.2,9.6) -- (5.4,9.5);\draw[densely dotted] (5.2,9.4) -- (5.4,9.5);\draw[densely dotted] (5.7,9.5) -- (5.8,9.5);

 
 \computonComposite{0.6}{4.1}{5.6}{3.8};
 \computonComposite{2.4}{4.3}{3.6}{3.4};
 \computonComposite{2.8}{4.5}{1}{3};

 \computonPrimitive{0.8}{5.2}{0.6}{1.6}{}
 \node at (1.1,6){\scriptsize $\kappa_1$};\draw[densely dotted] (0.8,6) -- (0.9,6);\draw[densely dotted] (1.25,6) -- (1.4,6.3);\draw[densely dotted] (1.25,6) -- (1.4,5.7);
 \node at (1.1,5.3){$\lnot$};\draw[densely dotted] (0.8,5.3) -- (1,5.3);\draw[densely dotted] (1.25,5.3) -- (1.4,5.3);
 
 \qmatch{g1o1}{2.2}{6.3}{};\flow{{1.4,6.3}}{g1o1}{dashed}{};\flow{g1o1}{{3,6.3}}{dashed}{}; 
 \qmatch{g1o3}{2.2}{5.7}{};\flow{{1.4,5.7}}{g1o3}{dashed}{};\flow{g1o3}{{3,5.7}}{dashed}{};
 \dmatch{g1o4}{2.2}{5.3}{}{above};\flow{{1.4,5.3}}{g1o4}{}{};\flow{g1o4}{{3,5.3}}{}{};

 \computonPrimitive{3}{6.1}{0.6}{1.2}{}
 \node at (3.3,6.3){\scriptsize $\kappa_1$};\draw[densely dotted] (3,6.3) -- (3.1,6.3);\draw[densely dotted] (3.45,6.3) -- (3.6,6.3);
 \node at (3.3,6.7){$\lnot$};\draw[densely dotted] (3,6.7) -- (3.2,6.7);\draw[densely dotted] (3.45,6.7) -- (3.6,6.7);
 \computonPrimitive{3}{4.7}{0.6}{1.2}{}
 \node at (3.3,5.7){\scriptsize $\kappa_1$};\draw[densely dotted] (3,5.7) -- (3.1,5.7);\draw[densely dotted] (3.45,5.7) -- (3.6,5.7);
 \node at (3.3,5.3){$\lnot$};\draw[densely dotted] (3,5.3) -- (3.2,5.3);\draw[densely dotted] (3.45,5.3) -- (3.6,5.3);

 \qmatch{n2o1}{4.4}{6.3}{};\flow{{3.6,6.3}}{n2o1}{dashed}{};\flow{n2o1}{{5.2,6.3}}{dashed}{};
 \dmatch{n2o2}{4.4}{6.7}{}{above};\flow{{3.6,6.7}}{n2o2}{}{};\flowdiag{n2o2}{{5.2,5.5}}{}{}{pos=0.55,rotate=315};
 \qmatch{n3o1}{4.4}{5.7}{};\flow{{3.6,5.7}}{n3o1}{dashed}{};\flow{n3o1}{{5.2,5.7}}{dashed}{};
 \dmatch{n3o2}{4.4}{5.3}{}{above};\flow{{3.6,5.3}}{n3o2}{}{};\flow{n3o2}{{5.2,5.3}}{}{};

 \computonPrimitive{5.2}{5.2}{0.6}{1.6}{}
 \node at (5.5,6){\scriptsize $\kappa_2$};\draw[densely dotted] (5.2,6.3) -- (5.3,6);\draw[densely dotted] (5.2,5.7) -- (5.3,6);\draw[densely dotted] (5.7,6) -- (5.8,6);
 \node at (5.5,5.4){\scriptsize $\uparrow_2$};\draw[densely dotted] (5.2,5.5) -- (5.4,5.4);\draw[densely dotted] (5.2,5.3) -- (5.4,5.4);\draw[densely dotted] (5.7,5.4) -- (5.8,5.4);
 
 
 \computonComposite{-0.3}{-0.4}{7.8}{4.2};
 \computonComposite{-0.1}{-0.2}{5.8}{3.8};
 \computonComposite{1.8}{0}{3.7}{3.4};
 \computonComposite{2.1}{1.6}{3.2}{1.6};
 
 \computonPrimitive{0.1}{1}{0.6}{1.2}{} 
 \node at (0.35,1.6){\scriptsize $\kappa_1$};\draw[densely dotted] (0.1,1.6) -- (0.2,1.6);\draw[densely dotted] (0.5,1.6) -- (0.7,2);\draw[densely dotted] (0.5,1.6) -- (0.7,1.2);
 
 \qmatch{f1o1}{1.5}{2}{};\flow{{0.7,2}}{f1o1}{dashed}{};\flow{f1o1}{{2.3,2}}{dashed}{};
 \qmatch{f1o2}{1.5}{1.2}{};\flow{{0.7,1.2}}{f1o2}{dashed}{};\flow{f1o2}{{4,1.2}}{dashed}{};
 
 \computonPrimitive{2.3}{1.8}{0.6}{1.2}{}
 \node at (2.6,2){\scriptsize $\kappa_1$};\draw[densely dotted] (2.3,2) -- (2.4,2);\draw[densely dotted] (2.75,2) -- (2.9,2);
 \node at (2.6,2.8){$\lnot$};\draw[densely dotted] (2.3,2.8) -- (2.5,2.8);\draw[densely dotted] (2.75,2.8) -- (2.9,2.8);
 \qmatch{n0o1}{3.7}{2}{};\flow{{2.9,2}}{n0o1}{dashed}{};\flow{n0o1}{{4.5,2}}{dashed}{};
 \dmatch{n0o2}{3.7}{2.8}{}{above};\flow{{2.9,2.8}}{n0o2}{}{};\flow{n0o2}{{4.5,2.8}}{}{};
 \computonPrimitive{4.5}{1.8}{0.6}{1.2}{} 
 \node at (4.8,2){\scriptsize $\kappa_1$};\draw[densely dotted] (4.5,2) -- (4.6,2);\draw[densely dotted] (4.95,2) -- (5.1,2);
 \node at (4.8,2.8){$\lnot$};\draw[densely dotted] (4.5,2.8) -- (4.7,2.8);\draw[densely dotted] (4.95,2.8) -- (5.1,2.8);
 
 \computonPrimitive{3.4}{0.2}{0.6}{1.2}{} 
 \node at (3.7,1.2){\scriptsize $\kappa_1$};\draw[densely dotted] (3.4,1.2) -- (3.5,1.2);\draw[densely dotted] (3.85,1.2) -- (4.1,1.2);
 \node at (3.7,0.7){\scriptsize $\epsilon_1$};\draw[densely dotted] (3.4,0.7) -- (3.55,0.7);\draw[densely dotted] (3.8,0.7) -- (4,0.7);
 \qmatch{nxo1}{5.9}{2}{};\flow{{5.1,2}}{nxo1}{dashed}{};\flow{nxo1}{{6.7,2}}{dashed}{};
 \qmatch{e1o1}{5.9}{1.2}{};\flow{{4,1.2}}{e1o1}{dashed}{};\flow{e1o1}{{6.7,1.2}}{dashed}{};
 \qmatch{e1o2}{5.9}{0.7}{};\flow{{4,0.7}}{e1o2}{dashed}{};\flow{e1o2}{{6.7,0.7}}{dashed}{};
 
 \computonPrimitive{6.7}{0.5}{0.6}{1.7}{}
 \node at (7,1.2){\scriptsize $\kappa_1$};\draw[densely dotted] (6.7,2) -- (7,1.3);\draw[densely dotted] (6.7,1.2) -- (6.8,1.2);\draw[densely dotted] (6.7,0.7) -- (7,1.2);\draw[densely dotted] (7.15,1.2) -- (7.3,1.2);
 

 \dinplain{pin}{-3}{10.7}{$p53$};
 \flowdiag{pin}{{3.2,15.5}}{bend left=40}{}{pos=0.5,rotate=40};
 \flow{pin}{{{0.8,10.7}}}{}{};
 \flowdiag{pin}{{3,6.7}}{bend right=30}{}{pos=0.4,rotate=320};
 \flowdiag{pin}{{2.3,2.8}}{bend right=30}{}{pos=0.43,rotate=310}; 
 
 \qinplain{qin}{-3}{8.1}{}; 
 \flowdiag{qin}{{1,14.3}}{dashed}{}{pos=0.5,rotate=55};
 \flowdiag{qin}{{0.8,10.1}}{dashed}{}{pos=0.5,rotate=35};
 \flowdiag{qin}{{0.8,6}}{dashed}{}{pos=0.5,rotate=320};
 \flowdiag{qin}{{0.1,1.6}}{dashed,bend right=15}{}{pos=0.5,rotate=305};
 
 \dinplain{min}{-3}{5.3}{$Mdm2$};
 \flowdiag{min}{{3.2,13.4}}{bend left=30}{}{pos=0.92,rotate=35};
 \flowdiag{min}{{3,9.4}}{bend left=30}{}{pos=0.75,rotate=30}; 
 \flowdiag{min}{{0.8,5.3}}{}{}{pos=0.65};
 \flowdiag{min}{{3.4,0.7}}{bend right=65}{}{pos=0.45,rotate=320};

 \doutplain{pout}{9.2}{9.5}{$p53$};
 \flowdiag{{3.8,15.5}}{pout}{bend left=45}{}{rotate=320};
 \flow{5.8,9.5}{pout}{}{};
 \flowdiag{{5.8,5.4}}{pout}{bend right=30}{}{rotate=40};
 \flowdiag{{5.1,2.8}}{pout}{bend right=30}{}{rotate=50};

 \qoutplain{qout}{9.2}{7}{};
 \flowdiag{{6,13.9}}{qout}{dashed}{}{pos=0.4,rotate=305};
 \flowdiag{{5.8,10.1}}{qout}{dashed}{}{rotate=320};
 \flowdiag{{5.8,6}}{qout}{dashed}{}{pos=0.3,rotate=8};
 \flowdiag{{7.3,1.2}}{qout}{dashed}{}{rotate=55};
  
 \end{tikzpicture}
 \hspace{1cm}
 {
 \begin{tikzpicture}
\begin{scope}
\draw[draw=black,fill=white] (0,0.4) rectangle ++(0.12,0.25);
\node[inner sep=0pt,minimum size=0pt,label=right:{\scriptsize Computation unit}] at (0.2,0.5) {};
\node[draw=black,fill=white,inner sep=0pt,minimum size=3pt,label=right:{\scriptsize Control inport}] at (3.5,0.5) {};
\node[circle,draw=black,fill=white,inner sep=0pt,minimum size=3pt,label={right:{\scriptsize Boolean inport}}] at (6.8,0.5) {};
\end{scope}

\begin{scope}
\draw[dashed] (0,0) to node[pos=0.5]{\arrowflow} (0.6,0);
\node[inner sep=0pt,minimum size=3pt,label=right:{\scriptsize Control flow}] at (0.6,0) {};
\node[fill=black,inner sep=0pt,minimum size=3pt,label={right:{\scriptsize Control outport}}] at (3.5,0) {};
\node[circle,fill=black,inner sep=0pt,minimum size=3pt,label={right:{\scriptsize Boolean outport}}] at (6.8,0) {};
\end{scope}

\begin{scope}[yshift=-0.5cm]
\draw (0,0) to node[pos=0.5]{\arrowflow} (0.5,0);
\node[inner sep=0pt,minimum size=3pt,label=right:{\scriptsize Boolean flow}] at (0.5,0){};
\node[draw,fill=white,fill fraction={black}{0.5},inner sep=0pt,minimum size=3pt,label=right:{\scriptsize Control inoutport}] at (3.5,0) {};
\node[circle,draw,fill=white,fill fraction={black}{0.5},inner sep=0pt,minimum size=3pt,label=right:{\scriptsize Boolean inoutport}] at (6.8,0) {};
\end{scope}
\end{tikzpicture}
 }
 \caption{A closed branching circuit that captures the computational behaviour of the context-sensitive stochastic Boolean network presented in \cite{liang_stochastic_2012}.} 
 \label{fig:branching-example}
\end{figure}

The elementary functions presented in Figure \ref{fig:elementary-circuits} can serve as building blocks to form even more complex circuits. For example, Figure \ref{fig:branching-example} displays a closed branching composite for a context-sensitive stochastic Boolean network \cite{liang_stochastic_2012}, which captures the oscillatory dynamics of a p53-Mdm2 regulatory pathway that is subject to molecular and genetic noise. The purpose of this circuit is to determine the next state of a gene by randomly selecting one out of four alternative sub-circuits. 

For the first alternative, we first compose a parallel composite for the simultaneous activation of a NOT function and a ``Boolean eater''. In Section \ref{sec:model}, we mentioned that a Boolean eater is an $\epsilon$ computing device which takes any number of Boolean values and produces a single control signal. This class of circuits is useful to wrap interfaces.

After composing our parallel composite, we form a partial sequential circuit by taking a ``fork'' as the left operand and the parallel structure as the right one. The result is then composed with an atomic ``join'' into a partial sequential composite that captures the whole behaviour of the first alternative branch.\footnote{The first alternative branch demonstrates the expressivity of our model to define computation units that do not receive or produce any Boolean values, but just fork or join control signals. These units are formed out of $\kappa$ computing devices to yieldso-called glue circuits (see Definition \ref{def:glue}).}

The second and third alternatives are total sequential circuits for the invocation of an atomic NOT and an OR composite, in that order. The only difference is in the way they process Boolean information (cf. the leftmost NOT functions and the inner parallel composites). Finally, at the bottom-level, we have a composite which is similar to the first alternative. The only difference is that, rather than just computing NOT, the inner parallel composite activates a total sequential circuit of two atomic NOTs, effectively passing information unchanged. In other words, the last branch defines a buffer to ``echo'' a $p53$ Boolean value.

It might not seem entirely obvious that the process of choosing one alternative or another is entirely governed by control flow. To elucidate this, observe that the initial state of our closed branching composite has to inject Boolean values into the $p53$ and $Mdm2$ ports as well as a control signal into the only control inport. As control flow arrows run from the control inport to the forks and to the leftmost NOT functions, all these units are enabled initially (see Definition \ref{def:unit-status}), but only one of them will be non-deterministically chosen for evaluation (as per Definition \ref{def:ready-units}). In other words, the initial state triggers only one alternative composite in the next execution step. 

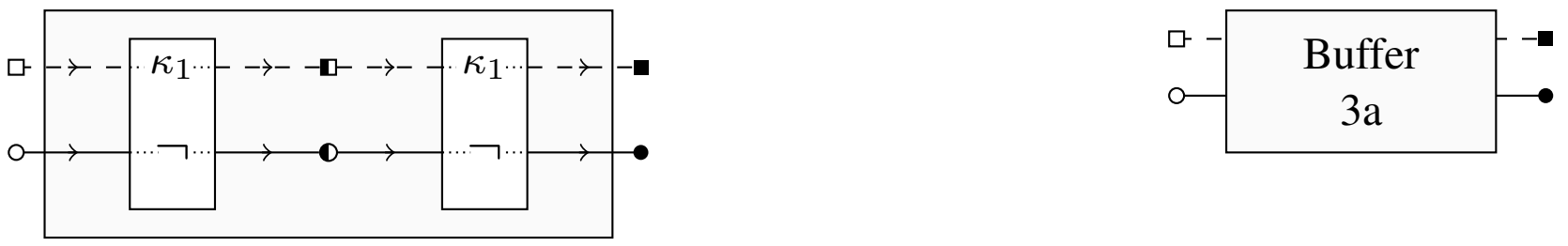
\begin{figure}[H]
 \captionsetup{position=above}
 \centering 
 \subcaptionbox{Internal details of a buffer.\label{fig:buffer}}
 {
 \begin{tikzpicture} 
 \computonComposite{1.7}{1.6}{4}{1.6}; 
 \qin{qin}{1.5}{2.8}{};
 \din{din}{1.5}{2.2}{};
 \computonPrimitive{2.3}{1.8}{0.6}{1.2}{}
 \node at (2.6,2.8){\scriptsize $\kappa_1$};\draw[densely dotted] (2.3,2.8) -- (2.4,2.8);\draw[densely dotted] (2.75,2.8) -- (2.9,2.8);
 \node at (2.6,2.2){$\lnot$};\draw[densely dotted] (2.3,2.2) -- (2.5,2.2);\draw[densely dotted] (2.75,2.2) -- (2.9,2.2);
 \qmatch{n0o1}{3.7}{2.8}{};\flow{{2.9,2.8}}{n0o1}{dashed}{};\flow{n0o1}{{4.5,2.8}}{dashed}{};
 \dmatch{n0o2}{3.7}{2.2}{}{above};\flow{{2.9,2.2}}{n0o2}{}{};\flow{n0o2}{{4.5,2.2}}{}{};
 \computonPrimitive{4.5}{1.8}{0.6}{1.2}{} 
 \node at (4.8,2.8){\scriptsize $\kappa_1$};\draw[densely dotted] (4.5,2.8) -- (4.6,2.8);\draw[densely dotted] (4.95,2.8) -- (5.1,2.8);
 \node at (4.8,2.2){$\lnot$};\draw[densely dotted] (4.5,2.2) -- (4.7,2.2);\draw[densely dotted] (4.95,2.2) -- (5.1,2.2);
 \qout{qout}{5.1}{2.8}{};
 \dout{dout}{5.1}{2.2}{};  
 \end{tikzpicture}
 }
 \hspace{1cm}
 \subcaptionbox{Buffer as a black box.}[0.45\textwidth]
 {
 \begin{tikzpicture} 
 \qin{qin}{1.5}{1.05}{};
 \din{din}{1.5}{0.65}{};
 \qout{qout}{3.3}{1.05}{};
 \dout{dout}{3.3}{0.65}{};  
 \computonComposite{1.85}{0.25}{1.9}{1};
 \node at (2.8,0.95){\footnotesize Buffer};
 \node at (2.8,0.55){\footnotesize\ref{fig:buffer}};
 \node at (2.8,0){};
 \end{tikzpicture}
 }
 \caption{A buffer is a total sequential circuit of two atomic NOTs. Here, we demonstrate the modularity property of control-driven circuits. A similar black-boxing process can be done with any of the composites displayed in Figures \ref{fig:elementary-circuits} and \ref{fig:branching-example}.} 
 \label{fig:modularity}
\end{figure}

Due to the modularity property of the proposed model, our branching composite can be used as a component to form even more complex circuits. To explain the modularity property in a more concrete manner, let us construct a tail-iterative circuit that performs the toggle action of a \emph{clocked set-reset flip-flop}. First, let us reuse the buffer we use in the last alternative of our branching circuit, which we depict in Figure \ref{fig:buffer} for convenience. 

This time we reuse the buffer structure to build the partial sequential composite displayed in Figure \ref{fig:composite-entry}. As \ref{fig:composite-entry} will be the entry point to our iterative computation, its purpose is to trigger three buffers in parallel for echoing the values of three ports, namely $R$ (reset), $Q_n$ (current state) and $S$ (set).

These three values are intended to be consumed by the composite from Figure \ref{fig:composite-iteration}, which captures the behaviour of our main cyclic computation. Like \ref{fig:composite-entry}, this building block is constructed from existing structures; in this case, the AND and OR composites from Figure \ref{fig:elementary-circuits}. Operationally, \ref{fig:composite-iteration} realises the following equation to determine the next flip-flop state:

\begin{equation}\label{eq:next-state}
 Q_{n+1}=S~\lor~(\lnot R~\land~Q_n)   
\end{equation}

Fully realising Equation \ref{eq:next-state} requires the negation of $R$, before applying AND on the result and the current state $Q_n$. In other words, a partial sequential circuit needs to be formed to compute NOT and AND, in that order (see Figure \ref{fig:composite-iteration}). 

\begin{figure}[!h]
 \centering 
 \subcaptionbox{A composite to be used as the entry to our iterative computation. This is formed by first composing three buffers in parallel and then using a fork and the result as left and right operands for a partial sequential circuit. The resulting sequential circuit and an atomic join are then used as left and right operands to form an even more complex partial sequential structure.\label{fig:composite-entry}}
 {
 \begin{tikzpicture}[scale=0.75]

 \computonComposite{-0.5}{-0.6}{7.6}{4.7};
 \computonComposite{-0.1}{-0.4}{5}{4.3};
 \computonComposite{2.1}{-0.2}{2.3}{3.9};

 \qin{fin}{-0.7}{1.8}{};
 \dinplain{rin}{-0.7}{3.3}{$R$};\flow{rin}{{2.3,3.3}}{}{};
 \dinplain{qin}{-0.7}{2.5}{$Q_n$};\flowdiag{qin}{{2.3,2.05}}{bend left=20}{}{};
 \dinplain{sin}{-0.7}{0.2}{$S$};\flow{sin}{{2.3,0.2}}{}{};
 
 \computonPrimitive{0.1}{1.2}{0.6}{1.2}{} 
 \node at (0.4,1.8){\scriptsize $\kappa_1$};\draw[densely dotted] (0.1,1.8) -- (0.3,1.8);\draw[densely dotted] (0.6,1.8) -- (0.7,2.1);\draw[densely dotted] (0.6,1.8) -- (0.7,1.8);\draw[densely dotted] (0.6,1.8) -- (0.7,1.5);
 
 \qmatch{f1o1}{1.5}{2.7}{};\flowdiag{{0.7,2.1}}{f1o1}{dashed}{}{rotate=45};\flow{f1o1}{{2.3,2.7}}{dashed}{};
 \qmatch{f1o2}{1.5}{1.8}{};\flow{{0.7,1.8}}{f1o2}{dashed}{};\flow{f1o2}{{2.3,1.8}}{dashed}{};
 \qmatch{f1o3}{1.5}{0.8}{};\flowdiag{{0.7,1.5}}{f1o3}{dashed}{}{rotate=315};\flow{f1o3}{{2.3,0.8}}{dashed}{};
 
 \qmatch{nxo1}{5.1}{2.7}{};\flow{{4.3,2.7}}{nxo1}{dashed}{};\flowdiag{nxo1}{{5.9,2.1}}{dashed}{}{rotate=315};
 \qmatch{e1o1}{5.1}{1.8}{};\flow{{4.3,1.8}}{e1o1}{dashed}{};\flow{e1o1}{{5.9,1.8}}{dashed}{};
 \qmatch{e1o2}{5.1}{0.8}{};\flow{{4.3,0.8}}{e1o2}{dashed}{};\flowdiag{e1o2}{{5.9,1.5}}{dashed}{}{rotate=45};
 
 \computonPrimitive{5.9}{1.2}{0.6}{1.2}{}
 \node at (6.2,1.8){\scriptsize $\kappa_3$};\draw[densely dotted] (5.9,2.1) -- (6,1.8);\draw[densely dotted] (5.9,1.8) -- (6,1.8);\draw[densely dotted] (5.9,1.5) -- (6,1.8);\draw[densely dotted] (6.4,1.8) -- (6.5,1.8);

 \begin{scope}[xshift=2.3cm,yshift=2.5cm] 
    \computonComposite{0}{0}{1.9}{1};
    \node at (0.95,0.7){\footnotesize Buffer};
    \node at (0.95,0.3){\footnotesize\ref{fig:buffer}};
 \end{scope}
 \begin{scope}[xshift=2.3cm,yshift=1.25cm] 
    \computonComposite{0}{0}{1.9}{1};
    \node at (0.95,0.7){\footnotesize Buffer};
    \node at (0.95,0.3){\footnotesize\ref{fig:buffer}};
 \end{scope}
 \begin{scope}[xshift=2.3cm,yshift=0cm] 
    \computonComposite{0}{0}{1.9}{1};
    \node at (0.95,0.7){\footnotesize Buffer};
    \node at (0.95,0.3){\footnotesize\ref{fig:buffer}};
 \end{scope}

 \qout{fout}{6.5}{1.8}{};
 \doutplain{rout}{6.5}{3.3}{$R$};\flow{{4.2,3.3}}{rout}{}{};
 \doutplain{qout}{6.5}{2.5}{$Q_n$};\flowdiag{{4.2,2.05}}{qout}{bend left=20}{}{};
 \doutplain{sout}{6.5}{0.2}{$S$};\flow{{4.2,0.2}}{sout}{}{};

 \begin{scope}[xshift=10.5cm,yshift=1.2cm] 
    \qin{qin}{-0.35}{1.4}{};
    \din{rin}{-0.35}{1}{$R$};
    \din{qin}{-0.35}{0.6}{$Q_n$};        
    \din{sin}{-0.35}{0.2}{$S$};
    \qout{qout}{1.45}{1.4}{};
    \dout{rout}{1.45}{1}{$R$};
    \dout{qout}{1.45}{0.6}{$Q_n$};        
    \dout{sout}{1.45}{0.2}{$S$};
    \computonComposite{0}{0}{1.9}{1.6};
    \node at (0.9,1){\footnotesize Entry};
    \node at (0.9,0.6){\footnotesize\ref{fig:composite-entry}};
 \end{scope}
 \end{tikzpicture}
 }
 \hspace{1cm}
 \subcaptionbox{A composite to be used as the main iterative computation, which is formed by first composing NOT and AND into a partial sequential circuit and then using the result and OR as left and right operands, respectively; thereby, forming a more complex partial sequential circuit.\label{fig:composite-iteration}}
 {
 \begin{tikzpicture}[scale=0.75]
 \computonComposite{0.2}{-0.4}{8.8}{2.8};
 \computonComposite{0.6}{0.2}{4.5}{2};
 
 \qin{cin}{0}{1.8}{};    
 \din{rin}{0}{1.4}{$R$};
 \dinplain{qin}{0}{0.6}{$Q_n$};\flow{qin}{{3,0.6}}{}{};
 \dinplain{sin}{0}{0}{$S$};\flowdiag{sin}{{6.5,0.6}}{bend right=16}{}{};
 
 \computonPrimitive{0.8}{0.8}{0.6}{1.2}{} 
 \node at (1.1,1.8){\scriptsize $\kappa_1$};\draw[densely dotted] (0.8,1.8) -- (0.93,1.8);\draw[densely dotted] (1.25,1.8) -- (1.4,1.8);
 \node at (1.1,1.4){$\lnot$};\draw[densely dotted] (0.8,1.4) -- (0.98,1.4);\draw[densely dotted] (1.25,1.4) -- (1.4,1.4);
 
 \qmatch{no1}{2.2}{1.8}{};\flow{{1.4,1.8}}{no1}{dashed}{};\flow{no1}{{3,1.8}}{dashed}{};
 \dmatch{no2}{2.2}{1.4}{}{below};\flow{{1.4,1.4}}{no2}{}{};\flow{no2}{{3,1.4}}{}{};
 \begin{scope}[xshift=3cm,yshift=0.4cm]     
    \computonComposite{0}{0}{1.9}{1.6};
    \node at (0.9,1){\small $\land$};
    \node at (0.9,0.6){\small\ref{fig:composite-and}};
 \end{scope}
 \qmatch{ao1}{5.7}{1.8}{};\flow{{4.9,1.8}}{ao1}{dashed}{};\flow{ao1}{{6.5,1.8}}{dashed}{};
 \dmatch{ao2}{5.7}{1.4}{}{below};\flow{{4.9,1.4}}{ao2}{}{};\flow{ao2}{{6.5,1.4}}{}{}; 
 \begin{scope}[xshift=6.5cm,yshift=0.4cm]     
    \computonComposite{0}{0}{1.9}{1.6};
    \node at (0.9,1){\small $\lor$};
    \node at (0.9,0.6){\small\ref{fig:composite-or}};
 \end{scope}
 \qout{cout}{8.4}{1.8}{};    
 \dout{qout}{8.4}{1.4}{$Q_{n+1}$};

 \begin{scope}[xshift=11.5cm,yshift=0.2cm] 
    \qin{qin}{-0.35}{1.4}{};    
    \din{rin}{-0.35}{1}{$R$};
    \din{qin}{-0.35}{0.6}{$Q_n$};        
    \din{sin}{-0.35}{0.2}{$S$};
    \qout{cout}{1.45}{1.4}{};
    \dout{qout}{1.45}{1}{$Q_{n+1}$};        
    \computonComposite{0}{0}{1.9}{1.6};
    \node at (0.95,1){\footnotesize Action};
    \node at (0.95,0.6){\footnotesize\ref{fig:composite-iteration}};
 \end{scope}
 \end{tikzpicture}
 }
 \hspace{1cm}
 \subcaptionbox{A composite to be used for the calculation of the next state, which simply is a partial sequential circuit for the execution of two atomic NOTs.\label{fig:composite-next}}
 {
 \begin{tikzpicture}[scale=0.75]
 \node at (0,1.3){};
 \computonComposite{3.2}{1.3}{4}{1.9}; 
 \qin{qin}{3}{2.8}{};
 \din{din}{3}{2.1}{$Q_{n+1}$};
 
 \computonPrimitive{3.8}{1.8}{0.6}{1.2}{}
 \node at (4.1,2.8){\scriptsize $\kappa_1$};\draw[densely dotted] (3.8,2.8) -- (3.93,2.8);\draw[densely dotted] (4.25,2.8) -- (4.4,2.8);
 \node at (4.1,2.1){$\lnot$};\draw[densely dotted] (3.8,2.1) -- (3.95,2.1);\draw[densely dotted] (4.25,2.1) -- (4.4,2);\draw[densely dotted] (4.25,2.1) -- (4.4,2.2);
 
 \qmatch{n0o1}{5.2}{2.8}{};\flow{{4.4,2.8}}{n0o1}{dashed}{};\flow{n0o1}{{6,2.8}}{dashed}{};
 \dmatch{n0o2}{5.2}{2.2}{}{below};\flow{{4.4,2.2}}{n0o2}{}{};\flow{n0o2}{{6,2.2}}{}{};
 
 \computonPrimitive{6}{1.8}{0.6}{1.2}{} 
 \node at (6.3,2.8){\scriptsize $\kappa_1$};\draw[densely dotted] (6,2.8) -- (6.13,2.8);\draw[densely dotted] (6.4,2.8) -- (6.6,2.8);
 \node at (6.3,2.2){$\lnot$};\draw[densely dotted] (6,2.2) -- (6.15,2.2);\draw[densely dotted] (6.45,2.2) -- (6.6,2.4);\draw[densely dotted] (6.45,2.2) -- (6.6,2);
 
 \qout{cout}{6.6}{2.8}{};
 \dout{rout}{6.6}{2.4}{$R_{n+1}$};
 \dout{qout}{6.6}{2}{$Q_{n+1}$};
 \doutplain{sout}{6.6}{1.6}{$S_{n+1}$};\flowdiag{{4.4,2}}{sout}{bend right=12}{}{}; 

 \begin{scope}[xshift=12.2cm,yshift=1.5cm] 
    \qin{cin}{-0.35}{1.4}{};    
    \din{rin}{-0.35}{1}{$Q_{n+1}$};
    \qout{cout}{1.45}{1.4}{};
    \dout{rout}{1.45}{1}{$R_{n+1}$};
    \dout{qout}{1.45}{0.6}{$Q_{n+1}$};
    \dout{sout}{1.45}{0.2}{$S_{n+1}$};
    \computonComposite{0}{0}{1.9}{1.6};
    \node at (0.95,1){\footnotesize Next};
    \node at (0.95,0.6){\footnotesize\ref{fig:composite-next}};
 \end{scope}
 \end{tikzpicture}
 }
 \caption{Composite building blocks for the tail-iterative circuit depicted in Figure \ref{fig:iterative-example}. Internal structures are shown on the left and their corresponding modular black boxes are displayed on the right.} 
 \label{fig:iterative-blocks}
\end{figure}

As OR cannot be applied before computing ${\lnot R~\land~Q_n}$, the newly created composite needs to be invoked before passing its result to the OR circuit which, in addition, receives an $S$-value from the external world. In effect, this constitutes another partial sequential circuit, which we fully depict in Figure \ref{fig:composite-iteration}.

The result of \ref{fig:composite-iteration}, $Q_{n+1}$, is intended to be passed on to another circuit to determine the set and reset values for the next iteration. As we aim to implement a cyclic toggle action, we define $R_{n+1}=Q_{n+1}$ and $S_{n+1}=\lnot Q_{n+1}$, and simply echo $Q_{n+1}$ to the next iteration. Accordingly, we construct the partial sequential circuit of Figure \ref{fig:composite-next}, which behaves almost like a buffer, with the difference that $\lnot Q_{n+1}$ is replicated to the external world immediately after computing it.

\begin{figure}[H]
 \centering 
 {
 \begin{tikzpicture}[scale=0.8]
 \computonComposite{0.2}{-0.2}{8.8}{4};
 \begin{scope}[xshift=0cm,yshift=1cm] 
    \qin{qin}{0}{1.4}{};    
    \din{rin}{0}{1}{$R$};
    \din{qin}{0}{0.6}{$Q_n$};    
    \din{sin}{0}{0.2}{$S$};
    \computonComposite{0.8}{0}{1.9}{1.6};
    \node at (1.7,1){\footnotesize Entry};
    \node at (1.7,0.6){\footnotesize\ref{fig:composite-entry}};
 \end{scope}
 \qmatch{eo3}{3.5}{2.4}{};\flow{{2.7,2.4}}{eo3}{dashed}{};\flowdiag{eo3}{{4.3,1.4}}{dashed}{}{rotate=310,pos=0.7}; 
 \dmatch{eo1}{3.5}{2}{}{below};\flow{{2.7,2}}{eo1}{}{};\flowdiag{eo1}{{4.3,1}}{}{}{rotate=310,pos=0.7};
 \dmatch{eo2}{3.5}{1.6}{}{below};\flow{{2.7,1.6}}{eo2}{}{};\flowdiag{eo2}{{4.3,0.6}}{}{}{rotate=310,pos=0.7};
 \dmatch{eo4}{3.5}{1.2}{}{below};\flow{{2.7,1.2}}{eo4}{}{};\flowdiag{eo4}{{4.3,0.2}}{}{}{rotate=310,pos=0.7};
 \begin{scope}[xshift=4.3cm,yshift=0cm]         
    \computonComposite{0}{0}{1.9}{1.6};
    \node at (0.95,1){\footnotesize Action};
    \node at (0.95,0.6){\footnotesize\ref{fig:composite-iteration}};
 \end{scope}
 \begin{scope}[xshift=4.3cm,yshift=2cm]
    \computonComposite{0}{0}{1.9}{1.6};
    \node at (0.95,1){\footnotesize Next};
    \node at (0.95,0.6){\footnotesize\ref{fig:composite-next}};
 \end{scope}
 \flowdiag{{4.3,3.4}}{eo3}{dashed}{}{rotate=225,pos=0.7};
 \flowdiag{{4.3,3}}{eo1}{}{}{rotate=225};
 \flowdiag{{4.3,2.6}}{eo2}{}{}{rotate=225,pos=0.25};
 \flowdiag{{4.3,2.2}}{eo4}{}{}{rotate=225,pos=0.25};

 \qmatch{ao1}{7}{2}{};
 \flowdiag{{6.2,1.2}}{ao1}{dashed}{}{rotate=45};
 \flowdiag{ao1}{{6.2,3}}{dashed}{}{rotate=135}; 
 \flow{ao1}{{7.8,2}}{dashed}{};
 \dmatch{ao2}{7}{1.6}{}{below};
 \flowdiag{{6.2,0.8}}{ao2}{}{}{rotate=45};
 \flowdiag{ao2}{{6.2,2.6}}{}{}{rotate=135};
 \flow{ao2}{{7.8,1.6}}{}{};
 
 \computonPrimitive{7.8}{1.2}{0.6}{1.2}{}
 \node at (8.1,2){\scriptsize $\kappa_1$};\draw[densely dotted] (7.8,2) -- (7.93,2);\draw[densely dotted] (8.25,2) -- (8.4,2);
 \node at (8.1,1.6){\scriptsize $\epsilon_1$};\draw[densely dotted] (7.8,1.6) -- (7.93,1.6);\draw[densely dotted] (8.2,1.6) -- (8.4,1.6);
 
 \qout{eo1}{8.4}{2}{};
 \qout{eo2}{8.4}{1.6}{};
 \end{tikzpicture}
 }
 {
 \begin{tikzpicture}
\begin{scope}
\draw[draw=black,fill=white] (0,0.4) rectangle ++(0.12,0.25);
\node[inner sep=0pt,minimum size=0pt,label=right:{\scriptsize Computation unit}] at (0.2,0.5) {};
\node[draw=black,fill=white,inner sep=0pt,minimum size=3pt,label=right:{\scriptsize Control inport}] at (3.5,0.5) {};
\node[circle,draw=black,fill=white,inner sep=0pt,minimum size=3pt,label={right:{\scriptsize Boolean inport}}] at (6.8,0.5) {};
\end{scope}

\begin{scope}
\draw[dashed] (0,0) to node[pos=0.5]{\arrowflow} (0.6,0);
\node[inner sep=0pt,minimum size=3pt,label=right:{\scriptsize Control flow}] at (0.6,0) {};
\node[fill=black,inner sep=0pt,minimum size=3pt,label={right:{\scriptsize Control outport}}] at (3.5,0) {};
\node[circle,fill=black,inner sep=0pt,minimum size=3pt,label={right:{\scriptsize Boolean outport}}] at (6.8,0) {};
\end{scope}

\begin{scope}[yshift=-0.5cm]
\draw (0,0) to node[pos=0.5]{\arrowflow} (0.5,0);
\node[inner sep=0pt,minimum size=3pt,label=right:{\scriptsize Boolean flow}] at (0.5,0){};
\node[draw,fill=white,fill fraction={black}{0.5},inner sep=0pt,minimum size=3pt,label=right:{\scriptsize Control inoutport}] at (3.5,0) {};
\node[circle,draw,fill=white,fill fraction={black}{0.5},inner sep=0pt,minimum size=3pt,label=right:{\scriptsize Boolean inoutport}] at (6.8,0) {};
\end{scope}
\end{tikzpicture}
 }
 \caption{A tail-iterative circuit to compute the toggle action of a clocked SR flip-flop. Like any other iterative composite, this one is not built from sequencing, parallelising or branching. Instead, a self-contained colimit operation based on pushout and coproduct constructions is used at once, as described in Definition \ref{def:tail}. For this reason, we only show the building blocks and the top-level composite. Relative to the notation used in Definition \ref{def:tail}, the building blocks \emph{Entry}, \emph{Next} and \emph{Action} correspond to the symbols $\lambda_2$, $\lambda_3$ and $\lambda$, respectively. The remaining atomic circuit corresponds to $\lambda_4$, which simply serves to absorb both a control signal and a Boolean value.} 
 \label{fig:iterative-example}
\end{figure}
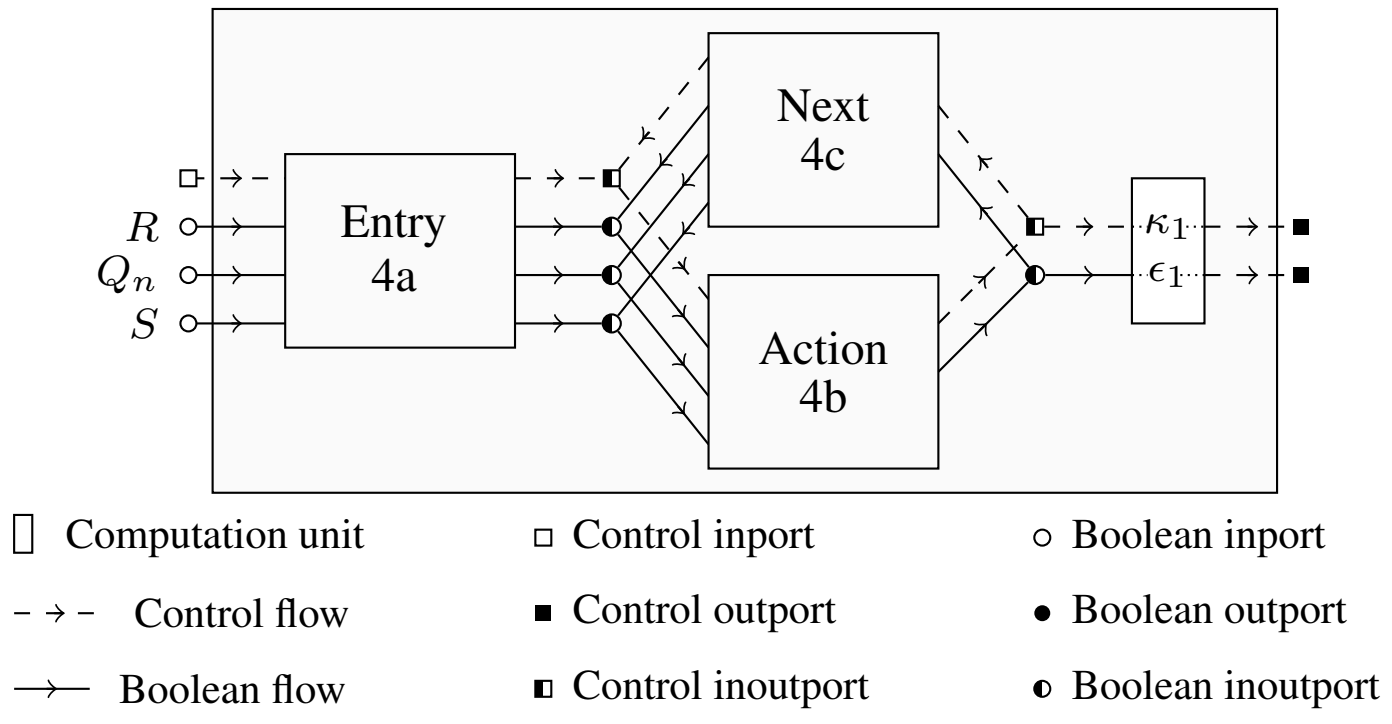

All the building blocks from Figure \ref{fig:iterative-blocks} and an atomic eater are used as operands for the colimit operation described in Definition \ref{def:tail}, effectively forming the tail-iterative circuit from Figure \ref{fig:iterative-example} which captures the behaviour of our cyclic toggle action. A glance at Definition \ref{def:tail} reveals that tail-iterative composites can only be created from connected circuits. By simply following the flows from inports to outports in each building block, one can easily verify connectivity. 

Below we show the characteristic table of our iterative circuit, where it is clear that, having ${Q_n=S=R=0}$ as the initial state, makes the circuit toggle from the second iteration by triggering a set operation. If we rather start with $Q_n=1$ and $S=R=0$, the composite will also toggle from the second iteration. But this time, from a reset operation. Starting with an invalid state will always trigger reset in the second step. Once a set or a reset state is reached, the iterative circuit toggles until termination. Termination occurs randomly at any iteration, effectively simulating the moment clock signals are stopped being produced. 

\begin{table}[H]
\centering
\resizebox{0.8\textwidth}{!}{
\begin{tblr}{
    colspec = {|c|c|c|c|c|c|c|}
  }
 \hline
 \footnotesize$S_n$ & \footnotesize$R_n$ & \footnotesize$Q_n$ & \footnotesize$S_{n+1}=\lnot Q_{n+1}$ & \footnotesize$R_{n+1}=Q_{n+1}$ & \footnotesize$Q_{n+1}=S\lor(\lnot R \land Q_n)$ & \\
 \hline
 \footnotesize0 & \footnotesize0 & \footnotesize0 & \footnotesize1 & \footnotesize0 & \footnotesize0 & \SetCell[r=2]{m}\footnotesize Hold\\
 \hline
 \footnotesize0 & \footnotesize0 & \footnotesize1 & \footnotesize0 & \footnotesize1 & \footnotesize1 & \\
 \hline
 \footnotesize0 & \footnotesize1 & \footnotesize0 & \footnotesize1 & \footnotesize0 & \footnotesize0 & \SetCell[r=2]{m} \footnotesize Reset\\
 \hline
 \footnotesize0 & \footnotesize1 & \footnotesize1 & \footnotesize1 & \footnotesize0 & \footnotesize0 & \\
 \hline
 \footnotesize1 & \footnotesize0 & \footnotesize0 & \footnotesize0 & \footnotesize1 & \footnotesize1 & \SetCell[r=2]{m}\footnotesize Set\\
 \hline
 \footnotesize1 & \footnotesize0 & \footnotesize1 & \footnotesize0 & \footnotesize1 & \footnotesize1 & \\
 \hline
 \footnotesize1 & \footnotesize1 & \footnotesize0 & \footnotesize0 & \footnotesize1 & \footnotesize1 & \SetCell[r=2]{m}\footnotesize Invalid\\
 \hline
 \footnotesize1 & \footnotesize1 & \footnotesize1 & \footnotesize0 & \footnotesize1 & \footnotesize1 & \\
 \hline
\end{tblr}
}
\caption{Characteristic table of the tail-iterative circuit from Figure \ref{fig:iterative-example}.}
\label{tab:iterative-table}
\end{table}

Although our tail-iterative circuit checks for termination immediately after executing \ref{fig:composite-iteration}, it is possible to define a head-iterative composite instead. To accomplish this, we can use the same building blocks as operands. But this time, on the operator described in Definition \ref{def:head}.

\section{Related Work}\label{sec:related-work}

In \cite{arellanes_compositional_2026}, we introduced the theoretical foundations of the computon model on which the present work builds upon. Here, we introduce a concrete instantiation of computons for simulating Boolean functions of arbitrary input length, moving beyond the abstract formulation of computation units towards an operationally defined model based on abstract computing devices. Unlike the original treatment \cite{arellanes_compositional_2026}, we specify units' behaviour in terms of particular data types and provide explicit, implementable execution semantics. Particularly, Definition \ref{def:circuit} characterises units as constructs that encapsulate families of devices for echoing control signals ($\kappa$), discarding Boolean data ($\epsilon$), performing NAND operations ($\uparrow$) or any combination thereof. Their operational semantics are described in Section \ref{sec:operation}. 

In \cite{arellanesboolean2026}, only a family of $\uparrow$-devices is considered so that a single $n$-ary device can be implicitly chosen. We address this expressivity limitation by introducing the set $D$ together with the function pipeline $I\xrightarrow{r}O\xrightarrow{f}D$ and the equation ${\sigma\circ r = \tau}$. In addition, we introduce the notion of \emph{atomic circuits} which generalise the so-called \emph{primitive circuits}; thereby, allowing more expressive computation units (e.g., now it is possible to replicate values within a single unit). Regarding compositionality, we revise the semantics of parallel and branching composition, eliminating the need for separate operators for synchronous parallelising and adding the operator for open branching for more expressive non-deterministic structures. Although our definition of open branching is equivalent to the one introduced in \cite{arellanescolimit2026}, we now provide explicit proofs to show that the associativity and commutativity properties are satisfied for both open and closed branching. Furthermore, we provide concrete examples of control-driven circuits in the domains of propositional logic, biomedicine and electronics, demonstrating the applicability of the proposal across diverse real-world scenarios. We also develop a more comprehensive transformation procedure from classical to control-driven circuits, establishing equivalence in terms of operational semantics and accommodating $n$-ary gates beyond binary ones (Section \ref{sec:equivalence}). In fact, the formulation of circuits is implementable, given that concrete computing devices, with input-output relationships have been introduced. The model has been implemented in both Idris \cite{arellanes2026idris} and Java \cite{arellanes2026java}, with the latter being the most stable version to date, capable of simulating arbitrary Boolean circuits.

Although classical Boolean circuits constitute a well-established model of computation, their composition capabilities are limited to implicit sequencing and implicit parallelising \cite{wegener_complexity_1987}. In this representation, control flow is not explicitly distinguished from data flow, making the order of execution difficult to identify independently of the circuit itself. A conventional mechanism for introducing branching is the multiplexer, or its dual, the demultiplexer, which selects between inputs or directs outputs according to a designated \emph{select line} \cite{lozhkin_complexity_2021,asadi_multiplexer_2024}. Other extensions introduce specialised components to capture stochastic or probabilistic behaviour. Examples include noisy gates \cite{pippenger_networks_1985,mozeika_reliable_2015}, random flip-flops \cite{stipcevic_quantum_2016} and stochastic multiplexers \cite{brown_stochastic_2001,zhu_stochastic_2014}. Such constructions can represent probability distributions over computational outcomes and support the simulation of probabilistic Turing machines or random Boolean functions. Probabilistic logic has also been investigated as a means of modelling non-determinism within Boolean circuits \cite{mansinghka_stochastic_2008,chakrapani_probabilistic_2008,flaminio_conditional_2023}. These approaches, however, typically rely on specialised circuit constructs rather than providing a general algebraic account of control flow.

Iteration presents a further challenge for the conventional circuit model. One approach is to introduce feedback by allowing gate outputs to be routed to the inputs of earlier gates, thereby obtaining cyclic Boolean circuits \cite{riedel_cyclic_2012}. When memory components (e.g., D-type flip-flops or latches) are incorporated, circuit outputs can depend on both present and past inputs. In these systems, feedback supports the storage and updating of internal states \cite{steinbach_sequential_2022,nishimura_classification_2023}. Beyond these hardware-oriented approaches, categorical methods have been developed to offer a theory of compositional cyclic circuits. In particular, circuits can be represented as morphisms in freely generated symmetric traced monoidal categories \cite{katis_feedback_2002,ghica_categorical_2016,ghica_diagrammatic_2017,di_lavore_canonical_2021,sprunger_differentiable_2025}. Their graphical syntax, given by string diagrams \cite{piedeleu_introduction_2025}, interprets trace as feedback from outputs to inputs. This representation yields an intuitive diagrammatic language for expressing circuit structure and reasoning about the associated equational theory \cite{ghica_complete_2025}. Nevertheless, the availability of a formal theory of feedback does not provide explicit composition operators for every form of control flow, particularly branching and partial sequential composition.

Equational theories offer a principled means of specifying the laws governing the construction and manipulation of algebraic structures. In the context of Boolean circuits, composition originates in Boolean algebra and permits logic gates with compatible interfaces to be combined into larger computational components. When the output of one gate feeds into another gate, the resulting composition defines total sequencing \cite{mansinghka_stochastic_2008}. The abstraction principle \cite{turner_new_1979} further supports modularity by allowing a composite to be treated as a single unit, without requiring access to its internal details. Although these principles underpin hierarchical circuit design, a comprehensive algebraic framework must account not only for sequential and parallel composition but also for branching, partial sequencing and iteration.

Several approaches have sought to formalise compositional circuit construction through algebra. Lafont \cite{lafont_towards_2003}, for example, develops an algebra generated by circuit gates, with operations corresponding to ordinary function composition and parallel composition over disjoint maps. While this framework captures total sequencing and parallelising, it does not provide algebraic constructions for branching, partial sequencing or iteration. Subsequent work on categorical and diagrammatic circuit semantics \cite{ghica_categorical_2016,ghica_diagrammatic_2017} takes a step towards a more general account, but relies on equations motivated by diagrammatic intuition rather than a fully developed formal treatment. More recently, \cite{ghica_complete_2025} constructs freely generated symmetric traced monoidal categories from circuit signatures, enabling arbitrary circuits to be assembled through morphism composition, tensor product and traces. These operations provide a unified treatment of total sequencing, parallelising and feedback-induced iteration, but do not explicitly support branching or partial sequencing.

The absence of partial sequencing in such frameworks is derived from the way symmetric monoidal categories represent computational structure. In particular, their conventional string-diagram syntax does not inherently distinguish control signals from data signals through separate classes of wires \cite{bonchi_string_2022}. Thus, limiting the direct representation of control-dependent execution for independent composition. An extension proposed in \cite{villoria_enriching_2025} enriches monoidal categories over Eilenberg--Moore algebras to associate probabilities with wires; thereby, representing probabilistic branching. However, this enrichment does not resolve the lack of explicit support for partial sequencing. Higher-dimensional automata offer an alternative framework capable of representing partial sequencing \cite{fahrenberg_languages_2021,fahrenberg_kleene_2022}. In these models, however, partial sequential execution is captured implicitly through state-transition activation rather than through dedicated composition operators. Since control and data flow are not independently represented, modular reasoning about these two dimensions remains limited.

The hierarchical organisation of circuits also resembles categorical approaches in cybernetics, where complex systems are assembled from smaller components. For example, \cite{ehresmann_memory_2007} employs general colimits to construct complex systems hierarchically, following a bottom-up approach to component integration. Although this supports structural system composition, it does not provide explicit control flow constructs for governing concrete causal and informational interactions. The approach in \cite{arellanes_composition_2024} addresses aspects of this limitation by supporting the dynamic formation of spaces of total sequential composites. In that setting, sequential composition emerges implicitly from the dynamics of self-organising spaces rather than being defined through explicit colimit constructions.

The approach developed in this paper differs from the current literature in the sense it treats control flow as a first-class compositional structure rather than encoding it through specialised gates or relying exclusively on the implicit dynamics of circuit execution. Sequential composition, parallelising, branching and iteration are expressed through algebraic, colimit-based operators that support the hierarchical construction of circuits from simpler ones. The resulting framework supports compositional (and hence modular) reasoning in the presence of feedback, while allowing deterministic and non-deterministic computations to coexist within a common framework. Crucially, the explicit orthogonality between control flow and data flow enables these aspects to be specified and analysed independently, providing a unified algebraic basis for Boolean circuits.

\section{Conclusions and Future Work}\label{sec:conclusion}

This paper introduces control-driven Boolean circuits as a non-uniform model of high-level computation in which Boolean functions are realised through composites that exhibit explicit control flow and data flow. The central contribution is a collection of elementary colimit-based operators for constructing circuits via partial sequencing, total sequencing, (asynchronous) parallelising, (non-deterministic) branching and iteration. Section \ref{sec:model} establishes some algebraic properties of these operators, showing that all but parallelising and open branching satisfy the identity law, whereas commutativity is established for parallelising and both forms of branching. Parallelising, (open and closed) branching and total sequencing satisfy the associativity property in addition. Together, these results provide an algebraic basis for constructing and reasoning about Boolean circuits. 

Section \ref{sec:operation} develops the operational semantics of our proposal by which control flow determines the order in which abstract computing devices are evaluated. From a high-level viewpoint, each unit acts as a passive encapsulation construct that coordinates control signals with Boolean data, which can be evaluated only when its input data becomes available. Evaluation entails triggering all the devices within a unit on the available inputs, before producing further Boolean values and control signals. Although the model describes a particular transition system, a general theory relating algebraic composition to operational semantics remains to be developed. Future work will investigate how colimit-based composition can be used to establish guarantees about execution, including studying the conditions under which termination can be ensured. Although deadlock freedom has already been considered in \cite{arellanes_compositional_2026}, termination needs further formal investigation, since the absence of deadlocks alone does not necessarily exclude non-termination.

In Section \ref{sec:equivalence}, we establish that classical Boolean circuits can be represented within the proposed framework by means of the construction given in the proof of Theorem \ref{th:classical-to-control} together with the behaviour equivalence presented in Theorem \ref{th:gate-firing}. Although these results demonstrate that our proposal can express the same computations as classical Boolean circuits, the hierarchical construction of resulting composites is not immediately apparent. A systematic approach to decompose composites into their constituent colimit operations is therefore an important direction for future research. We envisage introducing dedicated \emph{decomposition operators} to identify the operations required to construct a given composite and, where possible, to determine minimal decompositions.

Unfortunately, the construction from Theorem \ref{th:classical-to-control} introduces a number of control ports that exceeds the number of edges of the original classical circuit. This overhead motivates the development of alternative transformation procedures with the aim of reducing the size of resulting control-driven circuits while preserving computational behaviour. For the present work, the main aim is to establish expressive adequacy rather than translation efficiency. Specifically, Theorem \ref{th:computons-any-boolean} establishes that the model can represent arbitrary Boolean functions through suitable circuit families; thereby, demonstrating that its expressive power is at least that of classical families of Boolean circuits. This result should be understood in the non-uniform sense, i.e., we establish representability for each input length, without providing a uniform or efficient procedures for constructing the corresponding circuit family.

Although the use of the functionally complete set $\{\uparrow\}$ offers a simplified representation of computing devices, other functionally complete bases could be adopted without changing the underlying expressive-power argument, provided that the corresponding device semantics are defined appropriately. Beyond establishing computational power, we showed that the framework makes control flow explicit; thus, contrasting with classical circuits in which control is implicitly encoded within the network of logic gates. Control-driven circuits define control flow through dedicated composition operators. This distinction can support modular reasoning about the order in which (encapsulated) computing devices are activated, without requiring internal circuit inspection.

To the best of our knowledge, our framework provides a unified algebraic treatment of total and partial sequencing, asynchronous parallelising, non-deterministic (open and closed) branching and (head and tail) iteration within a single model of Boolean circuits. Its distinguishing feature is not merely the availability of these individual operators, but their integration into a formally specified composition framework grounded in finite colimit constructions. Although this integration offers a basis for hierarchical circuit construction and algebraic reasoning, identifying the classes of Boolean functions that can be represented through iteration is an important topic for further study.

The current treatment of branching also leaves open questions concerning the representation of input-dependent decision-making and probabilistic computation. Although branching provides a structural mechanism for representing non-deterministic computational paths, the ability to select paths according to external inputs requires an explicit account of how data values could constrain control transitions. We therefore plan to extend the semantics of computation units with pre- and post-conditions, enabling branching decisions that depend on user-defined conditions. Such an extension would support a more direct representation of conditional computation while preserving the separation of control and data flow. Although implementations are currently available \cite{arellanes2026idris,arellanes2026java,arellanes2026tutorial}, a complementary practical direction is the development of a visual programming environment for constructing and exploring compositional control-driven circuits. The Java implementation provides a first step by allowing the automatic generation of circuit diagrams \cite{arellanes2026java}.

Finally, extending the model with probabilistic choice would provide a principled way to represent randomised computation. Such an extension would require an explicit probabilistic semantics for selecting computational paths, rather than relying on non-deterministic branching alone. It could then support the representation and study of probabilistic Boolean computations and, subject to appropriate encoding, the simulation of probabilistic Turing machines. Together, these developments would extend the framework towards a more general composition theory of deterministic, non-deterministic and probabilistic computation.

\bibliographystyle{fundam}
\bibliography{refs}

@inproceedings{arellanes_composition_2024,
	title = {Composition {Machines}: {Programming} {Self}-organising {Software} {Models} for the {Emergence} of {Sequential} {Program} {Spaces}},
	shorttitle = {Composition {Machines}},
	booktitle = {18th {International} {Symposium} on {Theoretical} {Aspects} of {Software} {Engineering} ({TASE})},
	publisher = {Springer},
	author = {Arellanes, Damian},
	year = {2024},
	pages = {19--37},
}

@misc{ghica_complete_2025,
	title = {A {Complete} {Theory} of {Sequential} {Digital} {Circuits}: {Denotational}, {Operational} and {Algebraic} {Semantics}},
	shorttitle = {A {Complete} {Theory} of {Sequential} {Digital} {Circuits}},
	publisher = {arXiv},
	author = {Ghica, Dan R. and Kaye, George and Sprunger, David},
	year = {2025},
}

@incollection{giannakopoulou_compositional_2018,
	address = {Cham},
	title = {Compositional {Reasoning}},
	booktitle = {Handbook of {Model} {Checking}},
	publisher = {Springer},
	author = {Giannakopoulou, Dimitra and Namjoshi, Kedar S. and Păsăreanu, Corina S.},
	editor = {Clarke, Edmund M. and Henzinger, Thomas A. and Veith, Helmut and Bloem, Roderick},
	year = {2018},
	pages = {345--383},
}

@inproceedings{dardik_recomposition_2024,
	title = {Recomposition: {A} {New} {Technique} for {Efficient} {Compositional} {Verification}},
	booktitle = {24th {Conference} on {Formal} {Methods} in {Computer}-{Aided} {Design} ({FMCAD})},
	author = {Dardik, Ian and Porter, April and Kang, Eunsuk},
	year = {2024},
	pages = {130--141},
}

@article{arellanes_compositional_2026,
	title = {Compositional {Separation} of {Control} {Flow} and {Data} {Flow}},
	journal = {Journal of Logical and Algebraic Methods in Programming},
	author = {Arellanes, Damian},
	year = {2026},
	eid = {101125},
}

@article{lozhkin_complexity_2021,
	title = {The {Complexity} of the {Standard} {Multiplexer} {Function} in a {Class} of {Switching} {Circuits}},
	volume = {32},
	number = {4},
	journal = {Comput Math Model},
	author = {Lozhkin, S. A. and Khzmalyan, D. E.},
	year = {2021},
	pages = {478--489},
}

@incollection{asadi_multiplexer_2024,
	address = {Cham},
	series = {Synthesis {Lectures} on {Electrical} {Engineering}},
	title = {Multiplexer and {De}-multiplexer},
	booktitle = {Digital {Circuits} {Laboratory} {Manual}},
	publisher = {Springer},
	author = {Asadi, Farzin},
	year = {2024},
	pages = {81--104},
}

@book{wegener_complexity_1987,
	edition = {1st},
	title = {The complexity of {Boolean} functions},
	publisher = {John Wiley \& Sons},
	author = {Wegener, Ingo},
	year = {1987},
}

@book{clote_boolean_2002,
	address = {Berlin, Heidelberg},
	edition = {1st},
	series = {Texts in {Theoretical} {Computer} {Science}. {An} {EATCS} {Series}},
	title = {Boolean {Functions} and {Computation} {Models}},
	publisher = {Springer},
	author = {Clote, Peter and Kranakis, Evangelos},
	editor = {Brauer, Wilfried and Rozenberg, Grzegorz and Salomaa, Arto},
	year = {2002},
}

@article{mozeika_reliable_2015,
	title = {On {Reliable} {Computation} by {Noisy} {Random} {Boolean} {Formulas}},
	volume = {61},
	number = {1},
	journal = {IEEE Transactions on Information Theory},
	author = {Mozeika, Alexander and Saad, David},
	year = {2015},
	pages = {637--644},
}

@article{stipcevic_quantum_2016,
	title = {Quantum random flip-flop and its applications in random frequency synthesis and true random number generation},
	volume = {87},
	number = {3},
	journal = {Review of Scientific Instruments},
	author = {Stipčević, Mario},
	year = {2016},
	pages = {1--18},
}

@inproceedings{pippenger_networks_1985,
	title = {On networks of noisy gates},
	booktitle = {26th {Annual} {Symposium} on {Foundations} of {Computer} {Science} ({SFCS})},
	author = {Pippenger, Nicholas},
	year = {1985},
	pages = {30--38},
}

@article{zhu_stochastic_2014,
	title = {Stochastic {Multiple}-{Valued} {Gene} {Networks}},
	volume = {8},
	number = {1},
	journal = {IEEE Transactions on Biomedical Circuits and Systems},
	author = {Zhu, Peican and Han, Jie},
	year = {2014},
	pages = {42--53},
}

@article{brown_stochastic_2001,
	title = {Stochastic neural computation. {I}. {Computational} elements},
	volume = {50},
	number = {9},
	journal = {IEEE Transactions on Computers},
	author = {Brown, B.D. and Card, H.C.},
	year = {2001},
	pages = {891--905},
}

@techreport{mansinghka_stochastic_2008,
	address = {Cambridge, MA, USA},
	title = {Stochastic {Digital} {Circuits} for {Probabilistic} {Inference}},
	number = {MIT-CSAIL-TR-2008-069},
	institution = {Massachusetts Institute of Technology},
	author = {Mansinghka, Vikash and Jonas, Eric and Tenenbaum, Josh},
	year = {2008},
	pages = {1--10},
}

@techreport{chakrapani_probabilistic_2008,
	address = {Houston, Texas, USA},
	title = {A {Probabilistic} {Boolean} {Logic} and its {Meaning}},
	number = {TR08-05},
	institution = {Rice University, Department of Computer Science},
	author = {Chakrapani, Lakshmi N. B. and Palem, Krishna V},
	year = {2008},
}

@article{flaminio_conditional_2023,
	title = {On conditional probabilities and their canonical extensions to {Boolean} algebras of compound conditionals},
	volume = {159},
	journal = {International Journal of Approximate Reasoning},
	author = {Flaminio, Tommaso and Gilio, Angelo and Godo, Lluis and Sanfilippo, Giuseppe},
	year = {2023},
	pages = {1--31},
}

@article{riedel_cyclic_2012,
	title = {Cyclic {Boolean} circuits},
	volume = {160},
	number = {13},
	journal = {Discrete Applied Mathematics},
	author = {Riedel, Marc D. and Bruck, Jehoshua},
	year = {2012},
	pages = {1877--1900},
}

@article{nishimura_classification_2023,
	title = {Classification of {Sequential} {Circuits} as {Causal} {Functions}},
	volume = {2613},
	number = {1},
	journal = {Journal of Physics: Conference Series},
	author = {Nishimura, Shunji},
	year = {2023},
	pages = {012016},
}

@inproceedings{ghica_categorical_2016,
	title = {Categorical semantics of digital circuits},
	booktitle = {16th {Conference} on {Formal} {Methods} in {Computer}-{Aided} {Design} ({FMCAD})},
	author = {Ghica, Dan R. and Jung, Achim},
	year = {2016},
	pages = {41--48},
}

@inproceedings{ghica_diagrammatic_2017,
	address = {Dagstuhl, Germany},
	series = {Leibniz {International} {Proceedings} in {Informatics} ({LIPIcs})},
	title = {Diagrammatic {Semantics} for {Digital} {Circuits}},
	volume = {82},
	booktitle = {26th {EACSL} {Annual} {Conference} on {Computer} {Science} {Logic} ({CSL})},
	publisher = {Schloss Dagstuhl – Leibniz-Zentrum für Informatik},
	author = {Ghica, Dan R. and Jung, Achim and Lopez, Aliaume},
	editor = {Goranko, Valentin and Dam, Mads},
	year = {2017},
	pages = {1--16},
}

@article{katis_feedback_2002,
	series = {Fixed {Points} in {Computer} {Science}},
	title = {Feedback, trace and fixed-point semantics},
	volume = {36},
	number = {2},
	journal = {RAIRO - Theoretical Informatics and Applications},
	author = {Katis, P. and Sabadini, Nicoletta and Walters, Robert F. C.},
	year = {2002},
	pages = {181--194},
}

@inproceedings{di_lavore_canonical_2021,
	address = {Cham},
	title = {A {Canonical} {Algebra} of {Open} {Transition} {Systems}},
	booktitle = {17th {International} {Conference} on {Formal} {Aspects} of {Component} {Software} ({FACS})},
	publisher = {Springer},
	author = {Di Lavore, Elena and Gianola, Alessandro and Román, Mario and Sabadini, Nicoletta and Sobociński, Paweł},
	editor = {Salaün, Gwen and Wijs, Anton},
	year = {2021},
	pages = {63--81},
}

@article{turner_new_1979,
	title = {A {New} {Implementation} {Technique} for {Applicative} {Languages}},
	volume = {9},
	number = {1},
	journal = {Software - Practice and Experience},
	author = {Turner, D. A.},
	year = {1979},
	pages = {31--49},
}

@article{lafont_towards_2003,
	title = {Towards an algebraic theory of {Boolean} circuits},
	volume = {184},
	number = {2},
	journal = {Journal of Pure and Applied Algebra},
	author = {Lafont, Yves},
	year = {2003},
	pages = {257--310},
}

@incollection{steinbach_sequential_2022,
	address = {Cham},
	title = {Sequential {Circuits}},
	booktitle = {Logic {Functions} and {Equations}: {Fundamentals} and {Applications} using the {XBOOLE}-{Monitor}},
	publisher = {Springer},
	author = {Steinbach, Bernd and Posthoff, Christian},
	editor = {Steinbach, Bernd and Posthoff, Christian},
	year = {2022},
	pages = {741--794},
}

@book{piedeleu_introduction_2025,
	edition = {1st},
	series = {Elements in {Applied} {Category} {Theory}},
	title = {An {Introduction} to {String} {Diagrams} for {Computer} {Scientists}},
	publisher = {Cambridge University Press},
	author = {Piedeleu, Robin and Zanasi, Fabio},
	year = {2025},
}

@article{sprunger_differentiable_2025,
	title = {Differentiable causal computations via delayed trace (extended version)},
	volume = {35},
	journal = {Mathematical Structures in Computer Science},
	author = {Sprunger, David and Katsumata, Shin-ya},
	year = {2025},
	pages = {1--35},
}

@article{liang_stochastic_2012,
	title = {Stochastic {Boolean} networks: {An} efficient approach to modeling gene regulatory networks},
	volume = {6},
	shorttitle = {Stochastic {Boolean} networks},
	number = {1},
	journal = {BMC Systems Biology},
	author = {Liang, Jinghang and Han, Jie},
	year = {2012},
	pages = {1--20},
}

@article{arellanes_models_2025,
	title = {Models of {High}-{Level} {Computation}},
	volume = {7},
	journal = {Front. Comput. Sci.},
	author = {Arellanes, Damian},
	year = {2025},
	pages = {1--7},
}

@article{bonchi_string_2022,
	title = {String diagram rewrite theory {II}: {Rewriting} with symmetric monoidal structure},
	volume = {32},
	number = {4},
	journal = {Mathematical Structures in Computer Science},
	publisher = {Cambridge University Press},
	author = {Bonchi, Filippo and Gadducci, Fabio and Kissinger, Aleks and Sobocinski, Pawel and Zanasi, Fabio},
	year = {2022},
	pages = {511--541},
}

@inproceedings{villoria_enriching_2025,
	title = {Enriching {Diagrams} with {Algebraic} {Operations}},
	booktitle = {28th {International} {Conference} on {Foundations} of {Software} {Science} and {Computation} {Structures} ({FoSSaCS} 2024)},
	publisher = {Springer},
	author = {Villoria, Alejandro and Basold, Henning and Laarman, Alfons},
	year = {2025},
}

@article{fahrenberg_languages_2021,
	title = {Languages of higher-dimensional automata},
	volume = {31},
	number = {5},
	journal = {Mathematical Structures in Computer Science},
	author = {Fahrenberg, Uli and Johansen, Christian and Struth, Georg and Ziemianski, Krzysztof},
	year = {2021},
}

@inproceedings{fahrenberg_kleene_2022,
	title = {A {Kleene} {Theorem} for {Higher}-{Dimensional} {Automata}},
	booktitle = {33rd {International} {Conference} on {Concurrency} {Theory} ({CONCUR} 2022)},
	author = {Fahrenberg, Uli and Johansen, Christian and Struth, Georg and Ziemianski, Krzysztof},
	year = {2022},
	pages = {1--18},
}

@book{ehresmann_memory_2007,
	edition = {1st},
	title = {Memory {Evolutive} {Systems}; {Hierarchy}, {Emergence}, {Cognition}},
	volume = {4},
	publisher = {Elsevier},
	author = {Ehresmann, A. C. and Vanbremeersch, J. P.},
	year = {2007},
}

@inproceedings{arellanesboolean2026,
	title = {Compositional {Control}-{Driven} {Boolean} {Circuits}},
	author = {Arellanes, Damian},
	year = {2026},
    pages = {20--40},
    booktitle = {22nd {International} {Conference} on {Relational} and {Algebraic} {Methods} in {Computer} {Science} ({RAMICS})}, 
    publisher = {Springer},
    doi = {10.1007/978-3-032-22469-92},
}

@inproceedings{arellanescolimit2026,
	title = {{Colimit}-{Based} {Composition} {of} {High}-{Level} {Computing} {Devices}},
	author = {Arellanes, Damian},
	year = {2026},
    pages = {37--54},
    booktitle = {21st {International} {Symposium} on {Logical} and {Semantic} {Frameworks} with {Applications} ({LSFA})}, 
    publisher = {EPTCS},
}

@book{AroraBarak2009,
  author    = {Sanjeev Arora and Boaz Barak},
  title     = {Computational Complexity: A Modern Approach},
  publisher = {Cambridge University Press},
  year      = {2009},
  address   = {Cambridge},
}

@software{arellanes2026idris,
  author       = {Arellanes, Damian},
  title        = {Idris Implementation of The Computon Model},
  year         = {2026},
  howpublished = {https://github.com/damianarellanes/computons-idris},
}

@software{arellanes2026java,
  author       = {Arellanes, Damian},
  title        = {Java Implementation of The Computon Model},
  year         = {2026},
  howpublished = {https://github.com/damianarellanes/computons-java},
}

@software{arellanes2026tutorial,
  author       = {Arellanes, Damian},
  title        = {Computon Model Tutorial},
  year         = {2026},
  howpublished = {https://github.com/damianarellanes/computons-demo},
}
\end{document}